\documentclass[preprint,12pt]{elsarticle}

\usepackage{amssymb}
\usepackage{setspace}
\usepackage[mathscr]{euscript}
\usepackage[section]{placeins}
\usepackage{amsthm, amsmath, amssymb, mathrsfs, multirow, url, cases}
\usepackage{graphicx} 
\usepackage{ifthen}
\usepackage{amsfonts}
\usepackage[usenames]{color}
\usepackage{fullpage}
\usepackage{makecell}
\usepackage{array, booktabs}
\usepackage{listings}
\usepackage{caption}
\usepackage{relsize}
\usepackage[labelformat=simple]{subcaption}
\usepackage{float}
\usepackage{changepage}
\usepackage{relsize}
\usepackage[modulo, mathlines]{lineno}
\usepackage[svgnames, table]{xcolor}
\usepackage{subfiles}
\usepackage{appendix}
\usepackage{rotating}
\usepackage{algorithm}
\usepackage{algpseudocode}

\newcommand{\V}[1][\textbf]{#1}

\newcommand{\boldbeta}{\boldsymbol{\beta}}

\newcommand{\logof}[1]{\textrm{log}\big(#1\big)}
\DeclareMathOperator*{\argmin}{arg\,min}

\newtheorem{theorem}{Theorem}
\newtheorem{lemma}{Lemma}

\renewenvironment{proof}[1][\proofname]{{\noindent \bfseries #1.}}{\qedsymbol}

\def\spacingset#1{\renewcommand{\baselinestretch}%
{#1}\small\normalsize} \spacingset{1}

\spacingset{1.5} 

\journal{Computational Statistics and Data Analysis}

\begin{document}

\begin{frontmatter}



\title{Tuning Parameter Selection for Penalized Estimation via $R^2$}


\author[inst1]{Julia C. Holter}

\affiliation[inst1]{organization={Department of Statistics},
            addressline={North Carolina State University},
            city={Raleigh},
            state={NC},
            postcode={27695},
            country={USA}}

\author[inst1]{Jonathan W. Stallrich}

\begin{abstract}
The tuning parameter selection strategy for penalized estimation is crucial to identify a model that is both interpretable and predictive. However, popular strategies (e.g.,  minimizing average squared prediction error via cross-validation) tend to select models with more predictors than necessary. This paper proposes a simple, yet powerful cross-validation strategy based on maximizing squared correlations between the observed and predicted values, rather than minimizing squared error loss for the purposes of support recovery. The strategy can be applied to all penalized least-squares estimators and we show that, under certain conditions, the metric implicitly performs a bias adjustment.  Specific attention is given to the Lasso estimator, in which our strategy is closely related to the Relaxed Lasso estimator. We demonstrate our approach on a functional variable selection problem to identify optimal placement of surface electromyogram sensors to control a robotic hand prosthesis.
\end{abstract}



\end{frontmatter}


\section{Introduction}
\label{sec:Intro}

Many statistical problems aim to build a predictive model from a large set of potential predictor variables. Variable selection is often performed to select a predictive model that depends on as few predictor variables as possible. For example, \cite{SAFE} discussed an important functional variable selection problem to develop a prosethesis controller (PC) for a robotic hand. Electromyogram (EMG) signals from surface sensors placed on the residual forearm muscles of an amputee were input into a PC and translated into movement of the robotic hand.  
For able-bodied subjects, it is known that certain movements are caused by contractions of only a few muscles, implying a predictive PC requires a few strategically-placed EMG sensors.




This paper concerns problems that are well approximated by a sparse linear model:
\begin{equation}
    \V{y}_{n\times 1} =  \V{X}_{n \times p} \boldbeta_{p \times 1}^* + \boldsymbol{\epsilon}_{n \times 1}
    \label{eq:LinearModel}
\end{equation}
where $E(\boldsymbol{\epsilon}) = \V{0}$, $V(\boldsymbol{\epsilon})=\boldsymbol{\Sigma}$, ${\boldbeta^*}^T = (\beta_1^*,...,\beta_{p^*}^*,0,...,0)$, and $p^*$ is the number of important variables.  Without loss of generality, assume $\V{y}$ and predictor variables, $\V{X}$, are centered and the diagonals of $\V{X}^T\V{X}$ equal $n$. Let $\mathcal{M}^* = \{j \, : \, \beta_j^* \neq 0\}$ denote the support of $\boldbeta^*$. A predictive model's estimate for $\boldbeta^*$, denoted $\hat{\boldbeta}$, will ideally also have support $\mathcal{M}^*$ and will be close to $\boldbeta^*$ in some other sense, such as $||\hat{\boldbeta}-\boldbeta^*||_2^2=\sum_j (\hat{\beta}_j-\beta_j^*)^2$. 

For high-dimensional data like that in \cite{SAFE}, simultaneous support recovery and parameter estimation can be performed via penalized estimation.  A penalized estimator is represented generally by $\hat{\boldbeta}_\lambda = \argmin L(\boldbeta) + P_\lambda(\boldbeta)$ where $L(\boldbeta)$ is a loss function comparing $\V{y}$ to its predicted values $\V{X}\boldbeta$, and $P_\lambda(\boldbeta)$ is a penalty function that depends on tuning parameter(s) $\lambda \geq 0$. Henceforth we let $L(\boldbeta)=(2n)^{-1}||\V{y} - \V{X}\boldbeta||_2^2$. Penalty functions can take myriad forms but we are interested in those that increase as $\boldbeta$ moves away from $\V{0}$.  The Lasso \citep{Tibshirani1996} penalty, $P_\lambda(\boldbeta) = \lambda ||\boldbeta||_1 = \lambda \sum_{j = 1}^p |\beta_j|$, is one such penalty that can force estimates to equal 0, thereby performing simultaneous variable selection and estimation. For such sparsity-inducing estimators, we are interested in comparing the estimated support, $\mathcal{M}_\lambda=\{j  \, : \, \hat{\beta}_{\lambda,j} \neq 0\}$, to $\mathcal{M}^*$.




The chosen $\lambda$ balances the importance of minimizing $P_\lambda(\boldbeta)$ relative to $L(\boldbeta)$, so it is recommended to explore the tuning parameter space to identify an ``optimal" value. Potential criteria for an optimal value include identifying a $\hat{\boldbeta}_\lambda$ that minimizes $||\hat{\boldbeta}_\lambda-\boldbeta^*||_2^2$, minimizes  $||\V{X}\hat{\boldbeta}_\lambda-\V{X}\boldbeta^*||_2^2$, or has $\mathcal{M}_\lambda = \mathcal{M}^*$. The latter criterion is referred to as support recovery and is the primary focus of this paper. Even if a $\lambda$ exists where $\mathcal{M}_\lambda = \mathcal{M}^*$, there is no guarantee that we will be able to correctly identify it.
Popular approaches, such as minimizing information criteria \citep{Akaike1974,Schwarz1978} or minimizing average squared prediction error from $K$-fold cross validation often choose a $\lambda$ that overselects the number of important variables \citep{ConsistentCV,Hastie2017}, i.e., $\mathcal{M}^* \subset \mathcal{M}_\lambda$.  Post-selection inference techniques \citep{CovarianceTest,DebiasedLasso} and multi-stage modifications \citep{AdaptiveLasso,Relaxo} can correct for this overselection, albeit with added computations and assumptions.

\begin{figure}[ht]
     \centering
     \begin{subfigure}[b]{0.22\textwidth}
         \centering
         \caption{\textbf{CV 1SE}
         }
         \includegraphics[width=\textwidth]{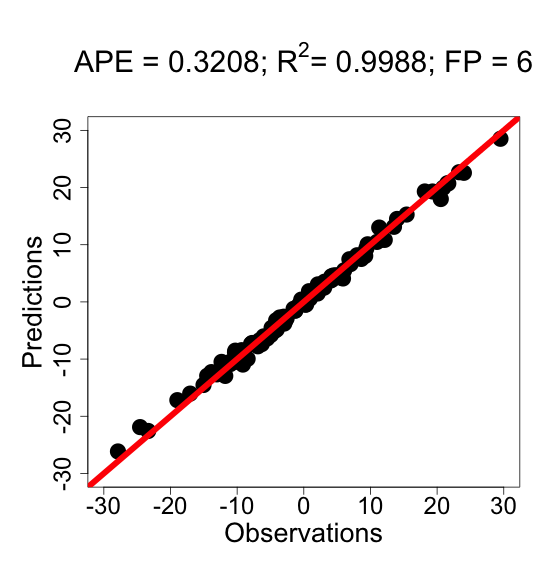}
         \label{fig:Motiv1SE}
     \end{subfigure}
    \hfill
     \begin{subfigure}[b]{0.22\textwidth}
         \centering
         \caption{\textbf{Min PB} 
         }
         \includegraphics[width=\textwidth]{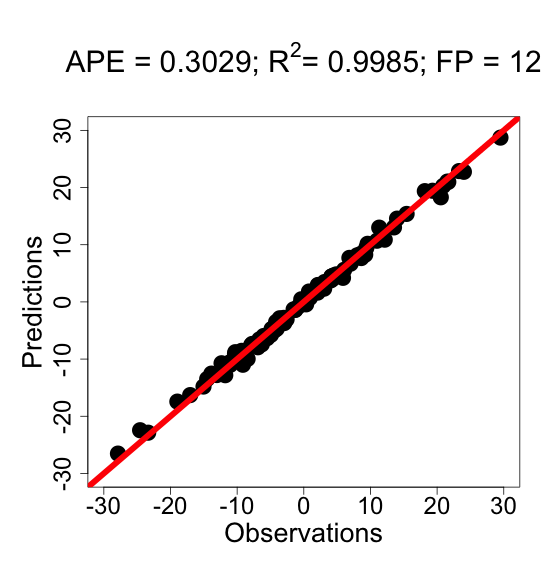}
         \label{fig:MotivMinPB}
     \end{subfigure}
     \hfill
     \begin{subfigure}[b]{0.22\textwidth}
         \centering
         \caption{\textbf{Max $\lambda$, $\mathcal{M}_\lambda=\mathcal{M}^*$} 
         }
         \includegraphics[width=\textwidth]{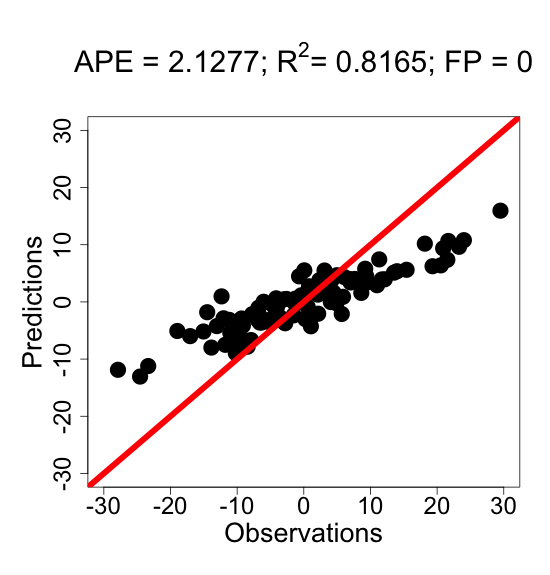}
         \label{fig:MotivPerfectLarge}
     \end{subfigure}
    \hfill
     \begin{subfigure}[b]{0.22\textwidth}
         \centering
         \caption{\textbf{Min $\lambda$, $\mathcal{M}_\lambda=\mathcal{M}^*$} 
         }
         \includegraphics[width=\textwidth]{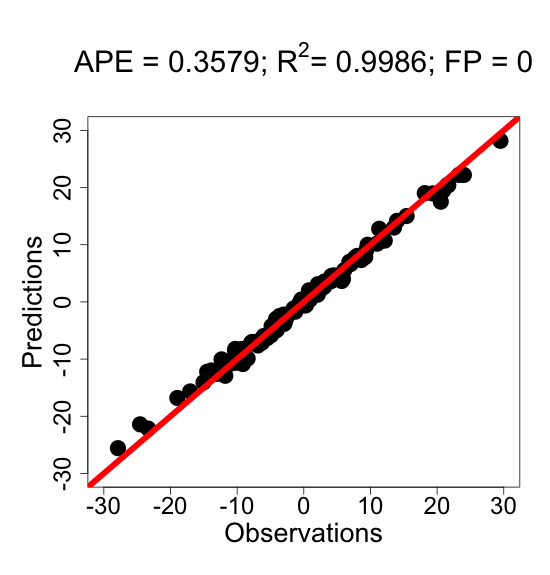}
         \label{fig:MotivPerfectSmall}
     \end{subfigure}
        \caption{In-sample observations ($y_i$) versus predictions ($\hat{y}_i$) for Lasso estimator at four values of $\lambda$ for a data set with $n=100$, $p = 100$, and $p^* = 5$. We also report average prediction error (APE) under 10-fold CV, in-sample squared  prediction correlation ($R^2$), and the number of false positives (FP). The minimum APE plus-or-minus one standard error is $0.2844 \pm 0.0262$ and all estimates have $\mathcal{M}^* \subseteq \mathcal{M}_\lambda$.}
        \label{fig:Motivation}
\end{figure}

This paper proposes a new $K$-fold cross validation strategy that assesses the predictive quality of $\hat{\boldbeta}_\lambda$ by maximizing average squared prediction correlation rather than minimizing average squared prediction error. The scale invariance of correlation reduces the impact of the potential shrinkage when estimating large $\beta^*_j$ that would burgeon squared prediction error.  To demonstrate, we simulated a dataset $\{\V{y},\V{X}\}$ with $n=p=100$, $p^* = 5$, and $\sigma^2 = 1$, and generated the entire solution path of the Lasso. The elements of $\V{X}$ were independently sampled from a standard Normal distribution and then appropriately centered and scaled. The active coefficients in $\boldbeta^*$ were $\{2.13 , 1.81, -2.46, -1.89, -2.51  \}$. 
Many $\lambda$ values had $\mathcal{M}_\lambda=\mathcal{M}^*$, so a wide range were optimal for support recovery.  Figure \ref{fig:Motivation} plots $\V{y}$ against their in-sample predictions, $\hat{\V{y}}_\lambda$, under four different $\lambda$ values, and summarizes the models' number of false positives (FP), average squared prediction error (APE) from 10-fold CV, and in-sample squared prediction correlation ($R^2$). Figure \ref{fig:Motiv1SE} corresponds to the $\lambda$ determined by a cross validation one-standard-error rule (CV 1SE) and Figure~\ref{fig:MotivMinPB} corresponds to the $\lambda$ that minimizes $||\V{X}\boldbeta^* - \V{X}\hat{\boldbeta}_\lambda||_2$ (Min PB). Among the $\lambda$ having $\mathcal{M}_\lambda=\mathcal{M}^*$, we consider the largest and smallest in Figures \ref{fig:MotivPerfectLarge} and  \ref{fig:MotivPerfectSmall}, respectively.


All models shown in Figure~1 include the 5 important variables. Both the CV 1SE and Min PB models have small APE and large $R^2$, but many false positives.
The model in Figure \ref{fig:MotivPerfectLarge} has no false positives, but has relatively large APE. The model under Figure \ref{fig:MotivPerfectSmall} also has no false positives and its in-sample $R^2$ approximately equals that of the Min PB estimate. This motivates a tuning parameter selection strategy based on an $R^2$ metric rather than APE to compromise between prediction and variable selection.

After justifying the $R^2$ metric, we highlight and investigate an equivalence between the metric and a multiplicative adjustment on $\hat{\boldbeta}_\lambda$, referred to here as the $\alpha$-modification. We argue that for $\hat{\boldbeta}_\lambda$ with certain statistical properties, the adjustment can reduce the bias of $\hat{\boldbeta}_\lambda$ thereby improving its predictive potential. We go on to study the $\alpha$-modification for the Lasso, highlighting its similarities to the Nonnegative Garrote \citep{NNGarotte} and Relaxed Lasso\citep{Relaxo}.  Unlike these two methods, the $\alpha$-modification can be applied to any penalized least-squares estimator, including non-convex penalties, without additional computational complexity.




This paper is organized as follows. In Section \ref{sec:Background} we review classes of penalized estimators, popular tuning parameter selection strategies, and post-selection inference methods. Section \ref{sec:Methodology} justifies the value in the $R^2$ metric and provides statistical properties for a general class of penalized estimators. Finite-sample properties are then derived for the $\alpha$-modification for the Lasso in Section \ref{sec:LassoResults}. Section \ref{sec:Simulations} presents a simulation study of the new approaches and Section \ref{sec:EMGApplication} applies our new methods to the EMG data of \cite{SAFE}. Section \ref{sec:Discussion} provides a discussion on the implications of our new framework for evaluating model fit and propose avenues of future research. Proofs of all results may be found in the  Supplementary Materials.

\section{Background}
\label{sec:Background}



\subsection{Classes of Penalized Estimators}


Consider the class of penalties $P_\lambda(\boldbeta)=||\boldbeta||_q^q=\sum_j |\beta_j|^q$ , $q>0$, corresponding to the so-called bridge estimators \citep{BridgeOrig}. It has been shown \citep{KnightFu2000} that for $q \leq 1$ and large enough $\lambda$, the penalized estimate will have some $\hat{\beta}_{\lambda,j}=0$, yielding a continuous approach to the intractable exploration of all submodels. Under appropriate regularity conditions, the limiting distributions of such $\hat{\beta}_{\lambda,j}$ whose corresponding $\beta_j^*=0$ can have positive probability mass at 0 when $p$ is fixed, meaning that they are capable of support recovery. The result has been shown to hold as $p$ and $n$ grow to infinity, under certain growth rate conditions \citep{BridgeResults}. However, the large $\lambda$ necessary for this to occur may cause $|\hat{\beta}_{\lambda,j}|<<|\beta^*_j|$ for $j \in \mathcal{M}^*$, thereby inflating criteria commonly used for tuning parameter selection. A $\lambda$ yielding support recovery may then be ignored by popular tuning parameter selection strategies.



To address this shrinkage problem, the Smoothly Clipped Absolute Deviation (SCAD) penalty \citep{SCAD} 
and 
the Minimax Concave penalty (MCP) give a continuous, nearly unbiased method of penalized estimation \citep{MinimaxConcavePenalty}. The SCAD estimator has been shown to be support recovery consistent (i.e., $P(\mathcal{M}_{\lambda_n} = \mathcal{M}^*)\to 1$ as $n \to \infty$) for appropriately chosen $\lambda$ \cite{SCAD}.  MCP shares a similar result but again under certain conditions on $\lambda$ \citep{MinimaxConcavePenalty}. 
These penalties are less likely to have $|\hat{\beta}_{\lambda,j}|<<|\beta^*_j|$ but their tuning parameter selection problem is further complicated by having to explore a multidimensional space.

Another approach to prevent $|\hat{\beta}_{\lambda,j}|<<|\beta^*_j|$ is to adjust bridge penalties. The Ridge penalty ($q = 2$) cannot shrink any coefficient estimates to exactly zero, so \cite{VSRidge} recently proposed $P_\lambda (\boldbeta) = \sum_{j = 1}^p \lambda_j \beta_j^2,$ to allow for this behavior. The Adaptive Lasso \citep{AdaptiveLasso} is a similar adjustment but for the Lasso penalty. It follows a two-stage process: first $\hat{\boldbeta}$---a consistent estimator for $\boldbeta^*$ such as Ordinary Least Squares (OLS)---is calculated. Then, with $\gamma > 0$, Adaptive Lasso estimates are found using the penalty $P_\lambda(\boldbeta) = \lambda \sum_{j=1}^p |\hat{\beta}_j|^{-\gamma} \lvert \beta_j \rvert.$
When $\gamma = 1$ and $\hat{\boldbeta} = \hat{\boldbeta}_{OLS}$, the objective function reduces to the Nonnegative Garotte \citep{NNGarotte}. This approach is support recovery consistent when $\lambda_n/\sqrt{n} \rightarrow 0$ and $\lambda_n n^{(\gamma - 1)/2}\rightarrow \infty$ as $n \rightarrow \infty$. The Adaptive Lasso is easily generalized to non-Lasso penalties, however the selection of $\gamma$ and consistent estimation of $\boldbeta^*$ for high dimensional data can be difficult to establish.

The Relaxed Lasso \citep{Relaxo} minimizes the objective function:
\begin{equation}
\frac{1}{2n}||\V{y}-\V{X}(\boldbeta \circ \V{1}_{\mathcal{M}_\lambda})||_2^2 + \lambda\phi ||\boldbeta||_1 \ . \ 
\label{eq:RLobj}
\end{equation}
where $\phi \in (0,1]$ and $\boldbeta \circ \V{1}_{\mathcal{M}_\lambda}$ is the Hadamard product of $\boldbeta$ with the support vector under $\mathcal{M}_\lambda$. In simulations, the Relaxed Lasso returns sparse models with low bias. Moreover, the expected value of the loss function of the Relaxed Lasso converges to 0 faster than the Lasso when $p$ increases quickly relative to $n$, meaning that Relaxed Lasso estimates tend to be closer to $\boldbeta^*$ for smaller $n$ than the traditional Lasso. This result is again achieved assuming that $\lambda$ is sufficiently large. The Relaxed Lasso is computationally efficient to calculate but its extensions to more complicated penalties can be computationally intensive, and to our knowledge has not been well-studied.

The Group Lasso \citep{GroupLasso} penalty assumes $E(y_i) = \sum_{j=1}^p \V{x}_{ij}^T \boldbeta_j$ where each $\V{x}_{ij}$ is a $d_j \times 1$ vector corresponding with the $i$th observation of the $j$th covariate group and shrinks coefficients at the group level.  This class of models includes general additive models (GAMs) and functional linear models. GAMs \citep{GAMs} have the form $y_i = \sum_{i = 1}^p f_j(\V{x}_{ij})+\epsilon_i$ where the $f_j$'s are functions of one or more covariates. Each $f(\cdot)$ is commonly approximated by a pre-specified basis expansion, such as B-splines, so that estimation of $f(\cdot)$ is equivalent to estimating the corresponding group of basis coefficients. The functional linear model \citep{RamsaySilverman}, an example of which may be found in Section \ref{sec:EMGApplication}, has functional covariates $x_{ij}(t)$ on domain $\mathcal{T}$ and models $y_i = \sum_{i = 1}^p \int x_{ij}(t)\beta_j(t)dt+\epsilon_i$. The model is often approximated by imposing a basis expansion of the $\beta_j(\cdot)$. 
The Group Lasso penalty for such models is $P_\lambda(\boldbeta) =  \sum_{j=1}^p \lambda_j \lvert \lvert \boldbeta_j \rvert \rvert_{K_j}$ where $\lVert \V{z} \rVert_K = (\V{z}^T 
\V{K} \V{z})^{\frac{1}{2}}$ and $\V{K}_1,...,\V{K}_J$ are known positive definite matrices. The group of coefficient estimates, $\hat{\boldbeta}_{\lambda,j}$, are then either all zero or all nonzero. For the linear model, and under some regularity conditions, \cite{GroupLassoAsymptotics} proved that the Group Lasso is support recovery consistent as long as  $\sqrt{n} \lambda_j \rightarrow \infty$ for all $j \notin \mathcal{M}^*$. 






\subsection{Tuning Parameter Selection Strategies}

Desirable statistical properties of penalized estimators hold under certain $\lambda$ values, making tuning parameter selection a pivotal step in the analysis. 
One popular approach is to choose $\lambda$ that minimizes an information criterion $ IC(\lambda) =  -2\log{L(\hat{\boldbeta}_\lambda)} + h(k_\lambda)$ where $L(\hat{\boldbeta}_\lambda)$ is the likelihood of the data under $\hat{\boldbeta}_\lambda$ and $h(\cdot)$ is a penalty to prevent overselection based on the $k_\lambda=|\mathcal{M}_\lambda|$. Two well-known criteria are AIC \citep{Akaike1974}, where $h(k_\lambda) = 2k_\lambda$ , and BIC \citep{Schwarz1978}, where $h(k_\lambda) = k_\lambda \textrm{log}(n)$. AIC often overselects, particularly for small sample sizes, so a corrected AIC (AICc), where $h(k_\lambda) = 2k_\lambda + \frac{2k_\lambda^2 + 2k_\lambda}{n - k_\lambda - 1}$ is recommended \citep{AICc}. BIC, unlike AIC, is support recovery consistent when $\epsilon_i \overset{iid}{\sim}N(0,\sigma^2)$ \citep{BICSelectionConsistent}. For Gaussian errors, $L(\hat{\boldbeta}_\lambda)$ involves $\sigma^2$ and for unknown $\sigma^2$, $-2\log{L(\hat{\boldbeta}_\lambda)} \propto n\logof{\hat{\sigma}^2_\lambda}$ where $\hat{\sigma}^2_\lambda$ is the estimated model variance at $\lambda$. This can lead to overselection when $\hat{\sigma}^2_\lambda < 1$ \cite{Buhlmann2011}. When possible, $\sigma^2$ is substituted with $\hat{\sigma}^2$ from a presumed low bias model \citep{Hastie2017}, but this may be challenging to identify for high-dimensional data. The Extended Regularized Information Criterion (ERIC), where $h(k_\lambda) = 2 \nu k_\lambda \textrm{log}(n\hat{\sigma}_\lambda^2/\lambda)$ and $\nu>0$, was proposed by \cite{ERIC} specifically for tuning parameter selection of penalized estimators. ERIC outperformed popular tuning parameter selection approaches in their simulations for the Adaptive Lasso, but the choice of $\nu$ is subjective and determines the balance between fit and sparsity.




Cross Validation (CV), is the process of splitting data into training and validation sets, in which the models are fitted on the training sets and overfitting is assessed by predicting observations in the validation sets. 
CV takes many forms, but $K$-fold CV \citep{Geisser1975, Allen1975, Stone1974} is arguably the most common. In $K$-fold CV, the data are partitioned into $K$ sets, or folds, of size $n_k$ each. For each $\lambda$, $K$ sets of estimates are generated using $K-1$ of the $K$ folds  and predictions are generated for the remaining fold, denoted $\hat{\V{y}}_{\lambda,k}$. Prediction error is calculated for each $\lambda$ and fold as $\frac{1}{n_k}||\V{y}_k - \hat{\V{y}}_{\lambda,k}||_2^2$ and is averaged across the $K$ folds to give the average prediction error (APE) for each $\lambda$. The $\lambda$ with the minimum APE is selected, or a 1SE rule---which chooses the largest $\lambda$ within one standard error of the minimum APE---is implemented. The use of a 1SE rule is most common for a one-dimensional $\lambda$, but \cite{SAFE} proposed a multidimensional extension where the $\hat{\boldbeta}_\lambda$ chosen lies withing one standard error of the minimum and also minimizes some penalty function. $K$-fold CV still has a tendency towards overselection, even with a 1SE rule \citep{Krstajic2014}. 

Generalized Cross Validation (GCV) is an efficient alternative to $n$-fold Cross Validation \citep{GCV1,GCV2}. GCV is appropriate when the estimation procedure admits linear predictions $\hat{\V{y}} = \V{S}\V{y}$ for some matrix $\V{S}$ \citep{Hastie2017}. For example, in Ridge Regression, $\V{S} = \V{X}((\V{X}^T\V{X})^{-1} + \lambda \V{I})\V{X}^T$. GCV minimizes a function that divides $L(\hat{\boldbeta}_\lambda)=||\V{y}-\hat{\V{y}}_\lambda||_2^2$ by a function of $k_\lambda$, the effective degrees of freedom. 
However, just like with other information criteria, 
GCV has been shown to lead to overselection \citep{Homrighausen2018}.

\subsection{Post-selection Inference}
Post-selection inference techniques carry out further variable selection after a tuning parameter value has been selected \citep{VanDeGeer2014, Javanmard2014, Taylor2015, Lee2016, Shi2020}. 
The Debiased Lasso estimator \citep{DebiasedLasso} is a linear adjustment to the Lasso estimate $\frac{1}{n} \boldsymbol{\Theta} \V{X}^T(\V{y} - \V{X}\hat{\boldbeta}_\lambda)$, where $\boldsymbol{\Theta}$ is an estimate of $(\V{X}^T \V{X})^{-1}$.
This method focuses on estimation bias rather than direct variable selection, so the use of an additive bias correction is coherent; estimates for zero effect coefficients are small and hence their bias is small. The correction leads to an approximate Normal distribution of $\hat{\boldbeta}_\lambda$ so one may perform hypothesis testing and construct confidence intervals. 

The Covariance Test  \citep{CovarianceTest} takes advantage of the LARS algorithm, which constructs the Lasso solution path by adding variables one at a time \citep{LARS}. It is distinctive in that it assesses model fit using covariance rather than squared error loss. 
The test requires estimation of $\sigma^2$ and its extension to non-Lasso penalties is not well studied, but it provides precedence for the use of correlation in the evaluation of model fit for variable selection.

In general, when the tuning parameter selection event can be written as $\{\V{A}\V{y} \leq \V{b}\}$ for some matrix $\V{A}$ and vector $\V{b}$, there exists  a general scheme for post-selection inference that gives exact confidence intervals and p-values for Gaussian errors. 
Choose $\boldsymbol{\eta}$ such that inference about $\boldsymbol{\eta}^T E[\V{y}]$ is of interest. Using the polyhedral lemma for Gaussian errors, \cite{Lee2016} and \cite{CovarianceTest} represent this event in terms of $\boldsymbol{\eta}^T\V{y}$ to perform conditional inference. This allows for inference upon multiple $\lambda$ or a fixed $\lambda$. When used for successive steps of LARS, it is known as the Spacing Test \citep{SpacingTest} and is a non-asymptotic version of the Covariance Test.

\section{Methodology}
\label{sec:Methodology}
\subsection{Why Correlation Over Squared Prediction Error?}
Distinguish the magnitude of $\boldbeta^*$, denoted $\alpha^*=||\boldbeta^*||_2$, from its direction, $\boldsymbol{\xi}^*=\boldbeta^*/\alpha^*$. Then $\boldsymbol{\xi}^*$ retains all information about $\mathcal{M}^*$. The same summaries may be computed from an estimate, $\hat{\boldbeta}$, denoted by $\tilde{\alpha}$ and $\tilde{\boldsymbol{\xi}}$, having support $\mathcal{M}$. Comparing $\mathcal{M}$ to $\mathcal{M}^*$ is equivalent to comparing the supports of $\tilde{\boldsymbol{\xi}}$ and $\boldsymbol{\xi}^*$. A tuning parameter selection strategy based on squared prediction error, however, concerns both magnitude and direction. Let $\{\V{y},\V{X}\}$ denote a holdout sample where $\V{y}$ and $\V{X}$ have been centered. The squared prediction error for $y_i = \V{x}_i^T\boldbeta^* + \epsilon_i$ is
\begin{equation}
   \sum_{i} (y_i - \V{x}_{i}^T \hat{\boldbeta})^2 = \sum_i (\epsilon_i + \V{x}_i^T(\alpha^*\boldsymbol{\xi}^*-\tilde{\alpha} \tilde{\boldsymbol{\xi}}))^2\ .\ 
   \label{eq:PE}
\end{equation}
In the ideal situation where $\tilde{\boldsymbol{\xi}}=\boldsymbol{\xi}^*$, \eqref{eq:PE} will be inflated when $\tilde{\alpha} \neq \alpha^*$. Indeed, for penalized estimators with large $\lambda$, typically $\tilde{\alpha} < \alpha^*$. Therefore, it is possible for an estimate having $\tilde{\boldsymbol{\xi}}=\boldsymbol{\xi}^*$ to have a large APE and so would be unlikely to be chosen by an APE-based tuning parameter selection strategy.







Consider now the correlation between $\V{y}$ and the predictions $\hat{\V{y}}=\V{X}\hat{\boldbeta}$. After some simplification, we get the expression
\begin{align}
    \text{Corr}(\V{y},\hat{\V{y}})
    =\frac{(\V{X}\boldsymbol{\xi}^*+\boldsymbol{\epsilon}^*)^T\V{X}\tilde{\boldsymbol{\xi}}}{||\V{X}\boldsymbol{\xi}^*+\boldsymbol{\epsilon}^*||_2 \  ||\V{X}\tilde{\boldsymbol{\xi}}||_2}
    \label{eq:PredictedCorrelation}
\end{align}
where $\boldsymbol{\epsilon}^*=\boldsymbol{\epsilon}/\alpha^*$ is a scaled error vector that does not depend on $\hat{\boldbeta}$. The $\tilde{\alpha}$ has no influence over this summary so this measure better compares the $\boldsymbol{\xi}^*$ and $\tilde{\boldsymbol{\xi}}$, and hence better assesses support recovery than squared prediction error.

Our proposed tuning parameter selection strategy, called AR2 CV, employs $K$-fold CV with folds $\{\V{y}_k,\V{X}_k\}$ but replaces APE with 
\begin{align}
    \text{AR2}=\frac{1}{K}\sum_{k=1}^K [1 - \text{Corr}(\V{y}_k, \hat{\V{y}}_{k})^2] \label{eq:AR2}\ .\
\end{align} 
The optimal $\lambda$ may be chosen as the one that minimizes AR2, but we have found significant improvements in support recovery under an analogous one-standard-error rule. Applying AR2 CV with a 1SE rule to the toy example in Section \ref{sec:Intro}, the optimal $\lambda$ is that given in Figure \ref{fig:MotivPerfectSmall}. The AR2 CV estimator was then able to compromise between support recovery and prediction error, while the APE-based CV prioritized prediction error. The coefficient estimates for the $j \in \mathcal{M}^*$ under the APE CV model exhibit less bias than those of the AR2 CV model. A potential drawback then of AR2 CV is that its indifference towards $\alpha^*$ may lead to a $\hat{\boldbeta}_\lambda$ that exhibits more shrinkage than is desired. A potential remedy is to follow selection of $\mathcal{M}_{\lambda}$ with unpenalized estimation for only predictors in $\mathcal{M}_{\lambda}$. For example, one could perform OLS on only the $j \in \mathcal{M}_\lambda$ to form the estimator $\hat{\boldbeta}^{\mathcal{M}_\lambda}_{OLS}$ where $\hat{\beta}^{\mathcal{M}_\lambda}_{OLS, j} = \hat{\beta}_{OLS,j}$ when $j \in \mathcal{M}_\lambda$ and zero otherwise. We next discuss an alternative strategy that is related to AR2 CV that adjusts the shrinkage of $\hat{\boldbeta}_\lambda$.

\subsection{The $\alpha$-Modification}
\label{sec:AlphaMod}

Consider now the training data's $\V{y}$ and their predictions, $\hat{\V{y}}$. If $\hat{\V{y}} \neq \V{0}$, calculate the least-squares estimate $\hat{\alpha} = \text{arg min}_\alpha || \V{y} - \alpha \hat{\V{y}} ||_2^2=(\hat{\V{y}}^T\hat{\V{y}})^{-1}\hat{\V{y}}^T\V{y}$. Note this $\hat{\alpha}$ likely differs from $\tilde{\alpha}=||\hat{\boldbeta}||_2$. By definition, the modified predictions $\hat{\alpha} \hat{\V{y}}$ will be as close or closer to $\V{y}$ as $\hat{\V{y}}$.  The modified prediction is also equivalent to prediction under the adjusted penalized estimate, $\hat{\alpha}\hat{\boldbeta}$. Calculating these $\alpha$-modified coefficient estimates and predictions is described in Algorithm \ref{alg:Alpha}. 

\begin{algorithm}
\caption{$\alpha$-Modification for the Linear Model}\label{alg:Alpha}
\begin{algorithmic}[1]
\State \textbf{Given:} Suppose $\V{y}$ and $\V{X}$ are centered such that an intercept term is unnecessary. Let $\hat{\boldbeta}$ be the vector of coefficient estimates. 
    \State Calculate $\hat{\alpha} = (\hat{\V{y}}^T\hat{\V{y}})^{-1}\hat{\V{y}}^T\V{y} = (\hat{\boldbeta}^T \V{X}^T \V{X}\hat{\boldbeta})^{-1}\hat{\boldbeta}^T \V{X}^T \V{y}$.
    \State Calculate $\alpha$-modified estimates, $\hat{\alpha} \hat{\boldbeta}$, and $\alpha$-modified predictions, $\hat{\alpha} \hat{\V{y}} = \hat{\alpha} \V{X}\hat{\boldbeta}$.
\end{algorithmic}
\end{algorithm}

For penalized estimators, the $\alpha$-modified estimate can be viewed as 
\begin{equation}
\argmin_{\alpha,\boldsymbol{\xi}}\frac{1}{2n}||\V{y}-\alpha\V{X}\boldsymbol{\xi}||_2^2 + \lambda P(\boldsymbol{\xi})
\label{eq:Alphaobj}
\end{equation}
by first fixing $\alpha = 1$ to get $\hat{\boldsymbol{\xi}} = \hat{\boldbeta}_\lambda$ and then minimizing the function for $\alpha$ given $\hat{\boldsymbol{\xi}}$. This is a slight abuse of notation because $\hat{\boldsymbol{\xi}}$ is not required to be a unit vector like $\boldsymbol{\xi}^*$, but it does help to show how the $\alpha$-modified estimate is related to separating $\boldbeta^*$ into its magnitude and direction. As $\lambda \to 0$, $\hat{\alpha}_\lambda\to 1$ because the objective function focuses most of its attention on the loss function. Thus the impact of the $\alpha$-Modification will be more pronounced for larger values of $\lambda$, and ideally the $\alpha$-modified estimate will correct the bias of $\hat{\boldbeta}_\lambda$ due to shrinkage.  

The $\alpha$-Modification is similar other existing modifications to penalized estimators.  First, one can view the modification as reversing the process of calculating the Nonnegative Garrote estimator, which starts with OLS estimates of $\boldsymbol{\beta}^*$ and then performs penalization. \cite{ElasticNet} also recommended a multiplicative adjustment to the Elastic Net estimator, although the adjustment only involves one of the tuning parameters.  
For penalties that satisfy $P(\alpha \boldsymbol{\xi}) = \alpha P(\boldsymbol{\xi})$ for $\alpha >0$, we may rewrite the penalty in \eqref{eq:Alphaobj} as $\lambda\alpha^{-1}P(\alpha\boldsymbol{\xi})$ which resembles the Relaxed Lasso penalty, so long as 
$\alpha^{-1} \in (0,1]$. Finally, the Debiased Lasso tries to reduce bias through an additive adjustment, but this will cause some or all $\hat{\beta}_{\lambda,j}=0$ to become nonzero, while a multiplicative adjustment will not change the support of $\hat{\boldbeta}_\lambda$.

Penalized estimates are typically shrunk towards zero so the $\alpha$-Modification will correct this type of bias only if $\hat{\alpha}_\lambda \geq 1$. This property is guaranteed for common penalties:
\begin{theorem}
Suppose $P_\lambda(\boldbeta)=\sum_{\ell=1}^L \lambda_\ell g_\ell(\boldbeta)$ where $g_\ell(\boldbeta)$ is convex and minimized at $\V{0}_p$. Then $\hat{\alpha}_\lambda \geq 1$ when $\hat{\boldbeta}_\lambda \neq \V{0}$.
\label{thm:AlphaGEQ1}
\end{theorem} 
\noindent Amplifying penalized estimates does not necessarily decrease bias. To evaluate the $\alpha$-Modification as a bias-reduction tool, we have the following result.
\begin{lemma}
If 
there exists a $\lambda$ where $P(\hat{\boldsymbol{\xi}}_\lambda=\boldsymbol{\xi}^*)=1$, then $E(\hat{\alpha}_\lambda \hat{\boldbeta}_\lambda)=\boldbeta^*$.\label{lem:Unbiased}
\end{lemma}
\noindent Lemma~\ref{lem:Unbiased} conditions on an event that may have probability 0. 
The following lemma considers a broader condition, whereby the penalized estimate recovers the direction of $\hat{\boldbeta}^{\mathcal{M}_\lambda}_{OLS}$, defined at the end of Section~3.1.
\begin{lemma}
If $\hat{\boldbeta}_\lambda \varpropto \hat{\boldbeta}^{\mathcal{M}_\lambda}_{OLS}$ then $\hat{\alpha}_\lambda \hat{\boldbeta}_\lambda=\hat{\boldbeta}^{\mathcal{M}_\lambda}_{OLS}$.
\label{lem:ScaledOLS}
\end{lemma}
There are multiple examples that satisfy the condition of Lemma~\ref{lem:ScaledOLS}. The OLS estimator itself qualifies as a scaled OLS estimator, where $\hat{\alpha}_\lambda = 1$. Ridge estimates when $\V{X}^T \V{X} = n\V{I}_p$ also take this form, having $\hat{\boldbeta}_\lambda = \frac{1}{1+\lambda}\hat{\boldbeta}_{OLS}$. 
Lemma \ref{lem:ScaledOLS} also applies whenever $\hat{\boldbeta}_\lambda$ contains exactly one non-zero entry. Finally, note the Relaxed Lasso always includes $\hat{\boldbeta}_{\lambda,\phi}=\hat{\boldbeta}^{\mathcal{M}_\lambda}_{OLS}$ among its solutions by setting $\phi=0$. Lemma \ref{lem:ScaledOLS} shows that this can sometimes occur for the $\alpha$-modified estimates as well.


The $\alpha$-Modification serves to improve predictions under a given $\hat{\boldbeta}_\lambda$ through a positive, multiplicative adjustment and so its ability to improve estimation depends on the properties of $\hat{\boldbeta}_\lambda$. While this paper is mainly concerned with tuning parameter selection for support recovery under finite sample sizes, properties of $\hat{\boldbeta}_\lambda$ are easier to study as $n \to \infty$ and the same is true for $\alpha$-modified estimators. Concerning support recovery consistency, since the support of $\hat{\alpha}_\lambda\hat{\boldbeta}_\lambda$ equals the support of $\hat{\boldbeta}_\lambda$, the $\alpha$-modified estimator is support recovery consistent if and only if $\hat{\boldbeta}_\lambda$ is support recovery consistent. 
Next, the following theorem establishes estimation consistency of $\alpha$-modified estimators.
\begin{theorem}
Fix $p$ and suppose there exists a positive definite matrix $\V{C}$ where
$\frac{1}{n} \V{X}_n^T \V{X}_n = \frac{1}{n} \V{C}_n \rightarrow \V{C}$  as $n \rightarrow \infty$. For a $P_\lambda(\cdot)$, if there exists a $\lambda_n$ where $\hat{\boldbeta}_{\lambda_n}$ converges in probability to $c\boldsymbol{\xi}^*$ for some $c>0$, then $\hat{\alpha}_{\lambda_n}\hat{\boldbeta}_{\lambda_n}$ converges in probability to $\boldbeta^*$.
\label{thm:Consistency}
\end{theorem}

\noindent
The case of $c=\alpha^*$ in Theorem 2 says that if $\hat{\boldbeta}_\lambda$ converges in probability to $\boldbeta^*$ then so will its corresponding $\alpha$-modified estimator. For example, \cite{KnightFu2000} determined certain conditions for which the Lasso exhibits this property. However, the rate of convergence for the $\alpha$-modified estimators may improve due to the relaxation of finding a $\lambda_n$ where $\hat{\boldbeta}_{\lambda_n}$ converges in probability to any $c\boldsymbol{\xi}^*$. Theorem 2 also suggests there may be opportunities for other penalized estimators that are not estimation consistent to have an $\alpha$-modified estimator that is estimation consistent.

\subsection{$\alpha$-Modified Cross Validation}


In addition to AR2 CV, we propose $\alpha$-modified CV based on average squared prediction error under the $\alpha$-modified estimates: 
\begin{align}
\text{Mod APE}=\frac{1}{K} \sum_{k=1}^K ||\V{y}_k-\hat{\alpha}_k\hat{\V{y}}_k||_2^2\ ,\    
\end{align}
where $\hat{\alpha}_k$ is calculated from the training data. Returning to the toy example from Section \ref{sec:Intro}, the $\lambda$ chosen through $\alpha$-modified CV with a 1SE rule returned a model with $\mathcal{M}_\lambda=\mathcal{M}^*$. Its APE was 0.3821, which is marginally larger than that of AR2 CV. This demonstrates the possibility that the two CV strategies may point to different optimal $\lambda$.


Another benefit of $\alpha$-modified CV over traditional CV surprisingly derives from the potential drawbacks of the modification. It is reasonable to say that $\hat{\boldbeta}_\lambda$ recovers the direction of $\boldsymbol{\xi}^*$ when $\hat{\boldbeta}_\lambda / \lVert \hat{\boldbeta}_\lambda \rVert_2 = \boldsymbol{\xi}^*$. Because the modification unshrinks the $\hat{\boldbeta}_\lambda$ without changing its direction, the modification may exacerbate an estimate of poor quality, causing the Mod APE to exceed APE based on $\hat{\boldbeta}_\lambda$. It is unlikely then for such a $\lambda$ to be selected as the optimal value. Similarly, when $\hat{\boldbeta}_\lambda$ recovers $\boldsymbol{\xi}^*$ but has small magnitude, Mod APE will decrease significantly over the traditional APE. Theorem \ref{thm:EModAPE} gives a theoretical justification for the latter situation.
\begin{theorem}
Suppose that $\hat{\boldbeta}_\lambda$ recovers the direction of $\boldbeta^*$ and assume all $\epsilon_i$ are independent with constant variance $\sigma^2$. Then the expected value of the $\alpha$-modified APE is less than the expected value of the traditional APE when:\label{thm:EModAPE}
\begin{equation}
\frac{{\alpha^*}^2}{E[(\lVert \hat{\boldbeta}_\lambda \rVert_2 - \alpha^*)^2]}  \leq \frac{\boldbeta^{*T}\V{X}^T \V{X} \boldbeta^*}{\sigma^2}\ .\
\end{equation}
\end{theorem}
\noindent Mod APE is then expected to be smaller than the traditional APE as long as the signal-to-noise ratio, represented by $\boldbeta^{*T}\V{X}^T \V{X} \boldbeta^*/\sigma^2$, is sufficiently large.



\section{The $\alpha$-Modified Lasso}
\label{sec:LassoResults}
The methods in Section~3 generalize to many types of penalties, but to better understand their properties we must focus on a specific penalty. 
Due to its popularity, we explore the properties for the Lasso penalty and finite $n$. First, note that the $\alpha$-modified Lasso may be thought of as a reverse Non-negative Garotte in that it starts with shrunken estimates and uses a least squares modification to readjust and ``un-shrink" them. The use of correlation to assess model quality is also consistent with the premise of the Covariance Test.  The $\alpha$-modified Lasso is most similar to the Relaxed Lasso.  The $\alpha$-Modification penalty is $\lambda \alpha^{-1} ||\alpha \boldsymbol{\xi}||_1$ and the Relaxed Lasso penalty is $\lambda\phi||\boldbeta||_1$ for $\phi \in (0,1]$. Theorem~\ref{thm:AlphaGEQ1} establishes $\alpha^{-1} \in (0,1]$ but we calculate the minimum $\alpha$ directly while the Relaxed Lasso treats $\phi$ as a tuning parameter. Our approach reduces computation, but the resulting estimators are less flexible than the Relaxed Lasso. Specifically, the $\alpha$-modified estimate cannot change the direction of $\hat{\boldbeta}_\lambda$. 

\begin{figure}[ht]
     \begin{subfigure}{0.49\textwidth}
         \includegraphics[width=\textwidth]{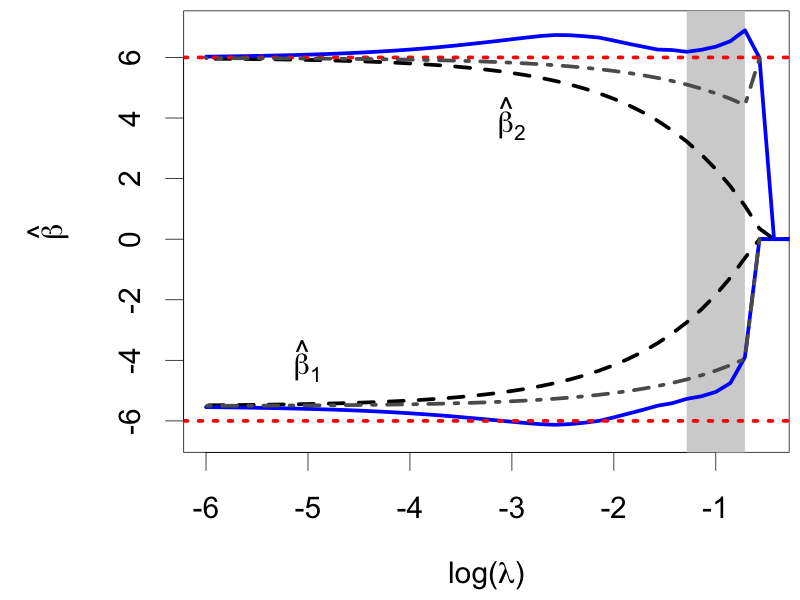}
     \end{subfigure}
     \hfill
     \begin{subfigure}{0.49\textwidth}
     \includegraphics[width=\textwidth]{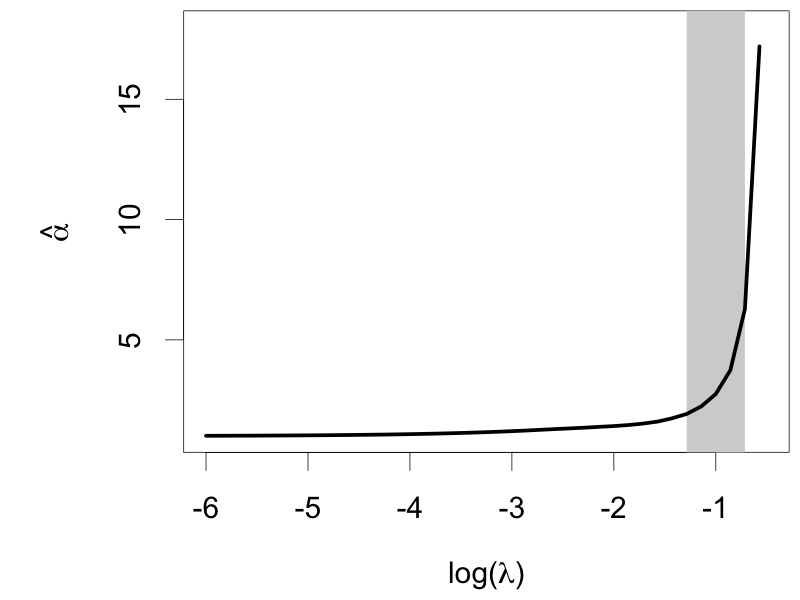}
     \end{subfigure}
     \caption{The leftmost plot shows solution path for the Lasso (dashed line), $\alpha$-modified Lasso (solid line), and the Relaxed Lasso at $\phi = 0.5$ (dashed and dotted line) for $\beta_1^*$ and $\beta_2^*$. The grey shaded area indicates the values of $\lambda$ for which support recovery has occurred. The horizontal dotted lines show the true values of $\beta_1^*$ and $\beta_2^*$. The plot on the right shows the $\hat{\alpha}_\lambda$ values.}
     \label{fig:LassoDir}
\end{figure}

Figure \ref{fig:LassoDir} illustrates the distinction between the $\alpha$-modified and Relaxed Lasso estimator for an orthogonal $\V{X}$ with $n=100$, $p = 50$ and $p^*=2$ where $(\beta_1^*,\beta_2^*)=(-6,6)$. The $\alpha$-modified Lasso and Relaxed Lasso solutions were generated for a set of 100 $\lambda$ and, for the Relaxed Lasso, 20 $\phi$ values. The solution paths for $\hat{\beta}_{\lambda,1}$ and $\hat{\beta}_{\lambda,2}$ across all values of $\lambda$ are shown for both the traditional and $\alpha$-modified Lasso. The Relaxed Lasso estimates for $\phi=0.5$ are also plotted. Figure \ref{fig:LassoDir} also includes a plot of the $\hat{\alpha}_\lambda$ for the full data. For the $\lambda$ exhibiting perfect support recovery, we see that $\hat{\alpha}_\lambda>1$, thereby demonstrating the value of the modification. 

\begin{figure}
    \centering
    \includegraphics[width = 0.8 \textwidth]{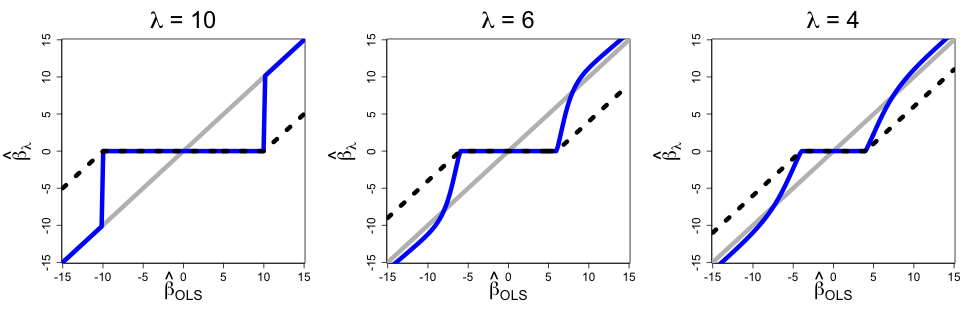}
    \caption{$\alpha$-modified estimates of $\alpha \beta_{1}$ as $\hat{\beta}_{OLS,1}$ changes, all other parameters being fixed. $\hat{\beta}_{OLS,j} = [-8, 5, -3, 1]$ for $j = [2,...,5]$. The dotted line gives the Lasso estimates and the solid blue line gives the $\alpha$-modified estimates.}
    \label{fig:FixOLS}
\end{figure}

We now derive properties of the $\alpha$-modified estimator for orthogonal $\V{X}$.  The Lasso estimator for such $\V{X}$ is $\hat{\beta}_{\lambda, j} = \textrm{sign}(\hat{\beta}_{OLS,j})(|\hat{\beta}_{OLS,j}| - \lambda)_+,$ where $\hat{\beta}_{OLS,j}$ is the OLS estimate of the $\beta^*_j$ and $(z)_+ = \textrm{max}(0, z)$. If the Lasso estimate for a $j^*$ is non-zero, the closed form expression for $\hat{\alpha}_\lambda \hat{\beta}_{\lambda,j^*}$ may be derived. 
\begin{lemma}
When \V{X} is orthogonal, the $\alpha$-modified Lasso estimator has elements:\begin{equation}
    \hat{\alpha}_\lambda \hat{\beta}_{\lambda, j^*} =w_1\hat{\beta}_{OLS,j*} + (1- w_1)\hat{\beta}_{\lambda,j^*} + w_2 \hat{\beta}_{\lambda,j^*}
    \label{eq:abLasso}
\end{equation} where $w_1=\frac{d_{j*}^2}{\sum_{j=1}^p d_j^2}$, $w_2= \frac{\lambda \sum_{j \neq j*} d_j}{\sum_{j = 1}^p d_j^2}$, and $d_j = (|\hat{\beta}_{OLS,j}|-\lambda)_+$. 
\label{thm:AlphaBetaExpression}
\end{lemma}
\noindent 
If $d_{j^*}=0$, then $\hat{\alpha}_\lambda \hat{\beta}_{\lambda, j^*}= \hat{\beta}_{\lambda, j^*}=0$. If $\hat{\alpha}_\lambda \hat{\beta}_{\lambda, j}=0$ for all $j \neq j^*$ and $d_{j^*}>0$ then $\hat{\alpha}_\lambda \hat{\beta}_{\lambda, j^*}=\hat{\beta}_{OLS,j}$ which is consistent with Lemma \ref{lem:ScaledOLS}. When $|\mathcal{M}_\lambda|>1$, (\ref{eq:abLasso}) involves a convex combination of the OLS and Lasso estimates, as well as an additive term, $w_2 \hat{\beta}_{j*,\lambda}$,
that may overcorrect the $\alpha$-modified Lasso beyond the OLS estimate. This behavior is illustrated for a simple example in Figure \ref{fig:FixOLS}. In that example, for $\lambda < 10$ there are potential values of $\hat{\beta}_{OLS,j^*}$ where the $\alpha$-modified Lasso exceeds $\hat{\beta}_{OLS,j^*}$. To better understand this behavior, Theorem \ref{thm:OrthAbsDiff} provides an upper bound for  $|\hat{\alpha}_\lambda \hat{\beta}_{\lambda,j^*} - \hat{\beta}_{OLS,j^*} |$. 
\begin{theorem}
Suppose $\V{X}$ is orthogonal and consider a given predictor, $j^*$, and $\lambda$ where at least one $j \neq j^*$ has $|\hat{\beta}_{OLS,j}| > \lambda$. Let $\V{x}_{j^*}$ denote column $j^*$ of $\V{X}$. Then $\hat{\beta}_{OLS, j*} = \beta^*_{j^*} + n^{-1}\V{x}_j^T\boldsymbol{\epsilon}$ where $n^{-1}\V{x}_j^T\boldsymbol{\epsilon}$ is fixed implies \begin{equation}
   |\hat{\alpha}_\lambda \hat{\beta}_{\lambda,j^*} - \hat{\beta}_{OLS,j^*} | \leq  \lambda \times \max \left( 1, \frac{\sqrt{u^2v + v^2} - v}{2v}\right)
\end{equation}  where $u = \sum_{j\neq j^*}d_j$ and $v = \sum_{j\neq j^*}d_j^2$. Moreover, 
$\lim_{|\beta_{j*}^*| \rightarrow \infty} |\hat{\alpha}_\lambda \hat{\beta}_{j*,\lambda} - \hat{\beta}_{OLS, j*}| \rightarrow 0$ .
\label{thm:OrthAbsDiff}
\end{theorem}

The result of Theorem \ref{thm:OrthAbsDiff} is demonstrated in Figure \ref{fig:FixOLS} for different values of $\lambda$. As $|\beta_1^*|$ increases in magnitude, thereby increasing $\hat{\beta}_{OLS,1}$, the $\alpha$-modified Lasso estimate moves away from the traditional Lasso estimate and towards the $\hat{\beta}_{OLS,1}$. This happens instantaneously for $\lambda=10$ and $|\hat{\beta}_{OLS,1}|>10$ because $\hat{\beta}_{OLS,1}$ is the only estimate that exceeds $\lambda$. For $\lambda=6$ and $4$, when $(\sqrt{u^2v + v^2} - v)/2v > 1$ there exists some $\beta^*_{1}$ in which the $\alpha$-modified Lasso estimate overshoots $\hat{\beta}_{OLS,1}$. This occurs as $u^2/v$ increases, meaning multiple $d_j$ are nonzero and are close to $0$. 

To generalize Theorem \ref{thm:OrthAbsDiff} to an arbitrary $\V{X}$ we condition on the event that the $\lambda$ recovers the sign vector of $\boldbeta^*$ and consider the oracle OLS estimator $\hat{\boldbeta}_{OLS}^{\mathcal{M}^*} =(\V{X}_{\mathcal{M}^*}^T\V{X}_{\mathcal{M}^*})^{-1}\V{X}_{\mathcal{M}^*}^T\V{y}$ where $\V{X}_{\mathcal{M}^*}$ is the subset of columns of $\V{X}$ corresponding to $\mathcal{M}^*$. Then $\hat{\boldbeta}_{OLS,j^*}^{\mathcal{M}^*}=\beta^*_{j^*} + (\V{X}_{\mathcal{M}^*}^T\V{X}_{\mathcal{M}^*})_{j^*}^{-1}\V{X}_{\mathcal{M}^*}^T\boldsymbol{\epsilon}$ where $(\V{X}_{\mathcal{M}^*}^T\V{X}_{\mathcal{M}^*})_{j^*}^{-1}$ denotes the $j$-th row of $(\V{X}_{\mathcal{M}^*}^T\V{X}_{\mathcal{M}^*})^{-1}$. Finally, let $\tilde{s}_{j^*}=(\V{X}_{\mathcal{M}^*}^T\V{X}_{\mathcal{M}^*})_{j^*}^{-1}\V{s}$ where $\V{s}$ is the $|\mathcal{M}^*| \times 1$ vector of signs for the $\beta_j^*$ in $\mathcal{M}^*$. 




\begin{theorem} Suppose the Lasso estimate recovers the correct sign vector of $\boldbeta^*$. Let $G_{j^*} = \lambda (s_{j^*} -n\tilde{s}_{j^*})$. Then for a fixed $\boldsymbol{\epsilon}$, $\lim_{\lvert \beta^*_{j^*} \rvert \rightarrow \infty} \lvert \hat{\alpha}_\lambda \hat{\beta}_{\lambda, j^*} - \hat{\beta}_{OLS, j^*} \rvert = |G_{j^*}|$. Moreover $|G_{j^*}|< | \hat{\beta}_{\lambda, j^*}-\hat{\beta}_{OLS, j^*}|$ if and only if
\begin{align}
     \left|1-\frac{s_{j^*}}{n\tilde{s}_{j^*}}\right|<1\ .\ 
\end{align} \label{thm:NonOrthAbsDiff} 
\end{theorem}
\noindent Note for an orthogonal design $n\tilde{s}_{j^*}=s_{j^*}$, making $G_{j^*}=0$. Theorem \ref{thm:NonOrthAbsDiff} shows that the absolute difference between the $\alpha$-modified Lasso estimate and the OLS estimate generally does not approach 0 as $|\beta_{j^*}^*| \to \infty$. Rather, it approaches a constant that is always smaller than $\lvert \hat{\beta}_{\lambda,j^*} - \hat{\beta}_{OLS,j^*} \rvert$ so long as $\tilde{s}_{j^*}$ is not close to 0. This indicates the $\alpha$-modified Lasso estimate generally improves the bias of $\hat{\beta}_{\lambda,j^*}$ for large $\beta^*_{j^*}$ when the sign is recovered.

\section{Numerical Results}
\label{sec:Simulations}

\subsection{$\alpha$-Modified CV Under Poor Bias Adjustment}

\begin{figure}
     \centering
     \begin{subfigure}[b]{0.48\textwidth}
         \centering
         \caption{$\boldbeta^* = (4.3,2.9,-5.8,-3.3,-6.1,0,...,0)^T$.}
         \includegraphics[width=0.9\textwidth]{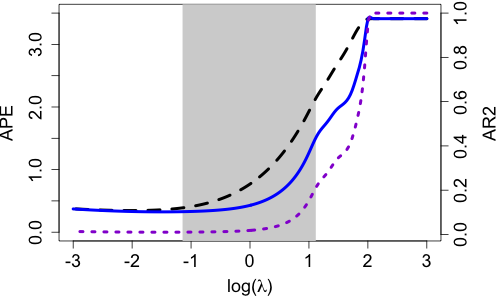}
         \label{fig:CVIllustration1}
     \end{subfigure}
     \hspace{0.25em}
          \begin{subfigure}[b]{0.48\textwidth}
         \centering
         \caption{$\boldbeta^* = (50,50,50,50,50,0,...,0)^T$.}
         \includegraphics[width=0.9\textwidth]{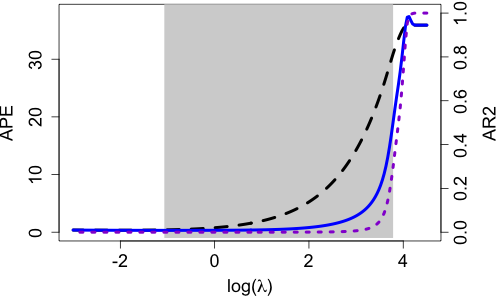}
         \label{fig:CVIllustration2}
     \end{subfigure}
     \caption{Average 10-fold Cross Validation results across 500 replications of linear data. $\V{X}$ was taken from the example in Section 1. Average values of $\lambda$ where the correct submodel was selected are represented by the grey box, the dashed line gives the APE, the solid line gives $\alpha$-modified APE, and the dotted line gives Average $R^2$. Standard errors from these simulations were too small to be depicted.}
     \label{fig:CVIllustration}
\end{figure}

Theorems \ref{thm:OrthAbsDiff} and \ref{thm:NonOrthAbsDiff} show that the $\alpha$-Modification cannot guarantee bias reduction for all $\lambda$. However, a poor bias adjustment may be detected by CV. To illustrate, we performed a simulation study for the $\V{X}$ used in Figure~\ref{fig:Motivation} and considered two $\boldbeta^*$ vectors, displayed in Figure~\ref{fig:CVIllustration}.  For each $\boldbeta^*$, we generated 500 $\boldsymbol{\epsilon} \sim N(\V{0},\V{I})$ and averaged the three 10-fold CV metrics (APE, AR2, and Mod APE) for a range of $\lambda$.  The results are presented in Figure \ref{fig:CVIllustration}. The grey shaded area indicates perfect support recovery, where values of $\lambda$ to the left of the grey area overselect predictors (false positives) and values of $\lambda$ to the right of the grey area underselect predictors (false negatives). 

As shown in Figure~\ref{fig:CVIllustration1}, the Mod APE was less than or equal to APE for all $\lambda$, and especially so for larger $\lambda$ that recover $\mathcal{M}^*$. The optimal $\lambda$'s according to the minimum AR2 and Mod APE resulted in sparser models than the minimum APE. With a 1SE rule, traditional CV recovered the support in only 24.4\% of the replications, whereas $\alpha$-modified and AR2 CV did so in 91.6\% and 93.6\% of the replications, respectively. 

Figure \ref{fig:CVIllustration2} corresponds to $\boldbeta^*$ with larger and equal magnitude coefficients. Generally, the Mod APE was equal to or smaller than regular APE except for $\log(\lambda)\approx 4$, which had only one nonzero coefficient estimate. Mod APE highlights the poor bias adjustment and so would not recommend choosing this $\lambda$. The support recovery percentages were 19.8\% for APE CV, 98.2\% for $\alpha$-modified CV, and 98.8\% for AR2 CV. Surprisingly, the support recovery percentage decreased for APE CV despite increasing magnitude of $\boldbeta^*$.

\subsection{Support Recovery Simulation Study}

To evaluate the new methods for the Lasso, a broader simulation study consisting of 100 replications of data from model (\ref{eq:LinearModel}) was conducted. 
Following \cite{Relaxo}, the rows of \V{X} were drawn from a multivariate Normal distribution with mean $\V{0}$ and covariance $\boldsymbol{\Sigma}_X$
. Nonzero elements of $\boldbeta^*$ were drawn from a Gamma distribution with shape 10 and scale 0.25 with negative and positive signs randomly assigned with equal probability, updating the  coefficient  vector with every replication. For each $\V{X}$ and $\boldbeta^*$, two independent error vectors were generated from $N(0,\sigma^2)$ distributions, with $\sigma$ determined by a fixed signal-to-noise ratio $\text{SNR} = \boldbeta^{*T} \boldsymbol{\Sigma}_X \boldbeta^*/\sigma^2$.  We considered $n \in \{100, 500, 1000\}$, $p \in \{50, 100, 200, 400, 800\}$, $SNR \in \{1.25, 5\}$ and $p^*=\{5,6,\dots,50\}$.  

Each simulated data  set was analyzed using 10-fold CV with and without a 1SE rule under traditional APE (CV Min, CV 1SE), AR2 (AR2 Min, AR2 1SE), Mod APE (Mod Min, Mod 1SE), and the Relaxed Lasso. For all methods, we considered 250 $\lambda$ values equally spaced on the exponential scale from $e^{-20}$ to $e^{10}$. For the Relaxed Lasso, we also considered 100 $\phi$ values equally spaced from $e^{-10}$ to 1. 
Selection under the Releaxed Lasso was done using minimum APE as well as a 1SE rule based on \cite{SAFE}, wherein the optimal combination of $\lambda$ and $\phi$ is the model with the smallest $||\hat{\boldbeta}_{\lambda,\phi}||_1$ among all models with an APE within one standard error of the minimum. Relaxed Lasso estimates were found using a coordinate descent algorithm, capable of admitting more than $n-1$ covariates into models, unlike LARS which was used by \cite{Relaxo}. Support recovery was evaluated using the Hamming Distance (HD) between the $0/1$ support vectors of $\boldbeta^*$ and $\hat{\boldbeta}_\lambda$, which is the sum of the number of false negatives and false positives. Additional metrics including the false discovery rate (FDR), average number of false positives and negatives, and the average prediction bias, where prediction bias is $||\V{X}\boldbeta^* - \V{X} \hat{\boldbeta}_\lambda||_2$ or $||\V{X}\boldbeta^* - \V{X} \hat{\alpha}_\lambda\hat{\boldbeta}_\lambda||_2$ for the $\alpha$-Modification, can be found in the Supplementary Materials.

We first consider the case of $\boldsymbol{\Sigma}_X=\boldsymbol{I}_p$. Tables \ref{tab:APEHDSmall}, \ref{tab:AR2HDSmall}, and \ref{tab:RelHDSmall} give the average HD between $\boldbeta^*$ and $\hat{\boldbeta}_\lambda$ for the traditional APE method with a 1SE Rule, AR2 with a 1SE rule, and the Relaxed Lasso with a minimum APE approach. AR2 and Mod APE perform fairly similarly; the latter results may be found in the supplementary materials.  In general, AR2 CV has better variable selection than APE CV and, for small $n$ and large $p^*$, the Relaxed Lasso.

\begin{table}[ht]
\centering
\caption{Average Hamming Distance for 100 replications of CV 1SE. Standard errors given in parentheses.} 
\label{tab:APEHDSmall}
\begin{tabular}{rcp{0.9cm}p{0.9cm}p{0.9cm}p{0.9cm}p{0.9cm}cp{0.9cm}p{0.9cm}p{0.9cm}p{0.9cm}p{0.9cm}}
  \hline
  \hline
 & $p$ & 50 & 100 & 200 & 400 & 800 &  & 50 & 100 & 200 & 400 & 800 \\ 
   \hline $n$ & & \multicolumn{5}{c}{$p^* = 5; SNR =  1.25 ; \bar{\sigma} =  5.19 $} & & \multicolumn{5}{c}{$p^* = 5; SNR =  5 ;  \bar{\sigma} =  2.59 $} \\
 \cmidrule(l){3-7} \cmidrule(l){9-13} 
100 &  & 3.1 (0.3) & 4.3 (0.4) & 6 (0.6) & 7.8 (1.1) & 9.1 (1.2) &  & 2.2 (0.2) & 4 (0.4) & 7.5 (1.2) & 9.3 (1) & 10.7 (1.2) \\ 
  500 &  & 0.6 (0.1) & 1 (0.2) & 1.3 (0.2) & 1.9 (0.5) & 2.9 (0.5) &  & 0.9 (0.2) & 0.9 (0.1) & 1.7 (0.3) & 1.7 (0.3) & 1.5 (0.3) \\ 
  1000 &  & 0.2 (0) & 0.4 (0.1) & 0.4 (0.1) & 0.6 (0.1) & 0.9 (0.2) &  & 0.1 (0) & 0.4 (0.1) & 0.5 (0.1) & 0.7 (0.1) & 0.9 (0.2) \\ 
   \hline $n$ & & \multicolumn{5}{c}{$p^* = 50; SNR =  1.25 ; \bar{\sigma} =  16.6 $} & & \multicolumn{5}{c}{$p^* = 50; SNR =  5 ;  \bar{\sigma} =  8.3 $} \\
 \cmidrule(l){3-7} \cmidrule(l){9-13} 
100 &  & 31.7 (1.3) & 43.5 (0.6) & 50.1 (0.5) & 51.1 (0.4) & 53.2 (1.1) &  & 5 (0.4) & 32 (0.6) & 47 (0.7) & 53.8 (1.1) & 54.5 (1.1) \\ 
  500 &  & 1.6 (0.2) & 18.2 (0.5) & 33.3 (0.8) & 48.6 (1.3) & 57.4 (1.8) &  & 0 (0) & 16.8 (0.5) & 35.3 (0.9) & 58.4 (1.6) & 81.2 (2) \\ 
  1000 &  & 0.4 (0.1) & 14.5 (0.4) & 28.5 (0.9) & 43.5 (1.4) & 57.6 (1.6) &  & 0 (0) & 14.2 (0.5) & 28 (0.8) & 44.8 (1.3) & 64.2 (1.8) \\ 
   \hline
\end{tabular}
\end{table}

\begin{table}[ht]
\centering
\caption{Average Hamming Distance for 100 replications of AR2 CV 1SE. Standard errors given in parentheses.} 
\label{tab:AR2HDSmall}
\begin{tabular}{rcp{0.9cm}p{0.9cm}p{0.9cm}p{0.9cm}p{0.9cm}cp{0.9cm}p{0.9cm}p{0.9cm}p{0.9cm}p{0.9cm}}
  \hline
  \hline
 & $p$ & 50 & 100 & 200 & 400 & 800 &  & 50 & 100 & 200 & 400 & 800 \\ 
   \hline $n$ & & \multicolumn{5}{c}{$p^* = 5; SNR =  1.25 ; \bar{\sigma} =  5.19 $} & & \multicolumn{5}{c}{$p^* = 5; SNR =  5 ;  \bar{\sigma} =  2.59 $} \\
 \cmidrule(l){3-7} \cmidrule(l){9-13} 
100 &  & 2.1 (0.3) & 2.8 (0.3) & 3.3 (0.3) & 4.9 (0.7) & 5 (0.5) &  & 0.6 (0.1) & 0.6 (0.1) & 1.9 (0.8) & 1.8 (0.4) & 2.6 (0.4) \\ 
  500 &  & 0.1 (0) & 0.2 (0) & 0.1 (0) & 0.2 (0) & 0.4 (0.1) &  & 0 (0) & 0 (0) & 0 (0) & 0 (0) & 0 (0) \\ 
  1000 &  & 0 (0) & 0 (0) & 0 (0) & 0 (0) & 0 (0) &  & 0 (0) & 0 (0) & 0 (0) & 0 (0) & 0 (0) \\ 
   \hline $n$ & & \multicolumn{5}{c}{$p^* = 50; SNR =  1.25 ; \bar{\sigma} =  16.6 $} & & \multicolumn{5}{c}{$p^* = 50; SNR =  5 ;  \bar{\sigma} =  8.3 $} \\
 \cmidrule(l){3-7} \cmidrule(l){9-13} 
100 &  & 23.6 (1.3) & 40.4 (0.5) & 51.4 (0.8) & 55.1 (1.2) & 54.8 (1.2) &  & 4.2 (0.4) & 31.9 (0.5) & 47.9 (0.8) & 54.6 (1.3) & 53.5 (0.9) \\ 
  500 &  & 1.8 (0.2) & 17.1 (0.5) & 28.7 (0.8) & 42.3 (1.3) & 51.9 (2.2) &  & 0.2 (0) & 9.1 (0.4) & 17.8 (0.7) & 30 (1) & 46.6 (1.3) \\ 
  1000 &  & 0.7 (0.1) & 9.1 (0.4) & 15.4 (0.6) & 23.1 (0.9) & 32.5 (1.1) &  & 0 (0) & 3.9 (0.3) & 6.3 (0.4) & 11.1 (0.6) & 16.9 (0.8) \\ 
   \hline
\end{tabular}
\end{table}

\clearpage

\begin{table}[ht]
\centering
\caption{Average Hamming Distance for 100 replications of Relaxed Lasso. Standard errors given in parentheses.} 
\label{tab:RelHDSmall}
\begin{tabular}{rcp{0.9cm}p{0.9cm}p{0.9cm}p{0.9cm}p{0.9cm}cp{0.9cm}p{0.9cm}p{0.9cm}p{0.9cm}p{0.9cm}}
  \hline
  \hline
 & $p$ & 50 & 100 & 200 & 400 & 800 &  & 50 & 100 & 200 & 400 & 800 \\ 
   \hline $n$ & & \multicolumn{5}{c}{$p^* = 5; SNR =  1.25 ; \bar{\sigma} =  5.19 $} & & \multicolumn{5}{c}{$p^* = 5; SNR =  5 ;  \bar{\sigma} =  2.59 $} \\
 \cmidrule(l){3-7} \cmidrule(l){9-13} 
100 &  & 3.7 (0.4) & 5.4 (0.8) & 7.4 (1.1) & 18.4 (3.4) & 28.1 (4.2) &  & 1.1 (0.3) & 1.8 (0.5) & 4.5 (1.7) & 6.6 (2.3) & 7.9 (2.6) \\ 
  500 &  & 0.8 (0.3) & 1.2 (0.3) & 0.5 (0.2) & 0.7 (0.4) & 1.5 (0.6) &  & 1.2 (0.3) & 0.8 (0.2) & 0.8 (0.4) & 0.3 (0.1) & 0.1 (0) \\ 
  1000 &  & 0.6 (0.2) & 0.5 (0.2) & 0.7 (0.3) & 0.4 (0.1) & 1.9 (1.5) &  & 0.6 (0.2) & 0.8 (0.3) & 0.4 (0.2) & 0.4 (0.2) & 0.4 (0.2) \\ 
   \hline $n$ & & \multicolumn{5}{c}{$p^* = 50; SNR =  1.25 ; \bar{\sigma} =  16.6 $} & & \multicolumn{5}{c}{$p^* = 50; SNR =  5 ;  \bar{\sigma} =  8.3 $} \\
 \cmidrule(l){3-7} \cmidrule(l){9-13} 
100 &  & 19.5 (1.3) & 39.9 (0.5) & 57.6 (1.7) & 70.8 (3.1) & 67.9 (3.2) &  & 2.1 (0.3) & 34.1 (0.5) & 75.9 (2.5) & 87.6 (3.3) & 93.6 (3.8) \\ 
  500 &  & 0.9 (0.1) & 22.9 (0.9) & 39.6 (1.8) & 66.2 (3.4) & 74.5 (4.2) &  & 0.1 (0) & 7.2 (0.5) & 14.8 (1.2) & 25.4 (2.3) & 36.8 (2.7) \\ 
  1000 &  & 0.2 (0) & 12.3 (0.9) & 17.5 (1.4) & 27.9 (2.7) & 32.7 (2.9) &  & 0 (0) & 2.4 (0.3) & 3.1 (0.3) & 4.9 (0.5) & 5.5 (0.3) \\ 
   \hline
\end{tabular}
\end{table}

Results from increasing values of $p^*$ were considered for $n = \{100, 500, 1000\}$ and $p = 100$ in Figure \ref{fig:SimPlotHD}. Additional methods and results may be found in the Supplementary Materials. 
For $n = 100$ all methods perform comparably, except for the Relaxed Lasso with a minimum APE rule, which tends to have a higher number of false positives than the other methods, leading to larger average HD. As $n$ increases, Mod APE and AR2 CV appear to perform comparably to the Relaxed Lasso with a 1SE Rule, and have a consistently smaller average HD than traditional CV with a 1SE Rule. The new methods of CV consistently perform similarly to Relaxed 1SE and out performs the CV 1SE in terms of support recovery. 

\begin{figure}[ht]
    \centering
    \includegraphics[width =  \textwidth]{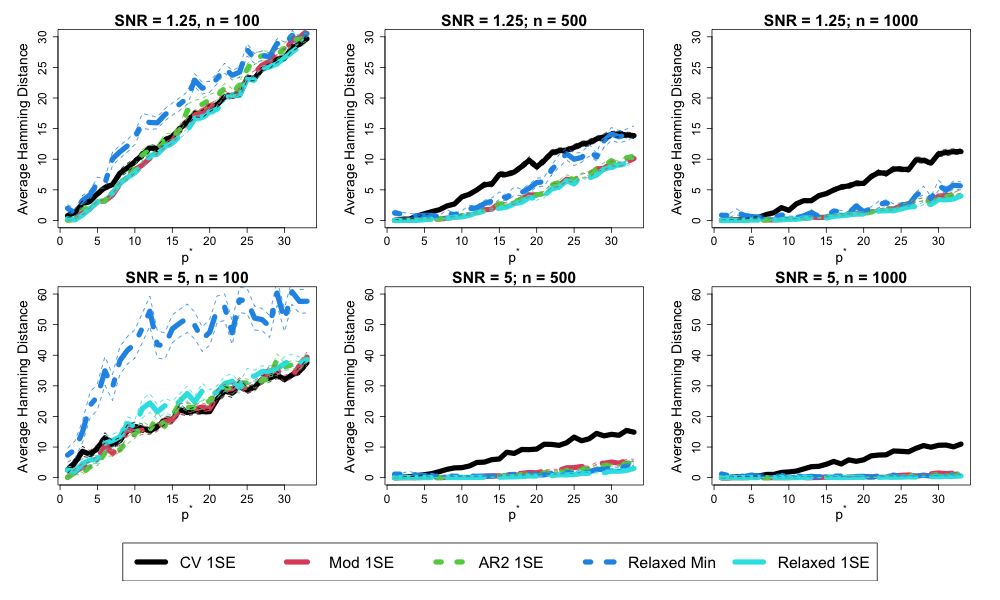}
    \caption{Average Hamming Distance between $\boldbeta^*$ and $\hat{\boldbeta}_\lambda$ from 100 replications of simulated data with $n = \{ 100, 500, 1000 \}$ and $p = 100$ with independent predictors for both SNR $= 1.25$ and SNR $= 5$. Thin dotted lines represent the mean $\pm$ one standard error.}
    \label{fig:SimPlotHD}
\end{figure}

\begin{figure}[ht]
    \centering
    \includegraphics[width = \textwidth]{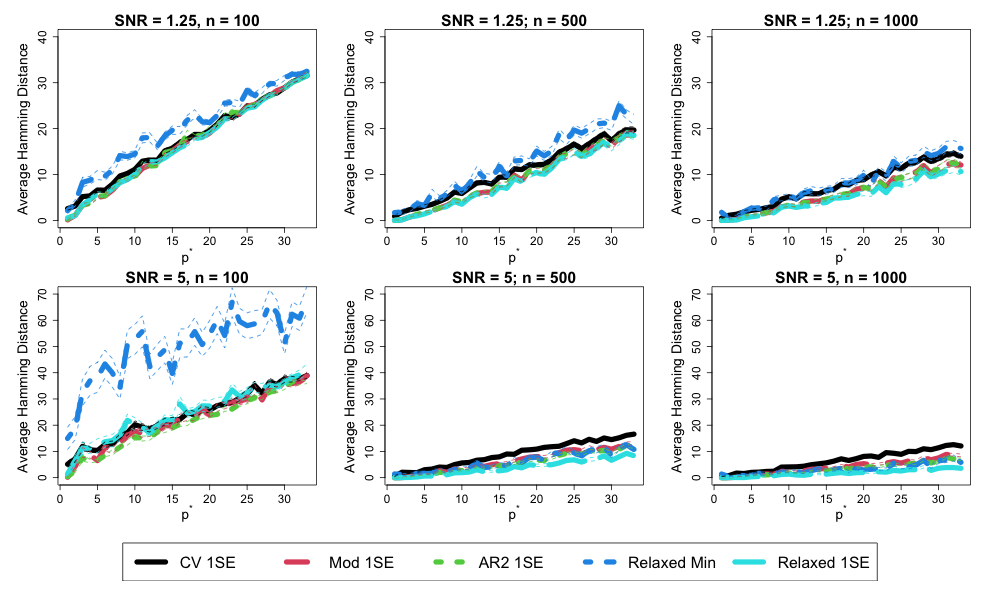}
    \caption{Average Hamming Distance between $\boldbeta^*$ and $\hat{\boldbeta}_\lambda$ from 100 replications of simulated data with $n = \{100, 500, 1000\}$ and $p = 100$ for correlated predictors. Thin dotted lines represent the mean $\pm$ one standard error.}
    \label{fig:SimPlotXCorrHD}
\end{figure}

We next considered correlated predictors where $\boldsymbol{\Sigma}_{X}$ satisfied $\Sigma_{ii} = 1$ and $\Sigma_{ij} = 0.75$ whenever $i \neq j$. 
Figure \ref{fig:SimPlotXCorrHD} gives the average HD  for 100 replications with $p = 100$ and $n = \{100,500,1000\}$, but further results can be found in the Supplementary Materials. Our two proposed CV methods are highly competitive with the Relaxed 1SE, whereas Relaxed Min has a tendency to overselect as $p^*$ increases, particularly for $n =100$. Similarly to the results from independent predictors, it is clear that as $n$ increases, the average Hamming distance for the Relaxed Lasso and the two new methods of CV decreases for all considered $p^*$. Relaxed Lasso with a minimum APE rule struggles when $n$ is small, even for a high signal-to-noise ratio, primarily due to overselection. Once again the new methods of CV generally have a smaller average HD than CV with a 1SE Rule and are similar in support recovery to the Relaxed Lasso with a 1SE Rule.

Unlike the Relaxed Lasso, the $\alpha$-modification is straightforward to apply to other penalties. The $\alpha$-modified Lasso was compared with the non-convex penalties SCAD and MCP, both traditionally and with an $\alpha$-modification. As expected, the $\alpha$-modification had a minimal effect on the predictive models. MCP and SCAD estimates were prone to underselection whereas the $\alpha$-modified Lasso estimates were more prone to overselection. Details may be found in the Supplementary Materials.

\section{Optimal EMG Placement for a Robotic Prosthesis Controller}
\label{sec:EMGApplication}
For the EMG application introduced in Section 1, we want to identify as few EMG sensors as needed to reliably predict hand movement. The data were collected from an able-bodied subject and consist of concurrent measurements of the subject's finger position and 16 EMG signals as predictors. A full description of the data, some of its challenges, and SAFE can be found in \cite{SAFE}. Six data sets were analyzed, corresponding to three consistent finger movement patterns (FC1, FC2, FC3) and three random patterns (FR1, FR2, FR3). As data were collected from an able-bodied subject, it was known that three of the 16 sensors, denoted $X_5,X_7,X_{12}$, targeted muscles known to fully explain finger movement. Sensors $X_5$ and $X_7$ collected information from the same muscle, however, and so only one of the pair is necessary to predict finger position. An ideal model would thus include $X_{12}$ and either $X_5$ or $X_7$, but recovery of all three sensors is also acceptable.

Due to known biomechanical features of hand movement, \cite{SAFE} use finger velocity as the response and treat the recent past EMG signals as functional covariates. The model is
\begin{equation}
    y_i = \sum_{j=1}^{16} \int_{-\delta}^0 X_{ij}(t)\gamma_j(t, z_i)dt + \epsilon_i\ ,\
    \label{eq:EMGModel}
\end{equation}
where $y_i$ is the velocity, $X_{ij}(t)$ represents past EMG signals, $t \in [-\delta,0]$ is the recent past time, and $z_i$ is the recent finger position. Although this model is not the linear model introduced in (\ref{eq:LinearModel}), its approximation via basis expansion allows it to be treated similarly to the linear model. 
Following \cite{Gertheiss2013}, \cite{SAFE} proposed a penalized estimation procedure
where the penalty accounted for both sparsity and smoothness of the $\gamma_j(\cdot,\cdot)$:
\begin{equation}
    P_{\lambda}(\gamma_j) = \lambda (f_j ||\gamma_j||_2^2 + g_j \lambda_t||\gamma_{j,t}''||_2^2 + h_j \lambda_z||\gamma_{j,z}''||_2^2)^{1/2}
    \label{eq:EMGPen}
\end{equation} 
where $||\gamma_j||_2^2=\int\int \gamma_j(t,z_i)^2 dtdz$ and $\gamma_{j,t}''=\partial^2 \gamma_j(t,z_i) /\partial t^2$. There are three tuning parameters, $\lambda, \lambda_t,$ and $\lambda_z$, and adaptive weights $f_k$, $g_k$, and $h_k$.  To facilitate estimation, the $\gamma_j(\cdot,\cdot)$ were written using a tensor product basis expansion, leading to a Group Lasso-type penalty; more details may be found in \cite{SAFE} and the supplementary materials. 

To perform variable selection, \cite{SAFE} proposed
Sequential Adaptive Functional Estimation (SAFE) that performs penalized estimation in stages. The first stage set $f_j=g_j=h_j=1$ and chose optimal tuning parameters based on an APE 1SE rule following 10-fold CV. Let $\mathcal{M}_{\lambda,1}$ denote the support of this estimator. Adaptive weights were updated based on the estimates of the nonzero effects and penalized estimation was performed again using these weights and only those $j \in \mathcal{M}_{\lambda,1}$. The process was repeated for up to 5 stages. While effective in identifying the correct submodel, the analysis can be very time consuming. We modified their method based on AR2 CV and modified APE for this application in hopes to reduce the number of stages required to perform variable selection.






Table \ref{tab:EMGFull} gives the variable selection results for the three CV methods based on the initial stage; results from subsequent stages of SAFE can be found in the supplementary materials. AR2 CV and Mod APE generally give smaller models than traditional APE CV, with one exception: the Average $R^2$ method has a large model size for FC3. On average, however, both new methods have fewer false positives and smaller model sizes. Mod APE gives very similar results to APE, suggesting that the $\alpha$-modified CV approach requires further study for the Group Lasso. Although AR2 CV is not perfect in this application, in general it reduces model size and decreases false positives at no additional computational cost.

\begin{table}[ht]
\centering
\caption{Variable selection results for EMG finger movements without adaptive weighting. FP indicates the total number of false positives in the model, and  Size is the total number of EMG signals contained in the model.} 
\begin{tabular}{rr|ccccccc}
  \hline
 &  & FC1 & FC2 & FC3 & FR1 & FR2 & FR3 & Mean \\ 
  \hline
\multirow{3}{*}{APE} & FP & 2 & 1 & 2 & 2 & 1 & 1 & 1.5 \\ 
   & Size & 4 & 3 & 4 & 5 & 4 & 4 & 4 \\ 
  \multirow{3}{*}{AR2} & FP & 0 & 1 & 3 & 0 & 1 & 1 & 1 \\ 
   & Size & 2 & 3 & 6 & 2 & 4 & 3 & 3.33 \\ 
  \multirow{3}{*}{Mod} & FP & 1 & 1 & 2 & 2 & 1 & 1 & 1.33 \\  
   & Size & 3 & 3 & 4 & 5 & 4 & 4 & 3.83 \\ 
   \hline
\end{tabular}
\label{tab:EMGFull}
\end{table}

\section{Discussion}
\label{sec:Discussion}
In this paper, we proposed AR2 CV to choose tuning parameters to balance support recovery and prediction performance. This led to the $\alpha$-Modification, a  multiplicative adjustment to predictions from penalized estimates which can also be used for  $\alpha$-modified CV. The $\alpha$-modification is simple and efficient, making it an attractive option when less flexible approaches are unavailable. 
A simulation study on the capabilities of AR2 and $\alpha$-modified CV found that their variable selection results were highly competitive with---or, in some cases, better than---the Relaxed Lasso. In order to ensure fair comparison, we introduced a 1SE Rule for the Relaxed Lasso. Finally, we applied the approaches to a functional data problem in a demonstration of their flexibility.

The $\alpha$-modification and the tuning parameter selection methods proposed here inspire several research questions.
First, further theoretical analysis of the methods is of interest. Because the two new methods of tuning parameter selection are CV-based approaches, the theoretical properties of CV are central.  Theoretical justifications for the use of CV for penalized estimators are relatively new and still evolving. \cite{CVLassoTheory} may be extended to show that $\alpha$-modified and AR2 CV lead to estimates that are low bias and appropriately sparse.
A second area of future research is the theoretical properties of $\alpha$-modified estimates themselves. In Theorem \ref{thm:Consistency}, it was shown that any consistent estimator will still be consistent with the $\alpha$-Modification. We posit that the rates of convergence for $\alpha$-modified estimators are faster than unmodified, but proof of this conjecture is the work of future research. Similarly, many of the results from this paper assume finite sample sizes. Further study is needed to determine more of the asymptotic properties of $\alpha$-modified estimators and to adapt the specific results given for the Lasso penalty to other penalties.

There are also a few extensions and adaptations of the $\alpha$-Modification that may prove fruitful. We are currently expanding the $\alpha$-modification to  Generalized Linear Models (GLMs). This extension will require an iterative algorithm to find estimates of $\alpha$ because closed form solutions do not exist and an accommodation for the inclusion of an intercept term will be necessary. Additionally, as noted in some of the theoretical results in this paper, the $\alpha$-modification does not always reduce bias. We are interested in exploring a further penalty on $\alpha$ itself to ensure bias reduction. Finally, the calculation of $\hat{\alpha}_\lambda$ described here uses in-sample predictions and observations. As overspecification is a particular concern, the question of whether out-of-sample data can be used to find estimates of $\alpha$ is another subject of further research. 

\section*{Acknowledgements}
\noindent
This work was partially supported by the National Science Foundation, grant IOS--2039226.

\section*{Supplementary Materials}

The supplementary material in the following document includes three sections. Section A gives more details of the proofs for the theorems and lemmas presented in the paper. Section B provides tables and figures that more fully explore the simulation studies discussed in the Numerical Results. This section also evaluates results from a simulation study of the non-convex SCAD and MC+ penalties. Finally, Section C describes the approximations used to achieve the estimates from the EMG data application and includes results from multiple stages of adaptive estimation.

\section*{A. Detailed Proofs for Theoretical Results}

\begin{proof}[Proof of Theorem~1]
	Suppose there is at least one element of $\hat{\boldbeta}_\lambda$ not equal to 0. Let $\lambda \V{g}(\boldbeta) = \sum_{\ell=1}^L \lambda_\ell g_\ell(\boldbeta)$ and $\nabla\V{g}^*$ be the subgradient vector of $\V{g}(\boldbeta)$. Then \begin{equation}
		-\frac{1}{n}\V{X}^T(\V{y} - \V{X}\hat{\boldbeta}_\lambda) + \nabla\V{g}^* = 0
	\end{equation} due to the KKT conditions. Further, we can say that there exists some generalized inverse of $\V{X}^T\V{X}$, $(\V{X}^T\V{X})^{-}$, such that $\hat{\boldbeta}_\lambda = (\V{X}^T\V{X})^{-}(\V{X}^T\V{y} -  n \nabla\V{g}^*)$. Therefore:
	\begin{equation}
		\hat{\V{y}}_\lambda^T\V{y} = \hat{\V{y}}_\lambda^T\hat{\V{y}}_\lambda + n \lambda \nabla\V{g}^{*T}(\V{X}^T\V{X})^{-}\V{X}^T\V{y} - n^2  \nabla\V{g}^{*T} (\V{X}^T\V{X})^{-} \nabla\V{g}^{*} \ . \
	\end{equation}
	Given this definition, $\hat{\alpha}_\lambda$ may be written:
	\begin{align}
		\hat{\alpha}_\lambda 
		&= \frac{\hat{\V{y}}_\lambda^T\hat{\V{y}}_\lambda + n  \nabla\V{g}^{*T}(\V{X}^T\V{X})^{-}(\V{X}^T\V{y} - n \nabla\V{g}^*) }{\hat{\V{y}}_\lambda^T\hat{\V{y}}_\lambda}\\
		&= \frac{\hat{\V{y}}_\lambda^T\hat{\V{y}}_\lambda + n  \nabla\V{g}^{*T}\hat{\boldbeta}_\lambda }{\hat{\V{y}}_\lambda^T\hat{\V{y}}_\lambda}\\
		&= 1 + \frac{n  \nabla\V{g}^{*T}\hat{\boldbeta}_\lambda}{\hat{\V{y}}_\lambda^T\hat{\V{y}}_\lambda}.
	\end{align}
	Because the denominator of the second term is greater than 0, as is $n$, this expression is greater than or equal to 1 whenever $\nabla\V{g}^{*T}\hat{\boldbeta}_\lambda$ is greater than or equal to zero. The vector $\nabla\V{g}^{*}$  the subgradient of a convex function at $\hat{\boldbeta}_\lambda$, therefore for any $\boldbeta$:
	\begin{align}
		g(\boldbeta) & \geq g(\hat{\boldbeta}_\lambda) + \nabla\V{g}^{*T}(\boldbeta - \hat{\boldbeta}_\lambda)\\
		\nabla\V{g}^{*T} \hat{\boldbeta}_\lambda & \geq g(\hat{\boldbeta}_\lambda) - g(\boldbeta) + \nabla\V{g}^{*T} \boldbeta.
	\end{align}
	Because this is true for all $\boldbeta$ vectors, it must be the case when $\boldbeta = \V{0}$. Hence, we have:
	\begin{align}
		\nabla\V{g}^{*T} \hat{\boldbeta}_\lambda & \geq g(\hat{\boldbeta}_\lambda) - g(\V{0}) \geq 0.
	\end{align}
	Thus $\hat{\alpha}_\lambda \geq 1.$
\end{proof}

%
\begin{proof}[Proof of Lemma~1]
	When $\lVert \hat{\boldbeta}_\lambda \rVert_2^{-1} \hat{\boldbeta}_\lambda = \boldsymbol{\xi}^*$, $\hat{\boldbeta}_\lambda = \lVert \hat{\boldbeta}_\lambda \rVert_2 \boldsymbol{\xi}^*.$ Write $\V{y} = \alpha^* \V{X} \boldsymbol{\xi}^* + \boldsymbol{\epsilon}$, to define:
	\begin{equation}
		\hat{\alpha}_\lambda = \frac{\alpha^* \boldsymbol{\xi}^{*T}\V{X}^T\V{X}\boldsymbol{\xi}^* + \boldsymbol{\xi}^*\V{X}^T\boldsymbol{\epsilon}}{\lVert \hat{\boldbeta}_\lambda \rVert_2 \boldsymbol{\xi}^{*T}\V{X}^T\V{X}\boldsymbol{\xi}^*} \ . \
	\end{equation}
	Therefore:
	\begin{align}
		\hat{\alpha}_\lambda \hat{\boldbeta}_\lambda &=  \hat{\alpha}_\lambda \lVert \hat{\boldbeta}_\lambda \rVert_2 \boldsymbol{\xi}^*\\
		&= \frac{1}{\lVert \hat{\boldbeta}_\lambda \rVert_2} \Big( \alpha^* + \frac{\boldsymbol{\xi}^{*T} \V{X}^T \boldsymbol{\epsilon}}{\boldsymbol{\xi}^{*T} \V{X}^T \V{X} \boldsymbol{\xi}^*} \Big)\lVert \hat{\boldbeta}_\lambda \rVert_2 \boldsymbol{\xi}^* \\
		&= \alpha^* \boldsymbol{\xi}^* + \Big(\frac{\boldsymbol{\xi}^{*T} \V{X}^T \boldsymbol{\epsilon}}{\boldsymbol{\xi}^{*T} \V{X}^T \V{X} \boldsymbol{\xi}^*} \Big) \boldsymbol{\xi}^* \ . \
	\end{align}
	Thus $E[\hat{\alpha}_\lambda \hat{\boldbeta}_\lambda] = \alpha^* \boldsymbol{\xi}^* = \boldbeta^*$ and the estimator is unbiased.
\end{proof}

\begin{proof}[Proof of Lemma~2]
	Fix $\lambda$ and let $\hat{\boldbeta}_\lambda = a \hat{\boldbeta}^{\mathcal{M}_\lambda}_{OLS}$ where $\hat{\boldbeta}^{\mathcal{M}_\lambda}_{OLS}$ is the OLS estimate for submodel $\mathcal{M}_\lambda$. Let $\V{X}_{\mathcal{M}_\lambda}$ be the submatrix of $\V{X}$ containing only the columns indexed by $\mathcal{M}_\lambda$. Then $\hat{\boldbeta}^{\mathcal{M}_\lambda}_{OLS} = (\V{X}_{\mathcal{M}_\lambda}^T\V{X}_{\mathcal{M}_\lambda})^{-1}\V{X}_{\mathcal{M}_\lambda}^T\V{y}$. Now: 
	\begin{equation}
		\hat{\alpha}_\lambda =  \frac{a \V{y}^T \V{X}_{\mathcal{M}_\lambda}(\V{X}_{\mathcal{M}_\lambda}^T\V{X}_{\mathcal{M}_\lambda})^{-1}\V{X}_{\mathcal{M}_\lambda}^T\V{y}}{a^2 \V{y}^T \V{X}_{\mathcal{M}_\lambda}(\V{X}_{\mathcal{M}_\lambda}^T\V{X}_{\mathcal{M}_\lambda})^{-1}\V{X}_{\mathcal{M}_\lambda}^T\V{X}_{\mathcal{M}_\lambda}(\V{X}_{\mathcal{M}_\lambda}^T\V{X}_{\mathcal{M}_\lambda})^{-1}\V{X}_{\mathcal{M}_\lambda}^T\V{y}} = \frac{1}{a} \ . \
	\end{equation}
	Therefore:
	\begin{equation}
		\hat{\alpha}_\lambda \hat{\boldbeta}_\lambda = \frac{1}{a}\times a \hat{\boldbeta}_{OLS}^{ \mathcal{M}_\lambda} = \hat{\boldbeta}_{OLS}^{ \mathcal{M}_\lambda} \ . \ 
	\end{equation}
\end{proof}


\begin{proof}[Proof of Theorem~2]
	Consider $\hat{\alpha}_{\lambda_n}:$
	\begin{equation}
		\hat{\alpha}_{\lambda_n} = \frac{\frac{1}{n}\hat{\boldbeta}_{\lambda_n}^T \V{X}_n^T \V{y}}{\frac{1}{n} \hat{\boldbeta}_{\lambda_n}^T \V{X}_n^T \V{X}_n \hat{\boldbeta}_{\lambda_n}} \ . \
	\end{equation}
	By Continuous Mapping Theorem, the denominator of this expression converges in probability to $c^2 \boldsymbol{\xi}^{*T} \V{C} \boldsymbol{\xi}^*$. The numerator is equal to:
	\begin{equation}
		\frac{1}{n}\hat{\boldbeta}_{\lambda_n}^T \V{X}_n^T \V{X}_n \boldbeta + \hat{\boldbeta}_{\lambda_n}^T \V{X}_n^T \boldsymbol{\epsilon} \rightarrow_p \alpha^* c \boldsymbol{\xi}^{*T} \V{C} \boldsymbol{\xi}^* \ . \
	\end{equation}
	Then $\hat{\alpha}_{\lambda_n} \rightarrow_p \frac{\alpha^*}{c}$ and, once again by Continuous Mapping Theorem, $\hat{\alpha}_{\lambda_n} \hat{\boldbeta}_{\lambda_n} \rightarrow \frac{\alpha^*}{c} c\boldsymbol{\xi}^* = \alpha^* \boldsymbol{\xi}^* = \boldbeta^*$. 
\end{proof}



\begin{proof}[Proof of Theorem~3]
	Average Prediction Error for $K$-fold CV is calculated as:
	\begin{equation}
		\frac{1}{K} \sum_k \frac{1}{n_k} \sum_{i} (y_{k_i} - \V{x}_{k_i}^T \hat{\boldbeta}_\lambda)^2 . \
	\end{equation}
	To show that the expected value of $\alpha$-Modified APE is less than that of regular APE, it is sufficient to show that $E[Mod PE] < E[PE]$ where $PE = \sum_{i} (y_{k_i} - \V{x}_{k_i}^T \hat{\boldbeta}_\lambda)^2$ and $Mod PE = \sum_{i} (y_{k_i} - \hat{\alpha}_\lambda \V{x}_{k_i}^T \hat{\boldbeta}_\lambda)^2$. When the direction of $\boldbeta^*$ has been captured by $\hat{\boldbeta}_\lambda$, we can write the estimate as $\lVert \hat{\boldbeta}_\lambda \rVert_2 \boldsymbol{\xi}^*.$ Expand $y_{k_i} = \alpha^* \V{x}_{k_i}^T \boldsymbol{\xi}^* + \epsilon_i$ to find:
	\begin{equation}
		PE = \sum_i \big( \epsilon_i - \V{x}_{k_i}^T \boldsymbol{\xi}^*(\lVert \hat{\boldbeta}_\lambda \rVert_2 - \alpha^*) \big)^2 \ . \
	\end{equation}
	Similarly,
	\begin{equation}
		Mod PE = \sum_i \Bigg( \epsilon_i - \V{x}_{k_i}^T \boldsymbol{\xi}^*\Big(\frac{\boldsymbol{\xi}^{*T}\V{X}^T \boldsymbol{\epsilon}}{\boldsymbol{\xi}^{*T}\V{X}^T \V{X} \boldsymbol{\xi}^*}\Big) \Bigg)^2 \ . \ \end{equation}
	Let $\boldsymbol{\Sigma}$ be the covariance matrix for $\boldsymbol{\epsilon}$. Therefore, $\boldsymbol{\Sigma} = \sigma^2 \V{I}$ and
	\begin{align}
		E[PE] &= n_k \sigma^2 + \sum_i (\V{x}_{k_i}^T \boldsymbol{\xi}^*)^2 E[(\lVert \hat{\boldbeta}_\lambda \rVert_2 - \alpha^*)^2]\\
		E[ModPE] &= E \Bigg[ \sum_i \Bigg( \epsilon_i - \V{x}_{k_i}^T \boldsymbol{\xi}^*\Big(\frac{\boldsymbol{\xi}^{*T}\V{X}^T \boldsymbol{\epsilon}}{\boldsymbol{\xi}^{*T}\V{X}^T \V{X} \boldsymbol{\xi}^*}\Big) \Bigg)^2 \Bigg]\\
		&= n_k \sigma^2 + \sum_i \frac{(\V{x}_{k_i}^T \boldsymbol{\xi}^*)^2 E[\boldsymbol{\epsilon}^T \V{X} \boldsymbol{\xi}^* \boldsymbol{\xi}^{*T} \V{X}^T \boldsymbol{\epsilon}]}{(\boldsymbol{\xi}^{*T}\V{X}^T \V{X} \boldsymbol{\xi}^*)^2} \\
		&= n_k \sigma^2 + \sum_i \frac{(\V{x}_{k_i}^T \boldsymbol{\xi}^*)^2 (tr(\boldsymbol{\xi}^{*T}\V{X}^T \Sigma \V{X} \boldsymbol{\xi}^*) + E[\boldsymbol{\epsilon}^T] \V{X} \boldsymbol{\xi}^* \boldsymbol{\xi}^{*T} \V{X}^T E[\boldsymbol{\epsilon}] )}{(\boldsymbol{\xi}^{*T}\V{X}^T \V{X} \boldsymbol{\xi}^*)^2} \\
		&= n_k \sigma^2 + \sum_i \frac{(\V{x}_{k_i}^T \boldsymbol{\xi}^*)^2 \sigma^2( \boldsymbol{\xi}^{*T}\V{X}^T  \V{X} \boldsymbol{\xi}^* )}{(\boldsymbol{\xi}^{*T}\V{X}^T \V{X} \boldsymbol{\xi}^*)^2} \\
		&= n_k \sigma^2 + \frac{\sigma^2}{\boldsymbol{\xi}^{*T}\V{X}^T \V{X} \boldsymbol{\xi}^*} \sum_i (\V{x}_{k_i}^T \boldsymbol{\xi}^*)^2
	\end{align}
	Therefore $E[PE] \geq E[ModPE]$ whenever
	\begin{align}
		&E[(\lVert \hat{\boldbeta}_\lambda \rVert_2 - \alpha^*)^2] \geq \frac{\sigma^2}{\boldsymbol{\xi}^{*T}\V{X}^T \V{X} \boldsymbol{\xi}^*}\\
		\Leftrightarrow &\frac{\alpha^{*2}}{E[(\lVert \hat{\boldbeta}_\lambda \rVert_2 - \alpha^*)^2]} \leq \frac{\boldbeta^{*T}\V{X}^T \V{X}  \boldbeta^*}{\sigma^2}
	\end{align}
	because $\boldbeta^* = \alpha^* \boldsymbol{\xi}^*$.
	Therefore we have the result.
\end{proof}

	\begin{proof}[Proof of Lemma~3]
		When $\V{X}$ is orthonormal,
		\begin{equation}
			\hat{\alpha}_\lambda =  \frac{\hat{\boldbeta}_\lambda^T \hat{\boldbeta}_{OLS}}{\hat{\boldbeta}_\lambda^T \hat{\boldbeta}_\lambda}= \frac{\sum_{j = 1}^p |\hat{\beta}_{OLS,j}|(|\hat{\beta}_{OLS,j}| - \lambda)_+}{\sum_{j = 1}^p (|\hat{\beta}_{OLS,j}| - \lambda)^2_+ } \ . \
		\end{equation}
		This gives the following expression:
		\begin{equation}
			\hat{\alpha}_\lambda \hat{\beta}_{\lambda, j^*} = \Bigg(\frac{\sum_{j = 1}^p |\hat{\beta}_{OLS,j}|(|\hat{\beta}_{OLS,j}| - \lambda)_+}{\sum_{j = 1}^p (|\hat{\beta}_{OLS,j}| - \lambda)^2_+  }\Bigg) \times \textrm{sign}(\hat{\beta}_{OLS,j^*}) \times  (|\hat{\beta}_{OLS,j^*}| - \lambda)_+ \ . \
		\end{equation}
		Letting $d_j = (|\hat{\beta}_{OLS,j}|-\lambda)_+$ and $s_{j*}=\textrm{sign}(\hat{\beta}_{OLS,j^*})$, the expression becomes:
		\begin{align}
			\hat{\alpha}_\lambda \hat{\beta}_{\lambda, j^*}
			&=s_{j*}|\hat{\beta}_{OLS,j*}|\frac{d_{j*}^2}{\sum_{j = 1}^p d_j^2 } + s_{j*}d_{j*}\frac{\sum_{j \neq j*}|\hat{\beta}_{OLS,j}|d_j}{\sum_{j = 1}^p d_j^2 }\\
			&=w_1\hat{\beta}_{OLS,j*} + \hat{\beta}_{\lambda, j^*}\frac{\sum_{j \neq j*}(d_j+\lambda)d_j}{\sum_{j = 1}^p d_j^2}\ , \
		\end{align}
		which uses $s_{j*}|\hat{\beta}_{OLS,j*}|=\hat{\beta}_{OLS,j*}$, $w_1=d_{j*}^2/\sum_j d_j^2$, $\hat{\beta}_{\lambda, j^*}=s_{j*}d_{j*}$, and $|\hat{\beta}_{OLS,j}|=(|\hat{\beta}_{OLS,j}|-\lambda)_++\lambda$ whenever $d_j > 0$.  Therefore:
		\begin{align}
			\hat{\alpha}_\lambda \hat{\beta}_{\lambda, j^*} &=w_1\hat{\beta}_{OLS,j*} + \hat{\beta}_{\lambda, j^*}\left(\frac{\sum_{j \neq j*} d_j^2}{\sum_{j = 1}^p d_j^2}
			+\lambda \frac{\sum_{j \neq j*} d_j}{\sum_{j = 1}^p d_j^2}\right)\\
			&=w_1\hat{\beta}_{OLS,j*} + (1- w_1)\hat{\beta}_{\lambda, j^*} + w_2 \hat{\beta}_{\lambda, j^*} \ ,\
		\end{align}
		where $w_2=\lambda \frac{\sum_{j \neq j*} d_j}{\sum_{j = 1}^p d_j^2}$. 
	\end{proof}
	
		\begin{proof}[Proof of Theorem~4]
			For $|\hat{\beta}_{OLS,j^*}| \leq \lambda$, we have $\hat{\alpha}_\lambda \hat{\beta}_{\lambda,j^*}=0$ so $|\hat{\alpha}_\lambda \hat{\beta}_{\lambda,j^*} - \hat{\beta}_{OLS,j^*}| = |\hat{\beta}_{OLS,j^*}| \leq \lambda$ so the theorem is satisfied for this range of values. Next we consider $\hat{\beta}_{OLS,j^*} > \lambda$. By Lemma~3,
			\begin{align}
				\hat{\alpha}_\lambda \hat{\beta}_{\lambda,j^*} - \hat{\beta}_{OLS,j^*} &= (w_1 - 1) \hat{\beta}_{OLS,j^*} + (1 - w_1 + w_2) \hat{\beta}_{\lambda,j^*}\\
				&= w_2 \hat{\beta}_{OLS,j^*} + (1 - w_1 + w_2)(-\lambda)\\
				&= \frac{\lambda (\sum_{j\neq j^*}d_j)\hat{\beta}_{OLS,j^*} - \lambda (\sum_{j\neq j^*}d_j^2 + \lambda \sum_{j\neq j^*}d_j)}{(\hat{\beta}_{OLS,j^*} - \lambda)^2 + \sum_{j\neq j^*}d_j^2}
			\end{align}
			Let $u = \sum_{j\neq j^*}d_j$, $v = \sum_{j\neq j^*}d_j^2$, and $x = \hat{\beta}_{OLS,j^*}=\beta^*_{j^*} + n^{-1}\V{x}_j^T\boldsymbol{\epsilon}$. This gives a function, $f(x)$, that is differentiable for $x \geq \lambda$:
			\begin{align}
				f(x) &= \frac{\lambda u x - \lambda(v + \lambda u)}{(x - \lambda)^2 + v}\\
				\frac{\partial f(x)}{\partial x} &= \frac{u\lambda}{(x - \lambda)^2 + v} - \frac{2(x - \lambda)(u\lambda x - \lambda(u\lambda + v))}{((x - \lambda)^2 + v)^2} \ .\
			\end{align}
			Treating $n^{-1}\V{x}_j^T\boldsymbol{\epsilon}$ as a constant, changes in $x$ are due to changes in $\beta^*_{j^*}$ alone. The derivative $\frac{\partial f(x)}{\partial x}$ is equal to zero whenever $x = x^*$, where $x^*$ is defined:
			\begin{align}
				x^* = \frac{u \lambda + v \pm \sqrt{u^2v + v^2}}{u} \ . \
			\end{align}
			Note that $x^* < \lambda$ when the negative part of the square root is taken, therefore the root is restricted to $x^* = (u \lambda + v + \sqrt{u^2v + v^2})/u.$ The second derivative of $f(x)$ is:
			\begin{align}
				\frac{\partial^2f(x)}{\partial x^2} = \frac{2\lambda [-3v(x - \lambda)(u+x-\lambda) + u(x-\lambda)^3 + v^2]}{(v + (x - \lambda)^2)^3} \ . \
			\end{align}
			When $x = x^*$, the second derivative equals:
			\begin{align}
				-\frac{u^4 \lambda(u^2 + v)(\sqrt{u^2 v + v^2} + v)}{2v^2(\sqrt{u^2 v + v^2} + u^2 + v)^3} < 0
			\end{align}
			meaning that $x^*$ gives the maximum of $f(x)$ when $x > \lambda$. For $\lambda \leq x \leq x^*$, $f(x)$ is an increasing function bounded below by $-\lambda$ and bounded above by
			\begin{align}
				\frac{\lambda}{2}\Big(\sqrt{\frac{u^2}{v} + 1 } - 1\Big)\ .\
			\end{align}
			For $x > x^*$, $f(x)$ is decreasing so we consider $\lim_{x \to \infty} f(x)$, which is easily shown to equal $0> -\lambda$. Therefore,
			\begin{equation}
				|f(x)| \leq \lambda \times \max\left(1, \ \frac{1}{2}\Big(\sqrt{\frac{u^2}{v } + 1 } - 1\Big) \right)\ ,\
			\end{equation}
			and $\lim_{x \to \infty} |f(x)| = 0$. For $\hat{\beta}_{OLS,j^*} < -\lambda$, note that $\hat{\alpha}_\lambda \hat{\beta}_{\lambda,j^*} - \hat{\beta}_{OLS,j^*}=-(\hat{\alpha}_\lambda |\hat{\beta}_{\lambda,j^*}| - |\hat{\beta}_{OLS,j^*}|)$ and so analogous arguments to the case of $\hat{\beta}_{OLS,j^*}$ also apply. This completes the proof. \end{proof}

		\begin{proof}[Proof of Theorem~5]
			If $\hat{\boldbeta}_\lambda$ recovers the true sign vector, $\V{s}$, then its nonzero estimates are $\hat{\boldbeta}_\lambda^{\mathcal{M}^*} = \hat{\boldbeta}_{OLS}^{\mathcal{M}^*} - n\lambda (\V{X}_{\mathcal{M}^*}^T \V{X}_{\mathcal{M}^*})^{-1} \V{s}$ and
			\begin{align}
				\hat{\alpha}_\lambda \hat{\boldbeta}_{\lambda} &= \left( \frac{\hat{\V{y}}_\lambda^T \V{y} }{\hat{\V{y}}_\lambda^T\hat{\V{y}}_\lambda} \right) \hat{\boldbeta}_{\lambda}.
			\end{align}
			Now, we can find definitions of these components in terms of $\hat{\boldbeta}_{OLS}^{\mathcal{M}^*}$:
			\begin{align}
				\hat{\V{y}}_\lambda^T \V{y} &= \hat{\boldbeta}_\lambda^{\mathcal{M}^*\,T} \V{X}_{\mathcal{M}^*}^T \V{y}\\
				&= \left(\hat{\boldbeta}_{OLS}^{\mathcal{M}^* \, T} - n\lambda \V{s}^T(\V{X}_{\mathcal{M}^*}^T \V{X}_{\mathcal{M}^*})^{-1}\right)\V{X}_{\mathcal{M}^*}^T
				\V{y}\\
				&= \hat{\boldbeta}_{OLS}^{\mathcal{M}^* \, T}\V{X}_{\mathcal{M}^*}^T\V{y} 
				- n\lambda \V{s}^T(\V{X}_{\mathcal{M}^*}^T \V{X}_{\mathcal{M}^*})^{-1}\V{X}_{\mathcal{M}^*}^T\V{y}\\
				&= \hat{\boldbeta}_{OLS}^{\mathcal{M}^*\, T} \V{X}_{\mathcal{M}^*}^T \V{X}_{\mathcal{M}^*}(\V{X}_{\mathcal{M}^*}^T \V{X}_{\mathcal{M}^*})^{-1} \V{X}_{\mathcal{M}^*}^T\V{y} 
				- n\lambda \V{s}^T\hat{\boldbeta}_{OLS}^{\mathcal{M}^*}\\
				&=\hat{\boldbeta}_{OLS}^{\mathcal{M}^*\, T} \V{X}_{\mathcal{M}^*}^T \V{X}_{\mathcal{M}^*}\hat{\boldbeta}_{OLS}^{\mathcal{M}^*}
				- n\lambda \V{s}^T\hat{\boldbeta}_{OLS}^{\mathcal{M}^*}\\
				\hat{\V{y}}_\lambda^T \hat{\V{y}}_\lambda &= \hat{\boldbeta}_\lambda^{\mathcal{M}^*\,T} \V{X}_{\mathcal{M}^*}^T \V{X}_{\mathcal{M}^*}\hat{\boldbeta}_\lambda^{\mathcal{M}^*}\\
				&=\left(\hat{\boldbeta}_{OLS}^{\mathcal{M}^* \, T} - n\lambda \V{s}^T(\V{X}_{\mathcal{M}^*}^T \V{X}_{\mathcal{M}^*})^{-1}\right)\V{X}_{\mathcal{M}^*}^T \V{X}_{\mathcal{M}^*}\left(\hat{\boldbeta}_{OLS}^{\mathcal{M}^*} - n\lambda (\V{X}_{\mathcal{M}^*}^T \V{X}_{\mathcal{M}^*})^{-1}\V{s}\right)\\
				&=\hat{\boldbeta}_{OLS}^{\mathcal{M}^*\,T }\V{X}_{\mathcal{M}^*}^T \V{X}_{\mathcal{M}^*} \hat{\boldbeta}_{OLS}^{\mathcal{M}^*} - 2n\lambda \V{s}^T  \hat{\boldbeta}_{OLS}^{\mathcal{M}^*} + n^2 \lambda^2 \V{s}^T (\V{X}_{\mathcal{M}^*}^T \V{X}_{\mathcal{M}^*})^{-1} \V{s}\ .\
			\end{align}
			Then both $\hat{\V{y}}_\lambda^T \V{y}$ and $\hat{\V{y}}_\lambda^T \hat{\V{y}}_\lambda$ are quadratic polynomials with respect to $\hat{\beta}_{OLS,j^*}^{\mathcal{M}^*}$ with a quadratic coefficients of $n$ due to the scaling of the columns of $\boldsymbol{X}$. We also have
			\begin{align}
				\hat{\V{y}}_\lambda^T \V{y} - \hat{\V{y}}_\lambda^T \hat{\V{y}}_\lambda &= n\lambda \V{s}^T  \hat{\boldbeta}_{OLS}^{\mathcal{M}^*}-  n^2 \lambda^2 \V{s}^T (\V{X}_{\mathcal{M}^*}^T \V{X}_{\mathcal{M}^*})^{-1} \V{s}\\
				&=n\lambda s_{j^*} \hat{\beta}_{OLS,j^*}^{\mathcal{M}^*} + n\lambda \sum_{j\neq j^*} s_j \hat{\beta}_{OLS,j}^{\mathcal{M}^*} - n^2 \lambda^2 \V{s}^T (\V{X}_{\mathcal{M}^*}^T \V{X}_{\mathcal{M}^*})^{-1} \V{s}\ ,\
			\end{align}
			and $(\hat{\V{y}}_\lambda^T \V{y} - \hat{\V{y}}_\lambda^T \hat{\V{y}}_\lambda)\hat{\beta}_{OLS,j^*}^{\mathcal{M}^*}$ is a quadratic polynomial with respect to $\hat{\beta}_{OLS,j^*}^{\mathcal{M}^*}$ with quadratic coefficient $n\lambda s_{j^*}$. Let $\tilde{s}_{j^*}=[(\V{X}_{\mathcal{M}^*}^T \V{X}_{\mathcal{M}^*})^{-1}]_{j^*} \V{s}$ where $[(\V{X}_{\mathcal{M}^*}^T \V{X}_{\mathcal{M}^*})^{-1}]_{j^*}$ denotes the $j^*$-th row of $(\V{X}_{\mathcal{M}^*}^T \V{X}_{\mathcal{M}^*})^{-1}$. Then
			\begin{align}
				\hat{\alpha}_\lambda \hat{\beta}_{\lambda, j^*} - \hat{\beta}_{OLS, j^*}^{\mathcal{M}^*} &= \hat{\alpha}_\lambda \big(\hat{\beta}_{OLS, j^*}^{\mathcal{M}^*} -n\lambda \tilde{s}_{j^*} \big)- \hat{\beta}_{OLS, j^*}^{\mathcal{M}^*}\\
				&= \frac{\hat{\V{y}}_\lambda^T \V{y}(\hat{\beta}_{OLS,j^*}^{\mathcal{M}^*} - n\lambda \tilde{s}_{j^*}}{\hat{\V{y}}_\lambda^T \hat{\V{y}}_\lambda} - \frac{\hat{\V{y}}_\lambda^T \hat{\V{y}}_\lambda\hat{\beta}_{OLS, j^*}^{\mathcal{M}^*}}{\hat{\V{y}}_\lambda^T \hat{\V{y}}_\lambda}\\
				&= \frac{(\hat{\V{y}}_\lambda^T \V{y} - \V{y}_\lambda^T \hat{\V{y}}_\lambda)\hat{\beta}_{OLS, j^*}^{\mathcal{M}^*} - n\lambda\tilde{s}_{j^*}\hat{\V{y}}_\lambda^T \V{y}}{\hat{\V{y}}_\lambda^T \hat{\V{y}}_\lambda}\ ,\
			\end{align}
			and $\hat{\alpha}_\lambda \hat{\beta}_{\lambda, j^*} - \hat{\beta}_{OLS, j^*}^{\mathcal{M}^*}$ is a ratio of quadratic polynomials with respect to $\hat{\beta}_{OLS, j^*}^{\mathcal{M}^*}$ where the numerator and denominator quadratic coefficients are $n\lambda (s_{j*}-n\tilde{s}_{j^*})$ and $n$, respectively. It follows that $\lim_{|\beta_{j^*}^*|\to\infty} \hat{\alpha}_\lambda \hat{\beta}_{\lambda, j^*} - \hat{\beta}_{OLS, j^*}^{\mathcal{M}^*} = \lambda (s_{j*}-n\tilde{s}_{j^*}) = G_{j^*}$ and so
			\begin{align}
				\lim_{|\beta_{j^*}^*|\to\infty} |\hat{\alpha}_\lambda \hat{\beta}_{\lambda, j^*} - \hat{\beta}_{OLS, j^*}^{\mathcal{M}^*}| = |G_{j^*}|\ .\
			\end{align}
			Clearly $-n\lambda\tilde{s}_{j^*}=\hat{\beta}_{\lambda, j^*} - \hat{\beta}_{OLS, j^*}^{\mathcal{M}^*}$ so $|G_{j^*}|=|\lambda s_{j^*} + \hat{\beta}_{\lambda, j^*} - \hat{\beta}_{OLS, j^*}^{\mathcal{M}^*}|$. If $\tilde{s}_{j^*}=0$ then $\hat{\beta}_{\lambda, j^*} = \hat{\beta}_{OLS, j^*}^{\mathcal{M}^*}$ so $|G_{j^*}|=\lambda > 0= |\hat{\beta}_{\lambda, j^*} - \hat{\beta}_{OLS, j^*}^{\mathcal{M}^*}|$. If $\tilde{s}_{j^*}\neq 0$ then 
			\begin{align}
				\lambda=\frac{-(\hat{\beta}_{\lambda, j^*} - \hat{\beta}_{OLS, j^*}^{\mathcal{M}^*})}{n\tilde{s}_{j^*}} 
			\end{align}
			which makes
			\begin{align}
				|G_{j^*}| = \left|\left(1-\frac{s_{j^*}}{n\tilde{s}_{j^*}}\right)(\hat{\beta}_{\lambda, j^*} - \hat{\beta}_{OLS, j^*}^{\mathcal{M}^*})\right|=\left|1-\frac{s_{j^*}}{n\tilde{s}_{j^*}}\right|\times |\hat{\beta}_{\lambda, j^*} - \hat{\beta}_{OLS, j^*}^{\mathcal{M}^*}|
			\end{align}
			Hence $|G_{j^*}| < |\hat{\beta}_{\lambda, j^*} - \hat{\beta}_{OLS, j^*}^{\mathcal{M}^*}|$ if and only if 
			\begin{align}
				\left|1-\frac{s_{j^*}}{n\tilde{s}_{j^*}}\right| < 1\ .\
			\end{align}
			
		\end{proof}

\section*{B. Further Numerical Results}
\subsection*{B.1. Simulation Study for Independent Predictors}
In this section we give additional results from the simulation studies conducted in the main document. Tables \ref{tab:APEHD} through \ref{tab:RelPB} give the average model size, average False Discovery Rates, average false negatives, and average prediction bias for the four methods considered in this part of the study: APE CV, AR2 CV, Mod CV, and the Relaxed Lasso for an additional $p^*$, $p^* = 15$. Mod CV was left out of the main document because of its similarity to AR2 CV. We can see from these results that all False negatives are close to zero. For smaller $p^*$, AR2 and Mod CV generally produce lower False Discovery Rates than APE CV. Average Prediction Bias, however generally appears lower for the two new approaches than APE CV, but higher than the Relaxed Lasso on average.
\begin{table}[ht]
\centering
\caption{Average Hamming Distance for 100 replications of CV 1SE. Standard errors given in parentheses.} 
\label{tab:APEHD}
\begin{tabular}{rcp{0.9cm}p{0.9cm}p{0.9cm}p{0.9cm}p{0.9cm}cp{0.9cm}p{0.9cm}p{0.9cm}p{0.9cm}p{0.9cm}}
  \hline
  \hline
 & p & 50 & 100 & 200 & 400 & 800 &  & 50 & 100 & 200 & 400 & 800 \\ 
   \hline n & & \multicolumn{5}{c}{$p^* = 5; SNR =  1.25 ; \bar{\sigma} =  5.19 $} & & \multicolumn{5}{c}{$p^* = 5; SNR =  5 ;  \bar{\sigma} =  2.59 $} \\
 \cmidrule(l){3-7} \cmidrule(l){9-13} 
100 &  & 3.1 (0.3) & 4.3 (0.4) & 6 (0.6) & 7.8 (1.1) & 9.1 (1.2) &  & 2.2 (0.2) & 4 (0.4) & 7.5 (1.2) & 9.3 (1) & 10.7 (1.2) \\ 
  500 &  & 0.6 (0.1) & 1 (0.2) & 1.3 (0.2) & 1.9 (0.5) & 2.9 (0.5) &  & 0.9 (0.2) & 0.9 (0.1) & 1.7 (0.3) & 1.7 (0.3) & 1.5 (0.3) \\ 
  1000 &  & 0.2 (0) & 0.4 (0.1) & 0.4 (0.1) & 0.6 (0.1) & 0.9 (0.2) &  & 0.1 (0) & 0.4 (0.1) & 0.5 (0.1) & 0.7 (0.1) & 0.9 (0.2) \\ 
   \hline n & & \multicolumn{5}{c}{$p^* = 15; SNR =  1.25 ; \bar{\sigma} =  9.06 $} & & \multicolumn{5}{c}{$p^* = 15; SNR =  5 ;  \bar{\sigma} =  4.53 $} \\
 \cmidrule(l){3-7} \cmidrule(l){9-13} 
100 &  & 10.1 (0.3) & 14.1 (0.6) & 16.7 (0.8) & 19.6 (1.4) & 21.4 (1.6) &  & 9.3 (0.4) & 15.7 (0.8) & 22.5 (1.1) & 25.5 (1.4) & 30.9 (2) \\ 
  500 &  & 4.1 (0.3) & 7.5 (0.4) & 9.7 (0.6) & 11.4 (0.6) & 15.7 (1) &  & 3.9 (0.2) & 6.1 (0.4) & 10.4 (0.6) & 13.9 (0.8) & 16.6 (1) \\
\hline
\end{tabular}
\end{table}
\begin{table}[ht]
\centering
\caption{Average Hamming Distance for 100 replications of AR2 CV 1SE. Standard errors given in parentheses.} 
\label{tab:AR2HD}
\begin{tabular}{rcp{0.9cm}p{0.9cm}p{0.9cm}p{0.9cm}p{0.9cm}cp{0.9cm}p{0.9cm}p{0.9cm}p{0.9cm}p{0.9cm}}
  \hline
  \hline
 & p & 50 & 100 & 200 & 400 & 800 &  & 50 & 100 & 200 & 400 & 800 \\ 
   \hline n & & \multicolumn{5}{c}{$p^* = 5; SNR =  1.25 ; \bar{\sigma} =  5.19 $} & & \multicolumn{5}{c}{$p^* = 5; SNR =  5 ;  \bar{\sigma} =  2.59 $} \\
 \cmidrule(l){3-7} \cmidrule(l){9-13} 
100 &  & 2.1 (0.3) & 2.8 (0.3) & 3.3 (0.3) & 4.9 (0.7) & 5 (0.5) &  & 0.6 (0.1) & 0.6 (0.1) & 1.9 (0.8) & 1.8 (0.4) & 2.6 (0.4) \\ 
  500 &  & 0.1 (0) & 0.2 (0) & 0.1 (0) & 0.2 (0) & 0.4 (0.1) &  & 0 (0) & 0 (0) & 0 (0) & 0 (0) & 0 (0) \\ 
  1000 &  & 0 (0) & 0 (0) & 0 (0) & 0 (0) & 0 (0) &  & 0 (0) & 0 (0) & 0 (0) & 0 (0) & 0 (0) \\ 
   \hline n & & \multicolumn{5}{c}{$p^* = 15; SNR =  1.25 ; \bar{\sigma} =  9.06 $} & & \multicolumn{5}{c}{$p^* = 15; SNR =  5 ;  \bar{\sigma} =  4.53 $} \\
 \cmidrule(l){3-7} \cmidrule(l){9-13} 
100 &  & 10.1 (0.3) & 14.9 (0.7) & 18.1 (1.1) & 20.8 (1.5) & 22.9 (1.6) &  & 6.4 (0.4) & 10.2 (0.6) & 15.9 (1) & 17.7 (1) & 21.8 (1.4) \\ 
  500 &  & 1.8 (0.2) & 2.8 (0.3) & 3.2 (0.2) & 4.2 (0.3) & 5.3 (0.4) &  & 0.3 (0.1) & 0.3 (0.1) & 0.6 (0.1) & 1.4 (0.2) & 1.5 (0.2) \\ 
  1000 &  & 0.4 (0.1) & 0.6 (0.1) & 0.9 (0.2) & 1.2 (0.1) & 1.3 (0.1) &  & 0 (0) & 0 (0) & 0 (0) & 0.1 (0) & 0.1 (0) \\ 
   \hline n & & \multicolumn{5}{c}{$p^* = 50; SNR =  1.25 ; \bar{\sigma} =  16.6 $} & & \multicolumn{5}{c}{$p^* = 50; SNR =  5 ;  \bar{\sigma} =  8.3 $} \\
 \cmidrule(l){3-7} \cmidrule(l){9-13} 
100 &  & 23.6 (1.3) & 40.4 (0.5) & 51.4 (0.8) & 55.1 (1.2) & 54.8 (1.2) &  & 4.2 (0.4) & 31.9 (0.5) & 47.9 (0.8) & 54.6 (1.3) & 53.5 (0.9) \\ 
  500 &  & 1.8 (0.2) & 17.1 (0.5) & 28.7 (0.8) & 42.3 (1.3) & 51.9 (2.2) &  & 0.2 (0) & 9.1 (0.4) & 17.8 (0.7) & 30 (1) & 46.6 (1.3) \\ 
  1000 &  & 0.7 (0.1) & 9.1 (0.4) & 15.4 (0.6) & 23.1 (0.9) & 32.5 (1.1) &  & 0 (0) & 3.9 (0.3) & 6.3 (0.4) & 11.1 (0.6) & 16.9 (0.8) \\ 
   \hline
\end{tabular}
\end{table}
\begin{table}[ht]
\centering
\caption{Average Hamming Distance for 100 replications of $\alpha$-modified CV 1SE. Standard errors given in parentheses.} 
\label{tab:ModHD}
\begin{tabular}{rcp{0.9cm}p{0.9cm}p{0.9cm}p{0.9cm}p{0.9cm}cp{0.9cm}p{0.9cm}p{0.9cm}p{0.9cm}p{0.9cm}}
  \hline
  \hline
 & p & 50 & 100 & 200 & 400 & 800 &  & 50 & 100 & 200 & 400 & 800 \\ 
   \hline n & & \multicolumn{5}{c}{$p^* = 5; SNR =  1.25 ; \bar{\sigma} =  5.19 $} & & \multicolumn{5}{c}{$p^* = 5; SNR =  5 ;  \bar{\sigma} =  2.59 $} \\
 \cmidrule(l){3-7} \cmidrule(l){9-13} 
100 &  & 2.3 (0.3) & 2.7 (0.3) & 3.7 (0.4) & 5.9 (1) & 6.8 (1.3) &  & 0.5 (0.1) & 0.9 (0.1) & 2.7 (0.9) & 2.7 (0.6) & 3.1 (0.5) \\ 
  500 &  & 0 (0) & 0.3 (0.1) & 0.1 (0) & 0.3 (0.2) & 0.5 (0.3) &  & 0 (0) & 0 (0) & 0 (0) & 0 (0) & 0 (0) \\ 
  1000 &  & 0 (0) & 0 (0) & 0 (0) & 0 (0) & 0 (0) &  & 0 (0) & 0 (0) & 0 (0) & 0 (0) & 0 (0) \\ 
   \hline n & & \multicolumn{5}{c}{$p^* = 15; SNR =  1.25 ; \bar{\sigma} =  9.06 $} & & \multicolumn{5}{c}{$p^* = 15; SNR =  5 ;  \bar{\sigma} =  4.53 $} \\
 \cmidrule(l){3-7} \cmidrule(l){9-13} 
100 &  & 9.9 (0.3) & 14 (0.6) & 16.8 (0.9) & 19.1 (1.4) & 21.8 (1.8) &  & 7.2 (0.4) & 11.6 (0.6) & 18 (1.1) & 21.1 (1.4) & 27.9 (2) \\ 
  500 &  & 1.7 (0.2) & 2.8 (0.3) & 3.3 (0.2) & 4.2 (0.3) & 5 (0.3) &  & 0.6 (0.1) & 0.5 (0.1) & 0.9 (0.1) & 2.1 (0.3) & 2 (0.3) \\ 
  1000 &  & 0.4 (0.1) & 0.6 (0.1) & 0.9 (0.1) & 1.3 (0.1) & 1.4 (0.1) &  & 0.1 (0) & 0.1 (0) & 0.1 (0) & 0.1 (0) & 0.2 (0.1) \\ 
   \hline n & & \multicolumn{5}{c}{$p^* = 50; SNR =  1.25 ; \bar{\sigma} =  16.6 $} & & \multicolumn{5}{c}{$p^* = 50; SNR =  5 ;  \bar{\sigma} =  8.3 $} \\
 \cmidrule(l){3-7} \cmidrule(l){9-13} 
100 &  & 35.7 (1.5) & 45.2 (0.6) & 51.5 (0.5) & 51.5 (0.7) & 54.3 (1.4) &  & 4.9 (0.5) & 32.7 (0.6) & 48.7 (0.8) & 55.5 (1.2) & 55.6 (1.4) \\ 
  500 &  & 1.7 (0.2) & 16.4 (0.4) & 26.5 (0.6) & 36.8 (1) & 43.9 (1.5) &  & 0.1 (0) & 9.9 (0.4) & 20.3 (0.7) & 34.1 (1.1) & 51.5 (1.4) \\ 
  1000 &  & 0.7 (0.1) & 8.6 (0.4) & 14.5 (0.5) & 21.6 (0.7) & 27.9 (1) &  & 0 (0) & 4.6 (0.3) & 7.8 (0.4) & 13.6 (0.6) & 20 (1) \\ 
   \hline
\end{tabular}
\end{table}
\begin{table}[ht]
\centering
\caption{Average Hamming Distance for 100 replications of Relaxed Lasso. Standard errors given in parentheses.} 
\label{tab:RelHD}
\begin{tabular}{rcp{0.9cm}p{0.9cm}p{0.9cm}p{0.9cm}p{0.9cm}cp{0.9cm}p{0.9cm}p{0.9cm}p{0.9cm}p{0.9cm}}
  \hline
  \hline
 & p & 50 & 100 & 200 & 400 & 800 &  & 50 & 100 & 200 & 400 & 800 \\ 
   \hline n & & \multicolumn{5}{c}{$p^* = 5; SNR =  1.25 ; \bar{\sigma} =  5.19 $} & & \multicolumn{5}{c}{$p^* = 5; SNR =  5 ;  \bar{\sigma} =  2.59 $} \\
 \cmidrule(l){3-7} \cmidrule(l){9-13} 
100 &  & 3.7 (0.4) & 5.4 (0.8) & 7.4 (1.1) & 18.4 (3.4) & 28.1 (4.2) &  & 1.1 (0.3) & 1.8 (0.5) & 4.5 (1.7) & 6.6 (2.3) & 7.9 (2.6) \\ 
  500 &  & 0.8 (0.3) & 1.2 (0.3) & 0.5 (0.2) & 0.7 (0.4) & 1.5 (0.6) &  & 1.2 (0.3) & 0.8 (0.2) & 0.8 (0.4) & 0.3 (0.1) & 0.1 (0) \\ 
  1000 &  & 0.6 (0.2) & 0.5 (0.2) & 0.7 (0.3) & 0.4 (0.1) & 1.9 (1.5) &  & 0.6 (0.2) & 0.8 (0.3) & 0.4 (0.2) & 0.4 (0.2) & 0.4 (0.2) \\ 
   \hline n & & \multicolumn{5}{c}{$p^* = 15; SNR =  1.25 ; \bar{\sigma} =  9.06 $} & & \multicolumn{5}{c}{$p^* = 15; SNR =  5 ;  \bar{\sigma} =  4.53 $} \\
 \cmidrule(l){3-7} \cmidrule(l){9-13} 
100 &  & 12.3 (0.6) & 18.8 (1.1) & 32.2 (3) & 45.3 (4.3) & 55.9 (4.9) &  & 9.1 (0.7) & 14.2 (1) & 34.2 (3.4) & 81.1 (4.8) & 82.3 (4.7) \\ 
  500 &  & 2.9 (0.5) & 3.8 (0.6) & 4.4 (0.5) & 5.1 (0.8) & 5 (0.4) &  & 0.6 (0.1) & 0.7 (0.2) & 0.4 (0.1) & 0.8 (0.2) & 0.8 (0.1) \\
\hline
\end{tabular}
\end{table}
\begin{table}[ht]
\centering
\caption{Average number of selected variables 100 replications of CV 1SE. Standard errors given in parentheses.} 
\label{tab:APESize}
\begin{tabular}{rcp{0.9cm}p{0.9cm}p{0.9cm}p{0.9cm}p{0.9cm}cp{0.9cm}p{0.9cm}p{0.9cm}p{0.9cm}p{0.9cm}}
  \hline
  \hline
 & p & 50 & 100 & 200 & 400 & 800 &  & 50 & 100 & 200 & 400 & 800 \\ 
   \hline n & & \multicolumn{5}{c}{$p^* = 5; SNR =  1.25 ; \bar{\sigma} =  5.23 $} & & \multicolumn{5}{c}{$p^* = 5; SNR =  5 ;  \bar{\sigma} =  2.61 $} \\
 \cmidrule(l){3-7} \cmidrule(l){9-13} 
50 &  & 6.4 (0.6) & 10.6 (1.2) & 9.7 (1.3) & 11.5 (1.4) & 11.2 (1.6) &  & 11 (0.8) & 12.7 (0.8) & 13.9 (0.9) & 17.9 (1.2) & 18.6 (1.5) \\ 
  100 &  & 6.8 (0.3) & 8.7 (0.7) & 9.9 (1) & 10.3 (1.3) & 12 (1.5) &  & 7.8 (0.4) & 9.8 (0.6) & 11.6 (0.7) & 14.5 (1.1) & 15.7 (1.4) \\ 
  200 &  & 6.5 (0.2) & 6.8 (0.2) & 8 (0.4) & 9 (0.5) & 9.8 (0.6) &  & 6.6 (0.2) & 7.6 (0.3) & 8.6 (0.4) & 10.3 (0.9) & 11.3 (0.9) \\ 
   \hline n & & \multicolumn{5}{c}{$p^* = 15; SNR =  1.25 ; \bar{\sigma} =  9.04 $} & & \multicolumn{5}{c}{$p^* = 15; SNR =  5 ;  \bar{\sigma} =  4.52 $} \\
 \cmidrule(l){3-7} \cmidrule(l){9-13} 
50 &  & 8.9 (1) & 8 (1.2) & 8.5 (1.3) & 5.6 (1.2) & 6.6 (1.3) &  & 18.9 (1) & 20.6 (1.5) & 15.1 (1.4) & 13.7 (1.6) & 13.1 (1.6) \\ 
  100 &  & 15.9 (0.6) & 15.2 (1.1) & 13.5 (1.3) & 11 (1.6) & 11.5 (1.9) &  & 22.8 (0.4) & 28.8 (0.9) & 33.7 (1.4) & 35 (1.7) & 29.9 (1.8) \\
\hline
\end{tabular}
\end{table}
\begin{table}[ht]
\centering
\caption{Average number of selected variables 100 replications of AR2 CV 1SE. Standard errors given in parentheses.} 
\label{tab:AR2Size}
\begin{tabular}{rcp{0.9cm}p{0.9cm}p{0.9cm}p{0.9cm}p{0.9cm}cp{0.9cm}p{0.9cm}p{0.9cm}p{0.9cm}p{0.9cm}}
  \hline
  \hline
 & p & 50 & 100 & 200 & 400 & 800 &  & 50 & 100 & 200 & 400 & 800 \\ 
   \hline n & & \multicolumn{5}{c}{$p^* = 5; SNR =  1.25 ; \bar{\sigma} =  5.23 $} & & \multicolumn{5}{c}{$p^* = 5; SNR =  5 ;  \bar{\sigma} =  2.61 $} \\
 \cmidrule(l){3-7} \cmidrule(l){9-13} 
50 &  & 6.7 (0.9) & 7.3 (1) & 6 (0.9) & 6 (0.8) & 5.4 (0.9) &  & 7.4 (0.7) & 7.9 (0.7) & 6.8 (0.6) & 7.5 (0.6) & 7.8 (0.8) \\ 
  100 &  & 4.8 (0.2) & 5.2 (0.4) & 5.6 (0.7) & 6.2 (1.1) & 6.4 (0.8) &  & 5.3 (0.1) & 5.8 (0.2) & 6.3 (0.4) & 6.6 (0.3) & 6.7 (0.4) \\ 
  200 &  & 4.9 (0.1) & 5.1 (0.2) & 5.2 (0.2) & 5.3 (0.3) & 5.2 (0.3) &  & 5.1 (0.1) & 5.2 (0.1) & 5.2 (0.1) & 5.2 (0.1) & 5.2 (0.1) \\ 
   \hline n & & \multicolumn{5}{c}{$p^* = 15; SNR =  1.25 ; \bar{\sigma} =  9.04 $} & & \multicolumn{5}{c}{$p^* = 15; SNR =  5 ;  \bar{\sigma} =  4.52 $} \\
 \cmidrule(l){3-7} \cmidrule(l){9-13} 
50 &  & 10.8 (1.3) & 10.3 (1.2) & 9.1 (1.2) & 5.9 (0.8) & 9.2 (1.2) &  & 16.3 (1) & 13.5 (1.2) & 10.8 (1.2) & 8.4 (1.1) & 9.2 (1.2) \\ 
  100 &  & 15.6 (0.7) & 15 (1.4) & 15.1 (1.8) & 9.8 (1.3) & 12.8 (1.7) &  & 18.3 (0.4) & 22.9 (0.7) & 24.1 (1.1) & 21.9 (1.2) & 22.9 (1.6) \\
\hline
\end{tabular}
\end{table}
\begin{table}[ht]
\centering
\caption{Average number of selected variables 100 replications of $\alpha$-modified CV 1SE. Standard errors given in parentheses.} 
\label{tab:ModSize}
\begin{tabular}{rcp{0.9cm}p{0.9cm}p{0.9cm}p{0.9cm}p{0.9cm}cp{0.9cm}p{0.9cm}p{0.9cm}p{0.9cm}p{0.9cm}}
  \hline
  \hline
 & p & 50 & 100 & 200 & 400 & 800 &  & 50 & 100 & 200 & 400 & 800 \\ 
   \hline n & & \multicolumn{5}{c}{$p^* = 5; SNR =  1.25 ; \bar{\sigma} =  5.23 $} & & \multicolumn{5}{c}{$p^* = 5; SNR =  5 ;  \bar{\sigma} =  2.61 $} \\
 \cmidrule(l){3-7} \cmidrule(l){9-13} 
50 &  & 5.2 (0.7) & 8.6 (1.2) & 8.5 (1.3) & 9.6 (1.4) & 9.8 (1.5) &  & 8.3 (0.7) & 9.7 (0.7) & 9.3 (0.8) & 12 (1.1) & 12.2 (1.3) \\ 
  100 &  & 4.8 (0.2) & 5.2 (0.3) & 6.3 (0.8) & 6.5 (1.2) & 7.8 (1.3) &  & 5.5 (0.1) & 6.3 (0.3) & 6.7 (0.4) & 7.6 (0.4) & 8 (0.5) \\ 
  200 &  & 5 (0.1) & 5.2 (0.2) & 5 (0.1) & 5.5 (0.3) & 5.2 (0.2) &  & 5.1 (0) & 5.3 (0.1) & 5.2 (0.1) & 5.6 (0.3) & 5.8 (0.3) \\ 
   \hline n & & \multicolumn{5}{c}{$p^* = 15; SNR =  1.25 ; \bar{\sigma} =  9.04 $} & & \multicolumn{5}{c}{$p^* = 15; SNR =  5 ;  \bar{\sigma} =  4.52 $} \\
 \cmidrule(l){3-7} \cmidrule(l){9-13} 
50 &  & 8 (1.1) & 7.2 (1.3) & 9 (1.4) & 5.3 (1.2) & 6.6 (1.4) &  & 17.7 (1.1) & 18.8 (1.5) & 14.2 (1.5) & 13.5 (1.7) & 13.7 (1.7) \\ 
  100 &  & 12.4 (0.7) & 11.2 (1) & 9.5 (1.3) & 9.5 (2) & 10.3 (2.1) &  & 19.6 (0.4) & 24.1 (0.7) & 27.3 (1.3) & 28.3 (1.8) & 25.8 (1.7) \\
\hline
\end{tabular}
\end{table}
\begin{table}[ht]
\centering
\caption{Average number of selected variables 100 replications of Relaxed Lasso. Standard errors given in parentheses.} 
\label{tab:RelSize}
\begin{tabular}{rcp{0.9cm}p{0.9cm}p{0.9cm}p{0.9cm}p{0.9cm}cp{0.9cm}p{0.9cm}p{0.9cm}p{0.9cm}p{0.9cm}}
  \hline
  \hline
 & p & 50 & 100 & 200 & 400 & 800 &  & 50 & 100 & 200 & 400 & 800 \\ 
   \hline n & & \multicolumn{5}{c}{$p^* = 5; SNR =  1.25 ; \bar{\sigma} =  5.23 $} & & \multicolumn{5}{c}{$p^* = 5; SNR =  5 ;  \bar{\sigma} =  2.61 $} \\
 \cmidrule(l){3-7} \cmidrule(l){9-13} 
50 &  & 9.1 (0.8) & 22.4 (2.2) & 30.4 (2.7) & 31.8 (2.7) & 36.5 (2.9) &  & 12.2 (1.1) & 19.5 (2) & 27.5 (2.6) & 39.9 (2.7) & 43 (2.7) \\ 
  100 &  & 7.8 (0.6) & 10.5 (1.3) & 11.5 (1.6) & 18.6 (3.2) & 30.2 (4.4) &  & 6 (0.2) & 7.2 (0.6) & 8.2 (1) & 13.7 (2.8) & 13.4 (2.7) \\ 
  200 &  & 6.5 (0.3) & 6.5 (0.4) & 7.2 (0.9) & 8.3 (1.1) & 8.7 (1.7) &  & 5.6 (0.2) & 5.7 (0.2) & 6.8 (1) & 5.4 (0.2) & 5.6 (0.2) \\ 
   \hline n & & \multicolumn{5}{c}{$p^* = 15; SNR =  1.25 ; \bar{\sigma} =  9.04 $} & & \multicolumn{5}{c}{$p^* = 15; SNR =  5 ;  \bar{\sigma} =  4.52 $} \\
 \cmidrule(l){3-7} \cmidrule(l){9-13} 
50 &  & 15.2 (1.3) & 26.4 (2.4) & 31 (2.7) & 26.6 (2.7) & 30.8 (2.9) &  & 23.8 (1.1) & 47.1 (2.2) & 40 (2.6) & 44.8 (2.6) & 39.4 (2.7) \\ 
  100 &  & 21.1 (0.8) & 23.6 (1.6) & 40.1 (3.9) & 35.2 (4.3) & 46.2 (5) &  & 21.6 (0.7) & 25.9 (1.1) & 49.8 (4.2) & 78.6 (5.3) & 91.5 (5.1) \\
\hline
\end{tabular}
\end{table}
\begin{table}[ht]
\centering
\caption{Average false discovery rate from 100 replications of CV 1SE. Standard errors given in parentheses.} 
\label{tab:APEFDR}
\begin{tabular}{rcp{0.9cm}p{0.9cm}p{0.9cm}p{0.9cm}p{0.9cm}cp{0.9cm}p{0.9cm}p{0.9cm}p{0.9cm}p{0.9cm}}
  \hline
  \hline
 & p & 50 & 100 & 200 & 400 & 800 &  & 50 & 100 & 200 & 400 & 800 \\ 
   \hline n & & \multicolumn{5}{c}{$p^* = 5; SNR =  1.25 ; \bar{\sigma} =  5.23 $} & & \multicolumn{5}{c}{$p^* = 5; SNR =  5 ;  \bar{\sigma} =  2.61 $} \\
 \cmidrule(l){3-7} \cmidrule(l){9-13} 
50 &  & 0.3 (0) & 0.4 (0) & 0.4 (0) & 0.5 (0) & 0.4 (0) &  & 0.4 (0) & 0.5 (0) & 0.6 (0) & 0.6 (0) & 0.6 (0) \\ 
  100 &  & 0.3 (0) & 0.4 (0) & 0.4 (0) & 0.4 (0) & 0.4 (0) &  & 0.3 (0) & 0.4 (0) & 0.5 (0) & 0.5 (0) & 0.5 (0) \\ 
  200 &  & 0.2 (0) & 0.2 (0) & 0.3 (0) & 0.3 (0) & 0.4 (0) &  & 0.2 (0) & 0.3 (0) & 0.3 (0) & 0.4 (0) & 0.4 (0) \\ 
   \hline n & & \multicolumn{5}{c}{$p^* = 15; SNR =  1.25 ; \bar{\sigma} =  9.04 $} & & \multicolumn{5}{c}{$p^* = 15; SNR =  5 ;  \bar{\sigma} =  4.52 $} \\
 \cmidrule(l){3-7} \cmidrule(l){9-13} 
50 &  & 0.2 (0) & 0.3 (0) & 0.3 (0) & 0.2 (0) & 0.3 (0) &  & 0.4 (0) & 0.4 (0) & 0.4 (0) & 0.4 (0) & 0.4 (0) \\ 
  100 &  & 0.3 (0) & 0.3 (0) & 0.4 (0) & 0.3 (0) & 0.3 (0) &  & 0.3 (0) & 0.5 (0) & 0.6 (0) & 0.6 (0) & 0.6 (0) \\ 
  200 &  & 0.3 (0) & 0.4 (0) & 0.4 (0) & 0.5 (0) & 0.5 (0) &  & 0.3 (0) & 0.4 (0) & 0.5 (0) & 0.6 (0) & 0.6 (0) \\ 
   \hline n & & \multicolumn{5}{c}{$p^* = 50; SNR =  1.25 ; \bar{\sigma} =  16.52 $} & & \multicolumn{5}{c}{$p^* = 50; SNR =  5 ;  \bar{\sigma} =  8.26 $} \\
 \cmidrule(l){3-7} \cmidrule(l){9-13} 
50 &  & 0 (0) & 0.2 (0) & 0.2 (0) & 0.2 (0) & 0.2 (0) &  & 0 (0) & 0.2 (0) & 0.2 (0) & 0.3 (0) & 0.3 (0) \\ 
  100 &  & 0 (0) & 0.1 (0) & 0.2 (0) & 0.2 (0) & 0.2 (0) &  & 0 (0) & 0.3 (0) & 0.3 (0) & 0.3 (0) & 0.2 (0) \\ 
  200 &  & 0 (0) & 0.2 (0) & 0.3 (0) & 0.3 (0) & 0.3 (0) &  & 0 (0) & 0.3 (0) & 0.5 (0) & 0.5 (0) & 0.5 (0) \\ 
   \hline
\end{tabular}
\end{table}
\begin{table}[ht]
\centering
\caption{Average false discovery rate for 100 replications of AR2 CV 1SE. Standard errors given in parentheses.} 
\label{tab:AR2FDR}
\begin{tabular}{rcp{0.9cm}p{0.9cm}p{0.9cm}p{0.9cm}p{0.9cm}cp{0.9cm}p{0.9cm}p{0.9cm}p{0.9cm}p{0.9cm}}
  \hline
  \hline
 & p & 50 & 100 & 200 & 400 & 800 &  & 50 & 100 & 200 & 400 & 800 \\ 
   \hline n & & \multicolumn{5}{c}{$p^* = 5; SNR =  1.25 ; \bar{\sigma} =  5.23 $} & & \multicolumn{5}{c}{$p^* = 5; SNR =  5 ;  \bar{\sigma} =  2.61 $} \\
 \cmidrule(l){3-7} \cmidrule(l){9-13} 
50 &  & 0.3 (0) & 0.3 (0) & 0.4 (0) & 0.4 (0) & 0.4 (0) &  & 0.2 (0) & 0.3 (0) & 0.3 (0) & 0.3 (0) & 0.4 (0) \\ 
  100 &  & 0.1 (0) & 0.1 (0) & 0.2 (0) & 0.2 (0) & 0.2 (0) &  & 0.1 (0) & 0.1 (0) & 0.1 (0) & 0.2 (0) & 0.2 (0) \\ 
  200 &  & 0 (0) & 0.1 (0) & 0.1 (0) & 0.1 (0) & 0.1 (0) &  & 0 (0) & 0 (0) & 0 (0) & 0 (0) & 0 (0) \\ 
   \hline n & & \multicolumn{5}{c}{$p^* = 15; SNR =  1.25 ; \bar{\sigma} =  9.04 $} & & \multicolumn{5}{c}{$p^* = 15; SNR =  5 ;  \bar{\sigma} =  4.52 $} \\
 \cmidrule(l){3-7} \cmidrule(l){9-13} 
50 &  & 0.3 (0) & 0.4 (0) & 0.5 (0) & 0.6 (0) & 0.7 (0) &  & 0.3 (0) & 0.3 (0) & 0.4 (0) & 0.5 (0) & 0.5 (0) \\ 
  100 &  & 0.3 (0) & 0.3 (0) & 0.4 (0) & 0.3 (0) & 0.4 (0) &  & 0.2 (0) & 0.4 (0) & 0.4 (0) & 0.5 (0) & 0.5 (0) \\ 
  200 &  & 0.2 (0) & 0.2 (0) & 0.3 (0) & 0.3 (0) & 0.4 (0) &  & 0.1 (0) & 0.2 (0) & 0.3 (0) & 0.3 (0) & 0.4 (0) \\ 
   \hline n & & \multicolumn{5}{c}{$p^* = 50; SNR =  1.25 ; \bar{\sigma} =  16.52 $} & & \multicolumn{5}{c}{$p^* = 50; SNR =  5 ;  \bar{\sigma} =  8.26 $} \\
 \cmidrule(l){3-7} \cmidrule(l){9-13} 
50 &  & 0 (0) & 0.2 (0) & 0.5 (0) & 0.6 (0) & 0.7 (0) &  & 0 (0) & 0.2 (0) & 0.4 (0) & 0.5 (0) & 0.7 (0) \\ 
  100 &  & 0 (0) & 0.2 (0) & 0.4 (0) & 0.5 (0) & 0.6 (0) &  & 0 (0) & 0.3 (0) & 0.4 (0) & 0.4 (0) & 0.5 (0) \\ 
  200 &  & 0 (0) & 0.3 (0) & 0.4 (0) & 0.5 (0) & 0.5 (0) &  & 0 (0) & 0.3 (0) & 0.4 (0) & 0.5 (0) & 0.5 (0) \\ 
   \hline
\end{tabular}
\end{table}
\begin{table}[ht]
\centering
\caption{Average false discovery rate of 100 replications for $\alpha$-modified CV 1SE. Standard errors given in parentheses.} 
\label{tab:ModFDR}
\begin{tabular}{rcp{0.9cm}p{0.9cm}p{0.9cm}p{0.9cm}p{0.9cm}cp{0.9cm}p{0.9cm}p{0.9cm}p{0.9cm}p{0.9cm}}
  \hline
  \hline
 & p & 50 & 100 & 200 & 400 & 800 &  & 50 & 100 & 200 & 400 & 800 \\ 
   \hline n & & \multicolumn{5}{c}{$p^* = 5; SNR =  1.25 ; \bar{\sigma} =  5.23 $} & & \multicolumn{5}{c}{$p^* = 5; SNR =  5 ;  \bar{\sigma} =  2.61 $} \\
 \cmidrule(l){3-7} \cmidrule(l){9-13} 
50 &  & 0.2 (0) & 0.3 (0) & 0.3 (0) & 0.4 (0) & 0.3 (0) &  & 0.3 (0) & 0.4 (0) & 0.4 (0) & 0.4 (0) & 0.4 (0) \\ 
  100 &  & 0.1 (0) & 0.1 (0) & 0.2 (0) & 0.2 (0) & 0.2 (0) &  & 0.1 (0) & 0.1 (0) & 0.2 (0) & 0.2 (0) & 0.3 (0) \\ 
  200 &  & 0 (0) & 0.1 (0) & 0.1 (0) & 0.1 (0) & 0.1 (0) &  & 0 (0) & 0.1 (0) & 0 (0) & 0.1 (0) & 0.1 (0) \\ 
   \hline n & & \multicolumn{5}{c}{$p^* = 15; SNR =  1.25 ; \bar{\sigma} =  9.04 $} & & \multicolumn{5}{c}{$p^* = 15; SNR =  5 ;  \bar{\sigma} =  4.52 $} \\
 \cmidrule(l){3-7} \cmidrule(l){9-13} 
50 &  & 0.2 (0) & 0.2 (0) & 0.3 (0) & 0.2 (0) & 0.2 (0) &  & 0.3 (0) & 0.4 (0) & 0.4 (0) & 0.4 (0) & 0.4 (0) \\ 
  100 &  & 0.2 (0) & 0.2 (0) & 0.2 (0) & 0.2 (0) & 0.2 (0) &  & 0.2 (0) & 0.4 (0) & 0.5 (0) & 0.5 (0) & 0.5 (0) \\ 
  200 &  & 0.2 (0) & 0.2 (0) & 0.3 (0) & 0.3 (0) & 0.3 (0) &  & 0.1 (0) & 0.2 (0) & 0.3 (0) & 0.4 (0) & 0.4 (0) \\ 
   \hline n & & \multicolumn{5}{c}{$p^* = 50; SNR =  1.25 ; \bar{\sigma} =  16.52 $} & & \multicolumn{5}{c}{$p^* = 50; SNR =  5 ;  \bar{\sigma} =  8.26 $} \\
 \cmidrule(l){3-7} \cmidrule(l){9-13} 
50 &  & 0 (0) & 0.2 (0) & 0.1 (0) & 0.2 (0) & 0.2 (0) &  & 0 (0) & 0.2 (0) & 0.2 (0) & 0.3 (0) & 0.3 (0) \\ 
  100 &  & 0 (0) & 0.1 (0) & 0.1 (0) & 0.1 (0) & 0.2 (0) &  & 0 (0) & 0.3 (0) & 0.3 (0) & 0.3 (0) & 0.2 (0) \\ 
  200 &  & 0 (0) & 0.2 (0) & 0.2 (0) & 0.2 (0) & 0.1 (0) &  & 0 (0) & 0.3 (0) & 0.4 (0) & 0.5 (0) & 0.5 (0) \\ 
   \hline
\end{tabular}
\end{table}
\begin{table}[ht]
\centering
\caption{Average false discovery rate for 100 replications of Relaxed Lasso. Standard errors given in parentheses.} 
\label{tab:RelFDR}
\begin{tabular}{rcp{0.9cm}p{0.9cm}p{0.9cm}p{0.9cm}p{0.9cm}cp{0.9cm}p{0.9cm}p{0.9cm}p{0.9cm}p{0.9cm}}
  \hline
  \hline
 & p & 50 & 100 & 200 & 400 & 800 &  & 50 & 100 & 200 & 400 & 800 \\ 
   \hline n & & \multicolumn{5}{c}{$p^* = 5; SNR =  1.25 ; \bar{\sigma} =  5.23 $} & & \multicolumn{5}{c}{$p^* = 5; SNR =  5 ;  \bar{\sigma} =  2.61 $} \\
 \cmidrule(l){3-7} \cmidrule(l){9-13} 
50 &  & 0.4 (0) & 0.6 (0) & 0.6 (0) & 0.7 (0) & 0.7 (0) &  & 0.4 (0) & 0.5 (0) & 0.5 (0) & 0.7 (0) & 0.7 (0) \\ 
  100 &  & 0.3 (0) & 0.3 (0) & 0.3 (0) & 0.4 (0) & 0.5 (0) &  & 0.1 (0) & 0.1 (0) & 0.2 (0) & 0.2 (0) & 0.2 (0) \\ 
  200 &  & 0.1 (0) & 0.1 (0) & 0.1 (0) & 0.2 (0) & 0.2 (0) &  & 0.1 (0) & 0.1 (0) & 0.1 (0) & 0 (0) & 0 (0) \\ 
   \hline n & & \multicolumn{5}{c}{$p^* = 15; SNR =  1.25 ; \bar{\sigma} =  9.04 $} & & \multicolumn{5}{c}{$p^* = 15; SNR =  5 ;  \bar{\sigma} =  4.52 $} \\
 \cmidrule(l){3-7} \cmidrule(l){9-13} 
50 &  & 0.4 (0) & 0.5 (0) & 0.6 (0) & 0.7 (0) & 0.6 (0) &  & 0.4 (0) & 0.7 (0) & 0.7 (0) & 0.7 (0) & 0.7 (0) \\ 
  100 &  & 0.4 (0) & 0.5 (0) & 0.6 (0) & 0.5 (0) & 0.6 (0) &  & 0.3 (0) & 0.4 (0) & 0.6 (0) & 0.7 (0) & 0.8 (0) \\ 
  200 &  & 0.3 (0) & 0.4 (0) & 0.5 (0) & 0.5 (0) & 0.5 (0) &  & 0.1 (0) & 0.2 (0) & 0.2 (0) & 0.3 (0) & 0.3 (0) \\ 
   \hline n & & \multicolumn{5}{c}{$p^* = 50; SNR =  1.25 ; \bar{\sigma} =  16.52 $} & & \multicolumn{5}{c}{$p^* = 50; SNR =  5 ;  \bar{\sigma} =  8.26 $} \\
 \cmidrule(l){3-7} \cmidrule(l){9-13} 
50 &  & 0 (0) & 0.3 (0) & 0.4 (0) & 0.5 (0) & 0.6 (0) &  & 0 (0) & 0.3 (0) & 0.5 (0) & 0.6 (0) & 0.6 (0) \\ 
  100 &  & 0 (0) & 0.2 (0) & 0.4 (0) & 0.5 (0) & 0.5 (0) &  & 0 (0) & 0.3 (0) & 0.5 (0) & 0.6 (0) & 0.6 (0) \\ 
  200 &  & 0 (0) & 0.3 (0) & 0.5 (0) & 0.5 (0) & 0.5 (0) &  & 0 (0) & 0.3 (0) & 0.5 (0) & 0.7 (0) & 0.8 (0) \\ 
   \hline
\end{tabular}
\end{table}
\begin{table}[ht]
\centering
\caption{Average number of false negatives from 100 replications of CV 1SE. Standard errors given in parentheses.} 
\label{tab:APEFN}
\begin{tabular}{rcp{0.9cm}p{0.9cm}p{0.9cm}p{0.9cm}p{0.9cm}cp{0.9cm}p{0.9cm}p{0.9cm}p{0.9cm}p{0.9cm}}
  \hline
  \hline
 & p & 50 & 100 & 200 & 400 & 800 &  & 50 & 100 & 200 & 400 & 800 \\ 
   \hline n & & \multicolumn{5}{c}{$p^* = 5; SNR =  1.25 ; \bar{\sigma} =  5.23 $} & & \multicolumn{5}{c}{$p^* = 5; SNR =  5 ;  \bar{\sigma} =  2.61 $} \\
 \cmidrule(l){3-7} \cmidrule(l){9-13} 
50 &  & 1.9 (0.2) & 2.2 (0.2) & 2.9 (0.2) & 2.9 (0.2) & 3.5 (0.2) &  & 0.2 (0.1) & 0.3 (0.1) & 0.6 (0.1) & 0.6 (0.1) & 1.1 (0.1) \\ 
  100 &  & 0.6 (0.1) & 0.6 (0.1) & 0.9 (0.1) & 1.2 (0.1) & 1.4 (0.1) &  & 0 (0) & 0 (0) & 0 (0) & 0.1 (0) & 0.1 (0) \\ 
  200 &  & 0.2 (0) & 0.2 (0) & 0.2 (0) & 0.3 (0) & 0.2 (0) &  & 0 (0) & 0 (0) & 0 (0) & 0 (0) & 0 (0) \\ 
   \hline n & & \multicolumn{5}{c}{$p^* = 15; SNR =  1.25 ; \bar{\sigma} =  9.04 $} & & \multicolumn{5}{c}{$p^* = 15; SNR =  5 ;  \bar{\sigma} =  4.52 $} \\
 \cmidrule(l){3-7} \cmidrule(l){9-13} 
50 &  & 10.1 (0.4) & 12.2 (0.3) & 12.8 (0.3) & 14 (0.2) & 14.3 (0.1) &  & 4.2 (0.4) & 6.5 (0.5) & 9.7 (0.4) & 11.7 (0.3) & 12.7 (0.2) \\ 
  100 &  & 4.5 (0.3) & 6.9 (0.4) & 9.1 (0.4) & 11.4 (0.3) & 12.3 (0.3) &  & 0.4 (0.1) & 0.8 (0.1) & 1.6 (0.2) & 3.2 (0.2) & 5.5 (0.3) \\ 
  200 &  & 1 (0.1) & 2 (0.2) & 3 (0.2) & 4.2 (0.2) & 5.7 (0.3) &  & 0 (0) & 0.1 (0) & 0.2 (0) & 0.3 (0.1) & 0.4 (0.1) \\ 
   \hline n & & \multicolumn{5}{c}{$p^* = 50; SNR =  1.25 ; \bar{\sigma} =  16.52 $} & & \multicolumn{5}{c}{$p^* = 50; SNR =  5 ;  \bar{\sigma} =  8.26 $} \\
 \cmidrule(l){3-7} \cmidrule(l){9-13} 
50 &  & 42.2 (1.1) & 44.4 (0.8) & 47.9 (0.4) & 48.7 (0.3) & 49.2 (0.2) &  & 27.8 (1.4) & 40.4 (0.9) & 45.3 (0.6) & 47.3 (0.4) & 48.6 (0.3) \\ 
  100 &  & 30.8 (1.2) & 41.5 (1.1) & 45.8 (0.7) & 47.6 (0.5) & 48.5 (0.3) &  & 6.3 (0.5) & 19.3 (1.1) & 34.5 (1.1) & 42 (0.9) & 46.3 (0.6) \\ 
  200 &  & 10.9 (0.6) & 21.6 (0.9) & 29.8 (1.1) & 40 (0.9) & 44.4 (0.7) &  & 0.8 (0.1) & 2.1 (0.2) & 5.8 (0.3) & 13.6 (0.6) & 26.3 (0.8) \\ 
   \hline
\end{tabular}
\end{table}
\begin{table}[ht]
\centering
\caption{Average number of false negatives for 100 replications of AR2 CV 1SE. Standard errors given in parentheses.} 
\label{tab:AR2FN}
\begin{tabular}{rcp{0.9cm}p{0.9cm}p{0.9cm}p{0.9cm}p{0.9cm}cp{0.9cm}p{0.9cm}p{0.9cm}p{0.9cm}p{0.9cm}}
  \hline
  \hline
 & p & 50 & 100 & 200 & 400 & 800 &  & 50 & 100 & 200 & 400 & 800 \\ 
   \hline n & & \multicolumn{5}{c}{$p^* = 5; SNR =  1.25 ; \bar{\sigma} =  5.23 $} & & \multicolumn{5}{c}{$p^* = 5; SNR =  5 ;  \bar{\sigma} =  2.61 $} \\
 \cmidrule(l){3-7} \cmidrule(l){9-13} 
50 &  & 2.1 (0.1) & 2.4 (0.1) & 2.9 (0.1) & 3 (0.1) & 3.6 (0.1) &  & 0.6 (0.1) & 0.8 (0.1) & 1.4 (0.1) & 1.4 (0.1) & 1.9 (0.1) \\ 
  100 &  & 1.1 (0.1) & 1.2 (0.1) & 1.6 (0.1) & 1.8 (0.1) & 1.9 (0.1) &  & 0.1 (0) & 0.1 (0) & 0.1 (0) & 0.2 (0.1) & 0.3 (0.1) \\ 
  200 &  & 0.4 (0.1) & 0.4 (0.1) & 0.5 (0.1) & 0.6 (0.1) & 0.7 (0.1) &  & 0 (0) & 0 (0) & 0 (0) & 0.1 (0) & 0 (0) \\ 
   \hline n & & \multicolumn{5}{c}{$p^* = 15; SNR =  1.25 ; \bar{\sigma} =  9.04 $} & & \multicolumn{5}{c}{$p^* = 15; SNR =  5 ;  \bar{\sigma} =  4.52 $} \\
 \cmidrule(l){3-7} \cmidrule(l){9-13} 
50 &  & 9.5 (0.4) & 11.1 (0.3) & 12.3 (0.3) & 13.6 (0.1) & 13.7 (0.1) &  & 5.2 (0.4) & 8.4 (0.5) & 10.5 (0.3) & 12.4 (0.2) & 13.1 (0.2) \\ 
  100 &  & 4.9 (0.3) & 7.2 (0.4) & 9.1 (0.4) & 11.3 (0.3) & 11.6 (0.3) &  & 1 (0.1) & 1.4 (0.2) & 2.7 (0.2) & 4.8 (0.3) & 6.5 (0.4) \\ 
  200 &  & 1.9 (0.2) & 2.9 (0.2) & 4 (0.2) & 5.2 (0.3) & 6.8 (0.3) &  & 0.2 (0) & 0.4 (0.1) & 0.4 (0.1) & 0.7 (0.1) & 0.9 (0.1) \\ 
   \hline n & & \multicolumn{5}{c}{$p^* = 50; SNR =  1.25 ; \bar{\sigma} =  16.52 $} & & \multicolumn{5}{c}{$p^* = 50; SNR =  5 ;  \bar{\sigma} =  8.26 $} \\
 \cmidrule(l){3-7} \cmidrule(l){9-13} 
50 &  & 35.6 (1.3) & 43.7 (0.7) & 46.8 (0.4) & 48.2 (0.3) & 49 (0.1) &  & 27.5 (1.4) & 41.8 (0.8) & 45.2 (0.5) & 47.5 (0.3) & 48.7 (0.2) \\ 
  100 &  & 23.6 (1.3) & 34.7 (1.3) & 42.2 (0.9) & 45.1 (0.6) & 46.7 (0.4) &  & 6 (0.6) & 18.1 (1) & 31 (1) & 40.1 (0.8) & 45.1 (0.5) \\ 
  200 &  & 7.7 (0.4) & 16.1 (0.8) & 24.2 (0.9) & 32.9 (1) & 38.4 (0.8) &  & 1 (0.1) & 3 (0.2) & 7 (0.4) & 15.4 (0.7) & 26.9 (0.8) \\ 
   \hline
\end{tabular}
\end{table}
\begin{table}[ht]
\centering
\caption{Average number of false negatives of 100 replications for $\alpha$-modified CV 1SE. Standard errors given in parentheses.} 
\label{tab:ModFN}
\begin{tabular}{rcp{0.9cm}p{0.9cm}p{0.9cm}p{0.9cm}p{0.9cm}cp{0.9cm}p{0.9cm}p{0.9cm}p{0.9cm}p{0.9cm}}
  \hline
  \hline
 & p & 50 & 100 & 200 & 400 & 800 &  & 50 & 100 & 200 & 400 & 800 \\ 
   \hline n & & \multicolumn{5}{c}{$p^* = 5; SNR =  1.25 ; \bar{\sigma} =  5.23 $} & & \multicolumn{5}{c}{$p^* = 5; SNR =  5 ;  \bar{\sigma} =  2.61 $} \\
 \cmidrule(l){3-7} \cmidrule(l){9-13} 
50 &  & 2.5 (0.2) & 2.6 (0.2) & 3.2 (0.2) & 3.1 (0.2) & 3.6 (0.1) &  & 0.4 (0.1) & 0.5 (0.1) & 1 (0.1) & 1 (0.1) & 1.5 (0.2) \\ 
  100 &  & 1.1 (0.1) & 1.1 (0.1) & 1.4 (0.1) & 1.8 (0.1) & 1.9 (0.1) &  & 0 (0) & 0.1 (0) & 0.1 (0) & 0.1 (0) & 0.2 (0) \\ 
  200 &  & 0.3 (0.1) & 0.4 (0.1) & 0.5 (0.1) & 0.6 (0.1) & 0.7 (0.1) &  & 0 (0) & 0 (0) & 0 (0) & 0 (0) & 0 (0) \\ 
   \hline n & & \multicolumn{5}{c}{$p^* = 15; SNR =  1.25 ; \bar{\sigma} =  9.04 $} & & \multicolumn{5}{c}{$p^* = 15; SNR =  5 ;  \bar{\sigma} =  4.52 $} \\
 \cmidrule(l){3-7} \cmidrule(l){9-13} 
50 &  & 10.9 (0.5) & 12.6 (0.3) & 12.8 (0.3) & 14 (0.2) & 14.3 (0.1) &  & 4.7 (0.4) & 7.2 (0.5) & 10 (0.4) & 11.8 (0.3) & 12.6 (0.3) \\ 
  100 &  & 6.3 (0.4) & 8.5 (0.4) & 10.7 (0.4) & 12.4 (0.3) & 12.8 (0.3) &  & 0.7 (0.1) & 1.3 (0.1) & 2.4 (0.2) & 4.2 (0.3) & 6.1 (0.4) \\ 
  200 &  & 2.1 (0.2) & 3.2 (0.2) & 5 (0.3) & 6.1 (0.4) & 7.8 (0.4) &  & 0.2 (0) & 0.3 (0.1) & 0.4 (0.1) & 0.6 (0.1) & 0.7 (0.1) \\ 
   \hline n & & \multicolumn{5}{c}{$p^* = 50; SNR =  1.25 ; \bar{\sigma} =  16.52 $} & & \multicolumn{5}{c}{$p^* = 50; SNR =  5 ;  \bar{\sigma} =  8.26 $} \\
 \cmidrule(l){3-7} \cmidrule(l){9-13} 
50 &  & 42.8 (1.1) & 44.6 (0.8) & 48.1 (0.4) & 48.5 (0.3) & 49.1 (0.2) &  & 28.7 (1.5) & 40.4 (1) & 45.5 (0.7) & 47.3 (0.4) & 48.7 (0.2) \\ 
  100 &  & 33.9 (1.5) & 44.4 (1) & 47 (0.7) & 47.9 (0.5) & 48.4 (0.4) &  & 6.7 (0.6) & 20.3 (1.3) & 34.6 (1.2) & 41.6 (1) & 46.4 (0.6) \\ 
  200 &  & 9.9 (0.7) & 27.3 (1.4) & 35.9 (1.3) & 44.1 (0.9) & 46.8 (0.7) &  & 0.9 (0.1) & 2.7 (0.2) & 7 (0.4) & 15.2 (0.7) & 29 (1.1) \\ 
   \hline
\end{tabular}
\end{table}
\begin{table}[ht]
\centering
\caption{Average number of false negatives for 100 replications of Relaxed Lasso. Standard errors given in parentheses.} 
\label{tab:RelFN}
\begin{tabular}{rcp{0.9cm}p{0.9cm}p{0.9cm}p{0.9cm}p{0.9cm}cp{0.9cm}p{0.9cm}p{0.9cm}p{0.9cm}p{0.9cm}}
  \hline
  \hline
 & p & 50 & 100 & 200 & 400 & 800 &  & 50 & 100 & 200 & 400 & 800 \\ 
   \hline n & & \multicolumn{5}{c}{$p^* = 5; SNR =  1.25 ; \bar{\sigma} =  5.23 $} & & \multicolumn{5}{c}{$p^* = 5; SNR =  5 ;  \bar{\sigma} =  2.61 $} \\
 \cmidrule(l){3-7} \cmidrule(l){9-13} 
50 &  & 1.4 (0.1) & 1.4 (0.1) & 2 (0.1) & 2 (0.1) & 2.5 (0.1) &  & 0.2 (0) & 0.3 (0.1) & 0.6 (0.1) & 0.4 (0.1) & 0.7 (0.1) \\ 
  100 &  & 0.6 (0.1) & 0.7 (0.1) & 0.8 (0.1) & 1.1 (0.1) & 1.2 (0.1) &  & 0 (0) & 0.1 (0) & 0.1 (0) & 0.1 (0) & 0.2 (0) \\ 
  200 &  & 0.2 (0) & 0.1 (0) & 0.3 (0.1) & 0.4 (0.1) & 0.6 (0.1) &  & 0 (0) & 0 (0) & 0 (0) & 0 (0) & 0 (0) \\ 
   \hline n & & \multicolumn{5}{c}{$p^* = 15; SNR =  1.25 ; \bar{\sigma} =  9.04 $} & & \multicolumn{5}{c}{$p^* = 15; SNR =  5 ;  \bar{\sigma} =  4.52 $} \\
 \cmidrule(l){3-7} \cmidrule(l){9-13} 
50 &  & 7.8 (0.4) & 8.4 (0.4) & 10 (0.4) & 12 (0.3) & 12.8 (0.2) &  & 2.8 (0.3) & 3.2 (0.3) & 6.5 (0.4) & 8.3 (0.3) & 10.6 (0.3) \\ 
  100 &  & 3.2 (0.2) & 5 (0.3) & 5.7 (0.3) & 8.5 (0.4) & 9.5 (0.3) &  & 0.6 (0.1) & 1.2 (0.1) & 1.8 (0.2) & 2.4 (0.2) & 3.6 (0.3) \\ 
  200 &  & 0.9 (0.1) & 1.9 (0.2) & 2.7 (0.2) & 4.1 (0.3) & 5.3 (0.3) &  & 0.2 (0) & 0.4 (0.1) & 0.6 (0.1) & 0.9 (0.1) & 1.1 (0.1) \\ 
   \hline n & & \multicolumn{5}{c}{$p^* = 50; SNR =  1.25 ; \bar{\sigma} =  16.52 $} & & \multicolumn{5}{c}{$p^* = 50; SNR =  5 ;  \bar{\sigma} =  8.26 $} \\
 \cmidrule(l){3-7} \cmidrule(l){9-13} 
50 &  & 33.9 (1.3) & 34.6 (1.3) & 43.2 (0.8) & 44.9 (0.5) & 47.2 (0.3) &  & 22.1 (1.4) & 29.5 (1.3) & 34.9 (0.9) & 41.4 (0.6) & 46.1 (0.4) \\ 
  100 &  & 19 (1.1) & 32.9 (1.2) & 35.9 (1.3) & 40.1 (1) & 43.9 (0.6) &  & 3.3 (0.5) & 13.4 (0.9) & 19.8 (1.1) & 30.7 (1.1) & 38.7 (0.9) \\ 
  200 &  & 4.9 (0.4) & 12.6 (0.8) & 20.7 (0.9) & 29.8 (1) & 37 (1) &  & 0.7 (0.1) & 2.2 (0.3) & 5 (0.4) & 8.8 (0.6) & 15.2 (0.8) \\ 
   \hline
\end{tabular}
\end{table}
\begin{table}[ht]
\centering
\caption{Average prediction bias from 100 replications of CV 1SE. Standard errors given in parentheses.} 
\label{tab:APEPB}
\begin{tabular}{rcp{0.9cm}p{0.9cm}p{0.9cm}p{0.9cm}p{0.9cm}cp{0.9cm}p{0.9cm}p{0.9cm}p{0.9cm}p{0.9cm}}
  \hline
  \hline
 & p & 50 & 100 & 200 & 400 & 800 &  & 50 & 100 & 200 & 400 & 800 \\ 
   \hline n & & \multicolumn{5}{c}{$p^* = 5; SNR =  1.25 ; \bar{\sigma} =  5.23 $} & & \multicolumn{5}{c}{$p^* = 5; SNR =  5 ;  \bar{\sigma} =  2.61 $} \\
 \cmidrule(l){3-7} \cmidrule(l){9-13} 
50 &  & 28.8 (0.9) & 28.5 (0.8) & 31.4 (0.8) & 32.4 (0.7) & 34.4 (0.8) &  & 14.2 (0.4) & 15.3 (0.5) & 16.5 (0.5) & 18.1 (0.6) & 21.2 (0.7) \\ 
  100 &  & 28.2 (0.8) & 29.6 (0.9) & 32.1 (0.9) & 35.8 (0.9) & 38.3 (0.9) &  & 14.6 (0.4) & 15.1 (0.4) & 16 (0.4) & 17.8 (0.4) & 19.6 (0.5) \\ 
  200 &  & 29.9 (0.8) & 32.2 (0.9) & 33.3 (0.9) & 36.4 (0.8) & 37.7 (0.8) &  & 15.1 (0.3) & 16.2 (0.4) & 16.4 (0.4) & 18.2 (0.4) & 19.1 (0.4) \\ 
   \hline n & & \multicolumn{5}{c}{$p^* = 15; SNR =  1.25 ; \bar{\sigma} =  9.04 $} & & \multicolumn{5}{c}{$p^* = 15; SNR =  5 ;  \bar{\sigma} =  4.52 $} \\
 \cmidrule(l){3-7} \cmidrule(l){9-13} 
50 &  & 56.4 (1.4) & 61.4 (1.1) & 61.7 (1.1) & 65.2 (1) & 66.5 (1) &  & 33.6 (1.1) & 39.1 (1.4) & 46.8 (1.5) & 51.4 (1.7) & 54.2 (1.7) \\ 
  100 &  & 60 (1.2) & 69.4 (1.4) & 77.5 (1.3) & 86.4 (1.5) & 88.3 (1.5) &  & 30.3 (0.6) & 34.1 (0.6) & 40.3 (0.9) & 46.7 (1.2) & 55.8 (1.5) \\
\hline
\end{tabular}
\end{table}
\begin{table}[ht]
\centering
\caption{Average prediction bias for 100 replications of AR2 CV 1SE. Standard errors given in parentheses.} 
\label{tab:AR2PB}
\begin{tabular}{rcp{0.9cm}p{0.9cm}p{0.9cm}p{0.9cm}p{0.9cm}cp{0.9cm}p{0.9cm}p{0.9cm}p{0.9cm}p{0.9cm}}
  \hline
  \hline
 & p & 50 & 100 & 200 & 400 & 800 &  & 50 & 100 & 200 & 400 & 800 \\ 
   \hline n & & \multicolumn{5}{c}{$p^* = 5; SNR =  1.25 ; \bar{\sigma} =  5.23 $} & & \multicolumn{5}{c}{$p^* = 5; SNR =  5 ;  \bar{\sigma} =  2.61 $} \\
 \cmidrule(l){3-7} \cmidrule(l){9-13} 
50 &  & 27.4 (0.8) & 28.6 (0.6) & 29.5 (0.6) & 31.4 (0.7) & 33 (0.6) &  & 15.3 (0.6) & 15.4 (0.5) & 18.3 (0.6) & 19.4 (0.7) & 21 (0.7) \\ 
  100 &  & 27.5 (0.8) & 28.3 (0.9) & 29.5 (0.9) & 32.9 (0.8) & 34.9 (0.9) &  & 13.7 (0.5) & 14.8 (0.4) & 14.3 (0.4) & 15.9 (0.5) & 16.8 (0.6) \\ 
  200 &  & 27.6 (0.9) & 28.8 (0.8) & 28.9 (0.8) & 30.9 (0.9) & 32.1 (0.9) &  & 14 (0.5) & 15 (0.4) & 14.2 (0.4) & 15.9 (0.4) & 15.8 (0.5) \\ 
   \hline n & & \multicolumn{5}{c}{$p^* = 15; SNR =  1.25 ; \bar{\sigma} =  9.04 $} & & \multicolumn{5}{c}{$p^* = 15; SNR =  5 ;  \bar{\sigma} =  4.52 $} \\
 \cmidrule(l){3-7} \cmidrule(l){9-13} 
50 &  & 57.5 (1.1) & 60.6 (0.9) & 61.6 (0.9) & 63.7 (0.9) & 66.1 (0.8) &  & 35.7 (1) & 43.1 (1.3) & 46.2 (1.2) & 50.7 (1.2) & 52.3 (1.2) \\ 
  100 &  & 61.7 (1.1) & 71.1 (1.2) & 78.4 (1.2) & 83.3 (1.1) & 85.1 (1.1) &  & 32.2 (0.7) & 34.6 (0.6) & 41.5 (1) & 47.8 (1.2) & 53.8 (1.5) \\
\hline
\end{tabular}
\end{table}
\begin{table}[ht]
\centering
\caption{Average number of prediction bias of 100 replications for $\alpha$-modified CV 1SE. Standard errors given in parentheses.} 
\label{tab:ModPB}
\begin{tabular}{rcp{0.9cm}p{0.9cm}p{0.9cm}p{0.9cm}p{0.9cm}cp{0.9cm}p{0.9cm}p{0.9cm}p{0.9cm}p{0.9cm}}
  \hline
  \hline
 & p & 50 & 100 & 200 & 400 & 800 &  & 50 & 100 & 200 & 400 & 800 \\ 
   \hline n & & \multicolumn{5}{c}{$p^* = 5; SNR =  1.25 ; \bar{\sigma} =  5.23 $} & & \multicolumn{5}{c}{$p^* = 5; SNR =  5 ;  \bar{\sigma} =  2.61 $} \\
 \cmidrule(l){3-7} \cmidrule(l){9-13} 
50 &  & 28.5 (0.9) & 29.8 (0.8) & 32.1 (0.7) & 32.8 (0.8) & 34.7 (0.7) &  & 13.1 (0.4) & 14.1 (0.5) & 16.2 (0.5) & 17.4 (0.5) & 19.6 (0.7) \\ 
  100 &  & 26.7 (0.9) & 26.9 (0.9) & 28.9 (0.9) & 33.3 (0.9) & 35.2 (0.9) &  & 12.4 (0.4) & 13.3 (0.4) & 12.9 (0.4) & 14.6 (0.4) & 15.5 (0.5) \\ 
  200 &  & 26.6 (0.8) & 27.9 (0.8) & 28.5 (0.8) & 30.3 (0.9) & 31.6 (0.9) &  & 13.1 (0.4) & 13.7 (0.4) & 13.1 (0.4) & 14.4 (0.4) & 14.2 (0.4) \\ 
   \hline n & & \multicolumn{5}{c}{$p^* = 15; SNR =  1.25 ; \bar{\sigma} =  9.04 $} & & \multicolumn{5}{c}{$p^* = 15; SNR =  5 ;  \bar{\sigma} =  4.52 $} \\
 \cmidrule(l){3-7} \cmidrule(l){9-13} 
50 &  & 61 (1.3) & 65.3 (1) & 63.8 (1) & 66.4 (1) & 68.7 (0.9) &  & 34.5 (1) & 40.6 (1.4) & 45.5 (1.5) & 50.4 (1.6) & 51.9 (1.6) \\ 
  100 &  & 65 (1.4) & 73.9 (1.4) & 83.1 (1.4) & 89.7 (1.3) & 90.8 (1.4) &  & 30.5 (0.6) & 33.6 (0.6) & 39.9 (0.9) & 45.6 (1.2) & 51.7 (1.4) \\
\hline
\end{tabular}
\end{table}
\begin{table}[ht]
\centering
\caption{Average prediction bias for 100 replications of Relaxed Lasso. Standard errors given in parentheses.} 
\label{tab:RelPB}
\begin{tabular}{rcp{0.9cm}p{0.9cm}p{0.9cm}p{0.9cm}p{0.9cm}cp{0.9cm}p{0.9cm}p{0.9cm}p{0.9cm}p{0.9cm}}
  \hline
  \hline
 & p & 50 & 100 & 200 & 400 & 800 &  & 50 & 100 & 200 & 400 & 800 \\ 
   \hline n & & \multicolumn{5}{c}{$p^* = 5; SNR =  1.25 ; \bar{\sigma} =  5.23 $} & & \multicolumn{5}{c}{$p^* = 5; SNR =  5 ;  \bar{\sigma} =  2.61 $} \\
 \cmidrule(l){3-7} \cmidrule(l){9-13} 
50 &  & 24.5 (0.7) & 27.8 (0.8) & 31.8 (0.8) & 32.4 (0.8) & 34.1 (0.7) &  & 12.3 (0.5) & 13.4 (0.5) & 15.6 (0.5) & 17 (0.5) & 17.8 (0.5) \\ 
  100 &  & 21.4 (0.7) & 23.2 (0.9) & 24.4 (0.8) & 28.9 (1.1) & 34.5 (1.1) &  & 8.5 (0.3) & 9.8 (0.4) & 9.8 (0.5) & 11.6 (0.6) & 12.3 (0.6) \\ 
  200 &  & 18.2 (0.8) & 17.4 (0.8) & 18.9 (0.8) & 22.7 (1) & 23.7 (1) &  & 7.7 (0.3) & 8.3 (0.4) & 7.8 (0.4) & 8.1 (0.4) & 7.9 (0.4) \\ 
   \hline n & & \multicolumn{5}{c}{$p^* = 15; SNR =  1.25 ; \bar{\sigma} =  9.04 $} & & \multicolumn{5}{c}{$p^* = 15; SNR =  5 ;  \bar{\sigma} =  4.52 $} \\
 \cmidrule(l){3-7} \cmidrule(l){9-13} 
50 &  & 50.5 (1) & 58.1 (1.1) & 59.6 (1) & 61.8 (1) & 63.7 (1) &  & 28.3 (0.5) & 32.7 (0.5) & 34.8 (0.9) & 36.9 (1) & 39.9 (1.2) \\ 
  100 &  & 52 (0.9) & 61.1 (1) & 68.7 (1.3) & 75.4 (1.3) & 80.8 (1.4) &  & 25.9 (0.6) & 31.6 (0.5) & 38 (0.8) & 42.9 (0.7) & 47 (0.8) \\
\hline
\end{tabular}
\end{table}

We also evaluated the results for more limited data settings---fixing $n$ and $p$, but varying $p^*$ and SNR across more methods of tuning parameter selection, including AIC, BIC, ERIC with $\nu = 0.5$, and GCV. For AIC and BIC, fixed $\hat{\sigma}^2$ was calculated as the MSE of the model chosen by CV 1SE whereas ERIC was calculated assuming unknown variance for the likelihood. 

\begin{figure}[h]
	\centering
	\includegraphics[width =  \textwidth]{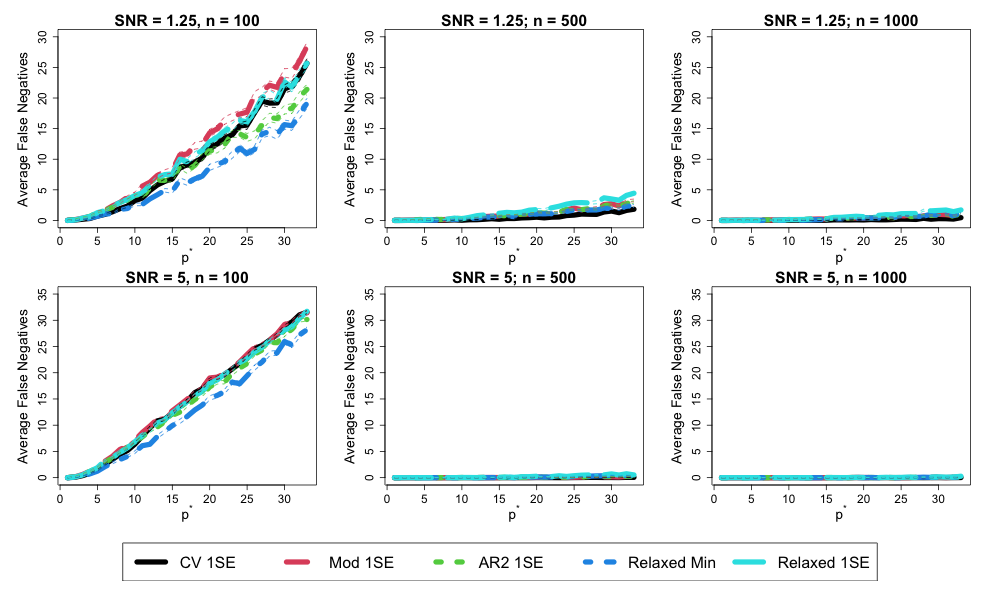}
	\caption{Average false negatives from 100 replications of simulated data with $n = 100, n = 500, n = 1000$ and $p = 100$ with independent predictors for both SNR $= 1.25$ and SNR $= 5$. Thin dotted lines represent the mean $\pm$ one standard error.}
	\label{fig:SimPlot100FN}
\end{figure}

\begin{figure}[h]
	\centering
	\includegraphics[width =  \textwidth]{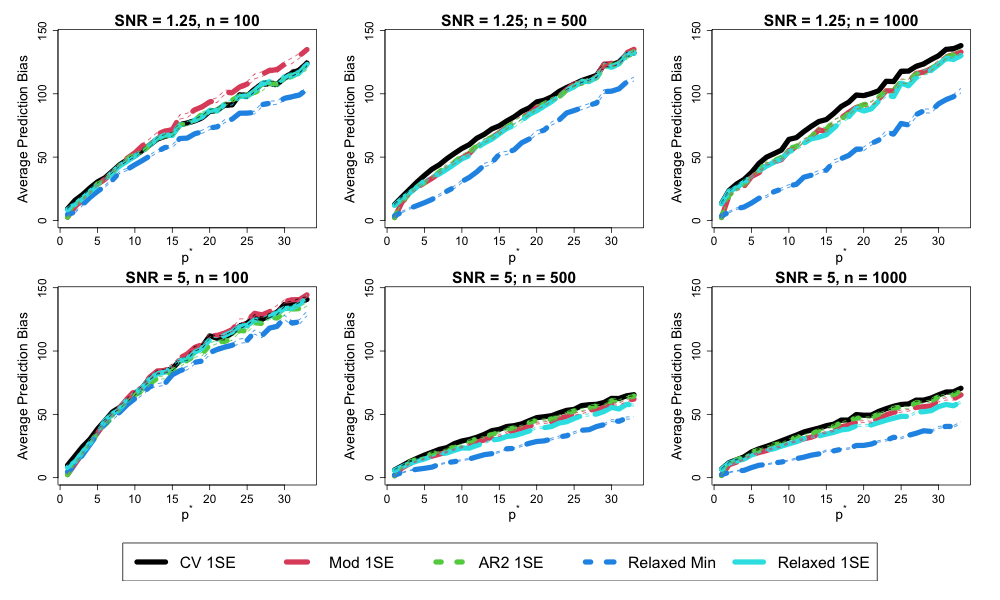}
	\caption{Average prediction bias from 100 replications of simulated data with $n = 100, n = 500, n = 1000$ and $p = 100$ with independent predictors for both SNR $= 1.25$ and SNR $= 5$. Thin dotted lines represent the mean $\pm$ one standard error.}
	\label{fig:SimPlot100PB}
\end{figure}

Figures \ref{fig:SimPlot100FN} and \ref{fig:SimPlot100PB} give the average false negatives and average prediction bias for the simulated data provided in the main document's Figure (5). For larger sample sizes, the number of false negatives is close to zero for most $p^*$, but for $n=100$, as noted in the main document, average false negatives increase with $p^*.$ 

\begin{figure}[h]
	\centering
	\includegraphics[width = 0.75 \textwidth]{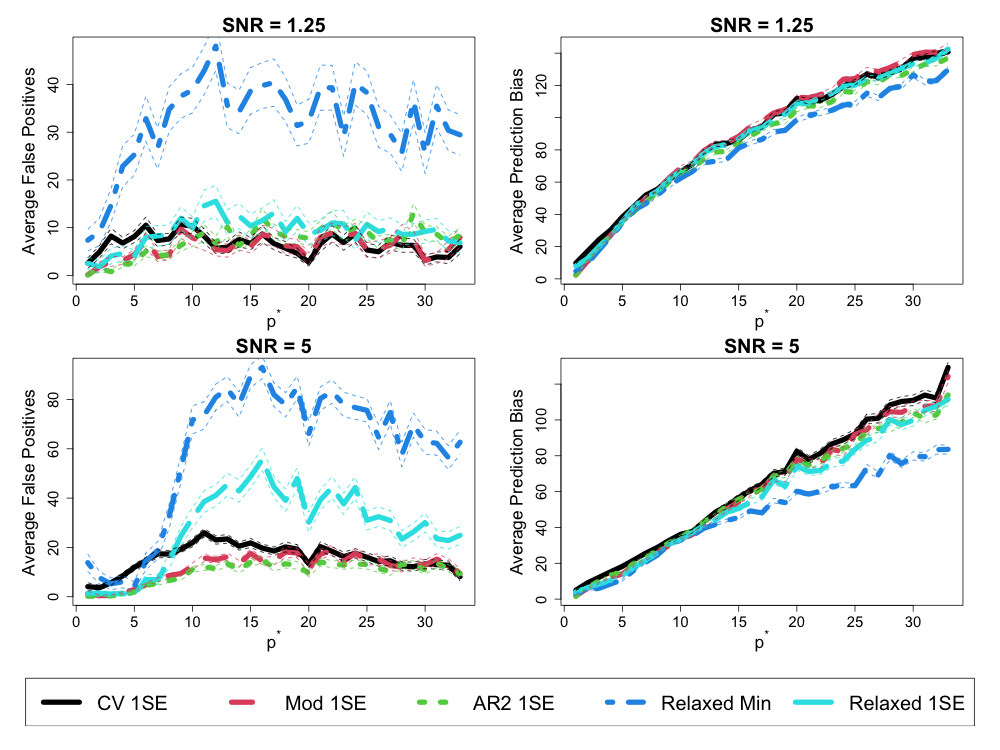}
	\caption{Average false positives and prediction error plots from 100 replications of simulated data with $n = 100$ and $p = 800$ for both SNR $= 0.5$ and SNR $= 2$. Thin dotted lines represent the mean $\pm$ one standard error.}
	\label{fig:SimPlot800}
\end{figure}

\begin{figure}[h]
	\centering
	\includegraphics[width = 0.75 \textwidth]{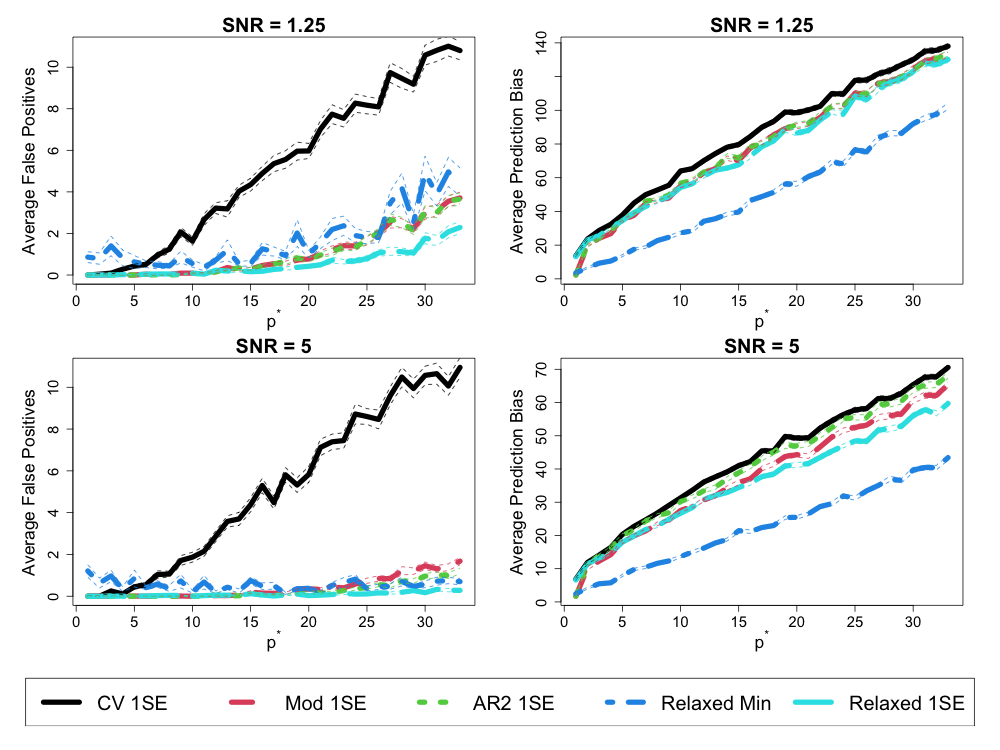}
	\caption{Average false positives and prediction error plots from 100 replications of simulated data with $n = 1000$ and $p = 100$ for both SNR $= 0.5$ and SNR $= 2$. Thin dotted lines represent the mean $\pm$ one standard error.}
	\label{fig:SimPlotLN}
\end{figure}

Table \ref{tab:n100FN} gives the average false negatives from the setting considered in the main document: $n = p = 100$. Figure \ref{fig:SimPlot800} gives the average false positives and prediction bias for data with $n = 100$ and $p = 800$. Table \ref{fig:SimPlotLN} gives average false positives and prediction bias for $n = 1000$ and $p = 100$. Tables \ref{tab:n1000FDR}, \ref{tab:n1000PB}, and \ref{tab:n1000FN} give the false discovery rates, prediction bias, and false negatives for this last setting.

\begin{table}[ht]
\centering
\caption{Average false negatives for 100 replications of simulated data with $n = 100$. Standard errors given in parentheses.} 
\label{tab:n100FN}
\begin{tabular}{r|p{1.3cm}p{1.4cm}p{1.3cm}p{1.4cm}|p{1.3cm}p{1.4cm}p{1.3cm}p{1.4cm}}
  \hline & \multicolumn{4}{c|}{$p = 100$} & \multicolumn{4}{c}{$p = 800$} \\

                      & \multicolumn{2}{c}{SNR = 1.25} &  \multicolumn{2}{c|}{SNR = 5} & \multicolumn{2}{c}{SNR = 1.25} & \multicolumn{2}{c}{SNR = 5}\\
 \hline
 & $p^* = 5$ & $p^* = 33$ & $p^* = 5$ & $p^* = 33$ & $p^* = 5$ & $p^* = 33$ & $p^* = 5$ & $p^* = 33$ \\ 
  \hline
CV Min & 0.22 (0.05) & 15.57 (0.67) & 0 (0) & 4.05 (0.28) & 0.69 (0.09) & 28.05 (0.49) & 0.02 (0.01) & 22.24 (0.62) \\ 
  CV 1SE & 0.71 (0.09) & 25.67 (0.74) & 0.01 (0.01) & 7.71 (0.46) & 1.45 (0.15) & 31.54 (0.3) & 0.04 (0.02) & 29.41 (0.46) \\ 
  Mod Min & 0.48 (0.07) & 20.78 (0.97) & 0.01 (0.01) & 4.75 (0.39) & 1.11 (0.12) & 28.87 (0.55) & 0.03 (0.02) & 24.36 (0.75) \\ 
  Mod 1SE & 1.4 (0.13) & 28.42 (0.7) & 0.09 (0.03) & 8.74 (0.57) & 1.96 (0.16) & 31.42 (0.36) & 0.14 (0.04) & 29.52 (0.5) \\ 
  AR2 Min & 0.48 (0.08) & 11.91 (0.82) & 0.01 (0.01) & 4.26 (0.4) & 1.02 (0.11) & 26.34 (0.48) & 0.05 (0.03) & 22.83 (0.6) \\
\hline
\end{tabular}
\end{table}

\begin{table}[ht]
\centering
\caption{Average false discovery rate for 100 replications of simulated data with $n = 1000$. Standard errors given in parentheses.} 
\label{tab:n1000FDR}
\begin{tabular}{r|cc|cc}
  \hline & \multicolumn{2}{c|}{SNR = 1.25} &  \multicolumn{2}{c}{SNR = 5}\\
 \hline
 & $p^* = 5$ & $p^* = 33$ & $p^* = 5$ & $p^* = 33$ \\ 
  \hline
CV Min & 0.68 (0.01) & 0.49 (0) & 0.69 (0.01) & 0.49 (0.01) \\ 
  CV 1SE & 0.06 (0.01) & 0.24 (0.01) & 0.06 (0.01) & 0.24 (0.01) \\ 
  Mod Min & 0.37 (0.03) & 0.36 (0.01) & 0.26 (0.03) & 0.32 (0.01) \\ 
  Mod 1SE & 0 (0) & 0.1 (0.01) & 0 (0) & 0.05 (0) \\ 
  AR2 Min & 0.38 (0.03) & 0.38 (0.01) & 0.27 (0.03) & 0.32 (0.01) \\ 
  AR2 1SE & 0 (0) & 0.1 (0.01) & 0 (0) & 0.03 (0) \\ 
  R2 Cut & 0 (0) & 0.09 (0.01) & 0 (0) & 0.05 (0.01) \\ 
  AIC & 0.68 (0.01) & 0.5 (0.01) & 0.69 (0.01) & 0.49 (0.01) \\ 
  BIC & 0.2 (0.01) & 0.22 (0.01) & 0.2 (0.02) & 0.22 (0.01) \\ 
  ERIC & 0.07 (0.01) & 0.11 (0.01) & 0.08 (0.01) & 0.12 (0.01) \\ 
  GCV & 0.7 (0.01) & 0.5 (0.01) & 0.73 (0.01) & 0.5 (0.01) \\ 
  Relaxed Min & 0.04 (0.01) & 0.1 (0.01) & 0.06 (0.02) & 0.02 (0) \\ 
  Relaxed 1SE & 0 (0) & 0.06 (0.01) & 0 (0) & 0.01 (0) \\ 
   \hline
\end{tabular}
\end{table}
\begin{table}[ht]
\centering
\caption{Average prediction bias for 100 replications of simulated data with $n = 1000$. Standard errors given in parentheses.} 
\label{tab:n1000PB}
\begin{tabular}{r|cc|cc}
  \hline & \multicolumn{2}{c|}{SNR = 1.25} &  \multicolumn{2}{c}{SNR = 5}\\
 \hline
 & $p^* = 5$ & $p^* = 33$ & $p^* = 5$ & $p^* = 33$ \\ 
  \hline
CV Min & 23.87 (0.67) & 111.79 (1.23) & 12.67 (0.34) & 56.79 (0.64) \\ 
  CV 1SE & 37.85 (0.96) & 138.02 (1.5) & 20.31 (0.56) & 70.56 (0.77) \\ 
  Mod Min & 18.13 (0.63) & 110.67 (1.3) & 9.14 (0.32) & 51.99 (0.66) \\ 
  Mod 1SE & 34.64 (0.96) & 132.97 (1.54) & 18.08 (0.54) & 65.4 (0.89) \\ 
  AR2 Min & 18.35 (0.63) & 110.69 (1.29) & 9.14 (0.32) & 52.05 (0.66) \\ 
  AR2 1SE & 35.3 (0.99) & 133.79 (1.56) & 19.99 (0.54) & 68.21 (0.91) \\ 
  R2 Cut & 77.07 (2.09) & 187.98 (2.28) & 50.31 (1.82) & 112.37 (1.51) \\ 
  AIC & 24.12 (0.64) & 112.62 (1.23) & 12.98 (0.35) & 57.37 (0.65) \\ 
  BIC & 29.12 (0.83) & 136.75 (1.81) & 15.13 (0.42) & 70.12 (0.87) \\ 
  ERIC & 32.23 (0.91) & 164.4 (2.56) & 16.47 (0.43) & 81.64 (1.11) \\ 
  GCV & 24.46 (0.66) & 112.66 (1.24) & 13.16 (0.36) & 57.28 (0.64) \\ 
  Relaxed Min & 13.75 (0.61) & 102.47 (1.77) & 7.96 (0.34) & 43.42 (0.75) \\ 
  Relaxed 1SE & 34.52 (0.96) & 130.12 (1.81) & 17.97 (0.47) & 59.71 (0.84) \\ 
   \hline
\end{tabular}
\end{table}
\begin{table}[ht]
\centering
\caption{Average false negatives for 100 replications of simulated data with $n = 1000$. Standard errors given in parentheses.} 
\label{tab:n1000FN}
\begin{tabular}{r|cc|cc}
  \hline & \multicolumn{2}{c|}{SNR = 1.25} &  \multicolumn{2}{c}{SNR = 5}\\
 \hline
 & $p^* = 5$ & $p^* = 33$ & $p^* = 5$ & $p^* = 33$ \\ 
  \hline
CV Min & 0 (0) & 0.09 (0.03) & 0 (0) & 0.01 (0.01) \\ 
  CV 1SE & 0.01 (0.01) & 0.45 (0.06) & 0 (0) & 0.02 (0.01) \\ 
  Mod Min & 0.01 (0.01) & 0.29 (0.05) & 0 (0) & 0.02 (0.01) \\ 
  Mod 1SE & 0.06 (0.02) & 1.05 (0.1) & 0 (0) & 0.07 (0.03) \\ 
  AR2 Min & 0.01 (0.01) & 0.24 (0.05) & 0 (0) & 0.02 (0.01) \\ 
  AR2 1SE & 0.08 (0.03) & 1.1 (0.11) & 0 (0) & 0.1 (0.03) \\ 
  R2 Cut & 0.07 (0.03) & 1.2 (0.12) & 0 (0) & 0.1 (0.04) \\ 
  AIC & 0 (0) & 0.09 (0.03) & 0 (0) & 0.01 (0.01) \\ 
  BIC & 0.01 (0.01) & 0.47 (0.07) & 0 (0) & 0.02 (0.01) \\ 
  ERIC & 0.01 (0.01) & 1.01 (0.11) & 0 (0) & 0.03 (0.02) \\ 
  GCV & 0 (0) & 0.07 (0.03) & 0 (0) & 0.01 (0.01) \\ 
  Relaxed Min & 0.01 (0.01) & 1.17 (0.12) & 0 (0) & 0.18 (0.05) \\ 
  Relaxed 1SE & 0.05 (0.02) & 1.71 (0.16) & 0 (0) & 0.27 (0.05) \\ 
   \hline
\end{tabular}
\end{table}

\FloatBarrier
\subsection*{B.2 Simulation Study for Correlated Predictors}
Additional results from the simulation study described in the main document for correlated predictors can be found in Tables \ref{tab:APEFDRXCorr} through \ref{tab:RelPBXCorr}. These results include average model size, average FDR, average false negatives, and average prediction bias for traditional CV, $\alpha$-modified and Average $R^2$ CV, and the Relaxed Lasso using a minimum APE rule. 
\input{XCorrAPEHD}
\input{XCorrAR2HD}
\input{XCorrModHD}
\input{XCorrRelHD}
\input{XCorrAPESize}
\input{XCorrAR2Size}
\input{XCorrModSize}
\input{XCorrRelSize}
\input{XCorrAPEFDR}
\input{XCorrAR2FDR}
\input{XCorrModFDR}
\input{XCorrRelFDR}
\input{XCorrAPEFN}
\input{XCorrAR2FN}
\input{XCorrModFN}
\input{XCorrRelFN}
\input{XCorrAPEPB}
\input{XCorrAR2PB}
\input{XCorrModPB}
\input{XCorrRelPB}

\clearpage

\begin{figure}[h]
	\centering
	\includegraphics[width =  \textwidth]{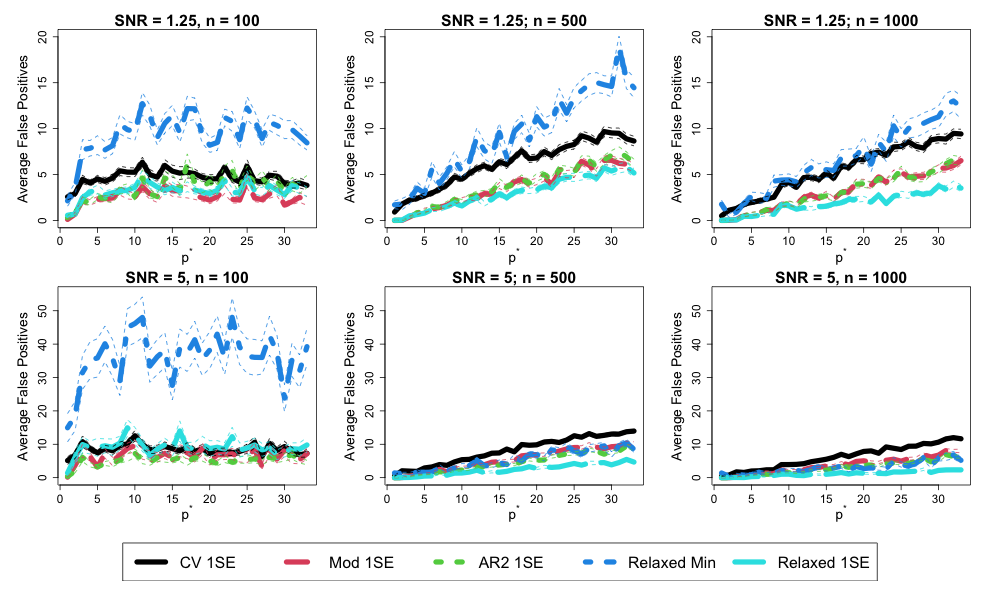}
	\caption{Average false positives from 100 replications of simulated data with $n = 100, n = 500, n = 1000$ and $p = 100$ with independent predictors for both SNR $= 1.25$ and SNR $= 5$. Thin dotted lines represent the mean $\pm$ one standard error.}
	\label{fig:SimPlot100FPCorrX}
\end{figure}

\begin{figure}[h]
	\centering
	\includegraphics[width =  \textwidth]{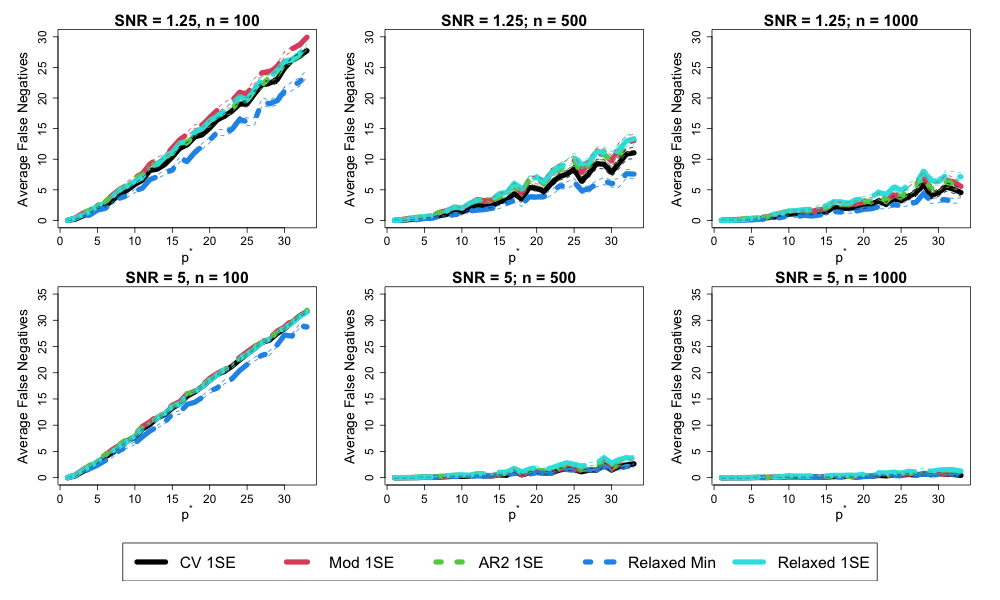}
	\caption{Average false negatives from 100 replications of simulated data with $n = 100, n = 500, n = 1000$ and $p = 100$ with independent predictors for both SNR $= 1.25$ and SNR $= 5$. Thin dotted lines represent the mean $\pm$ one standard error.}
	\label{fig:SimPlot100FNCorrX}
\end{figure}

\begin{figure}[h]
	\centering
	\includegraphics[width =  \textwidth]{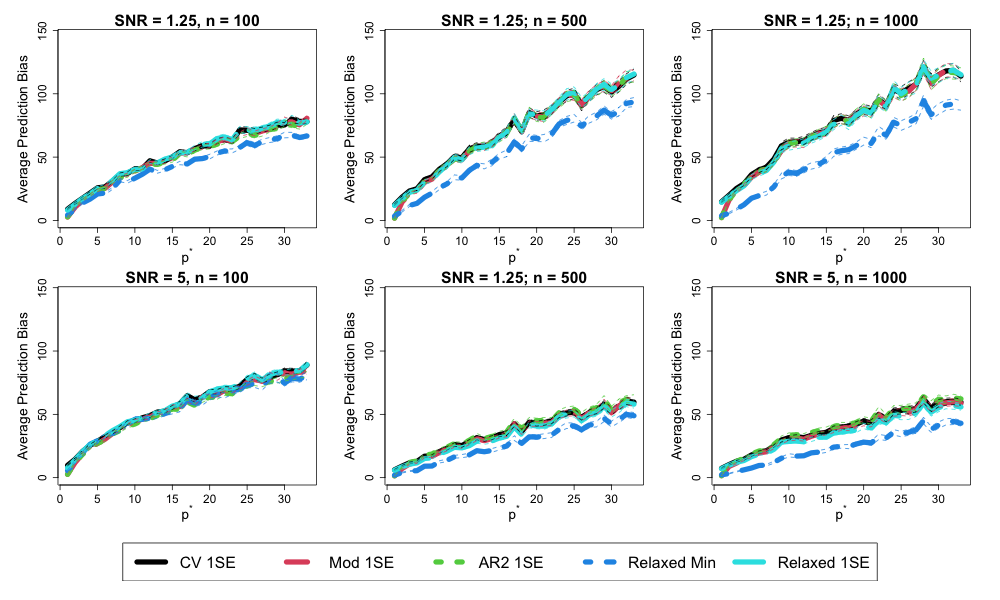}
	\caption{Average prediction bias from 100 replications of simulated data with $n = 100, n = 500, n = 1000$ and $p = 100$ with independent predictors for both SNR $= 1.25$ and SNR $= 5$. Thin dotted lines represent the mean $\pm$ one standard error.}
	\label{fig:SimPlot100PBCorrX}
\end{figure}

Figures \ref{fig:SimPlot100FPCorrX}, \ref{fig:SimPlot100FNCorrX}, and \ref{fig:SimPlot100PBCorrX} give the average false positives, false negatives, and prediction bias for the simulated data provided in the main document's Figure (6). Similar to independent predictors, larger sample sizes maintain low false negatives, but for $n=100$ and $n = 500$ with the smaller SNR, average false negatives increase with $p^*.$ 

\clearpage

\begin{figure}[h]
	\centering
	\includegraphics[width = 0.75 \textwidth]{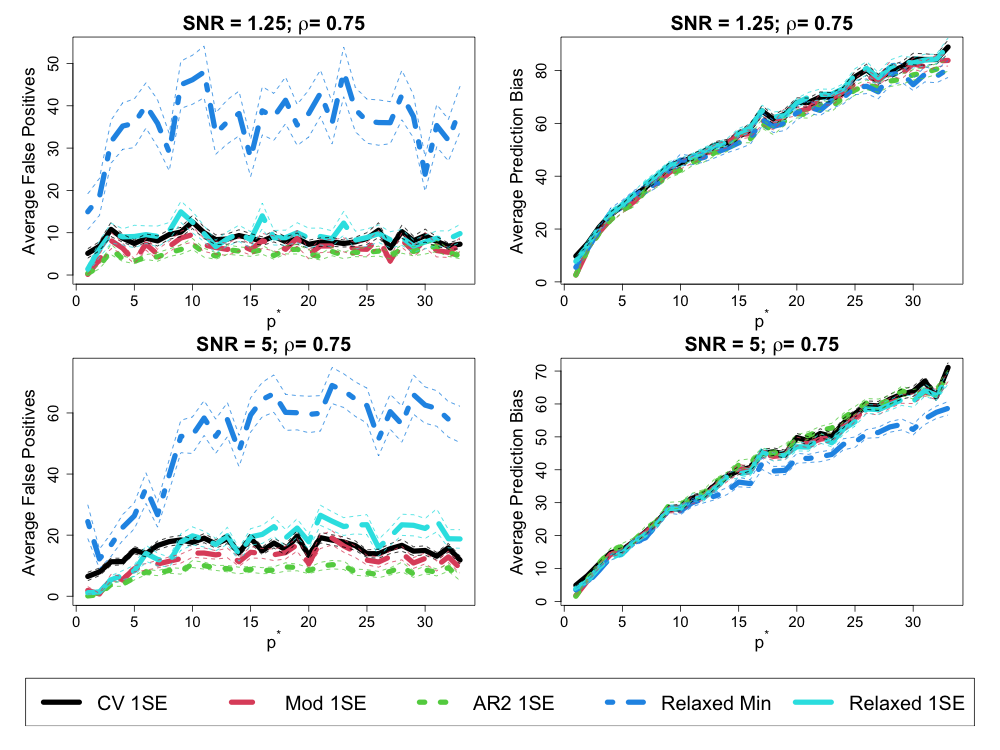}
	\caption{Average false positives and prediction error plots from 100 replications of simulated data with $n = 100$ and $p = 800$ for both SNR $= 0.5$ and SNR $= 2$ with correlated predictors. Thin dotted lines represent the mean $\pm$ one standard error.}
	\label{fig:SimPlot800CorrX}
\end{figure}

\begin{figure}[h]
	\centering
	\includegraphics[width = 0.75 \textwidth]{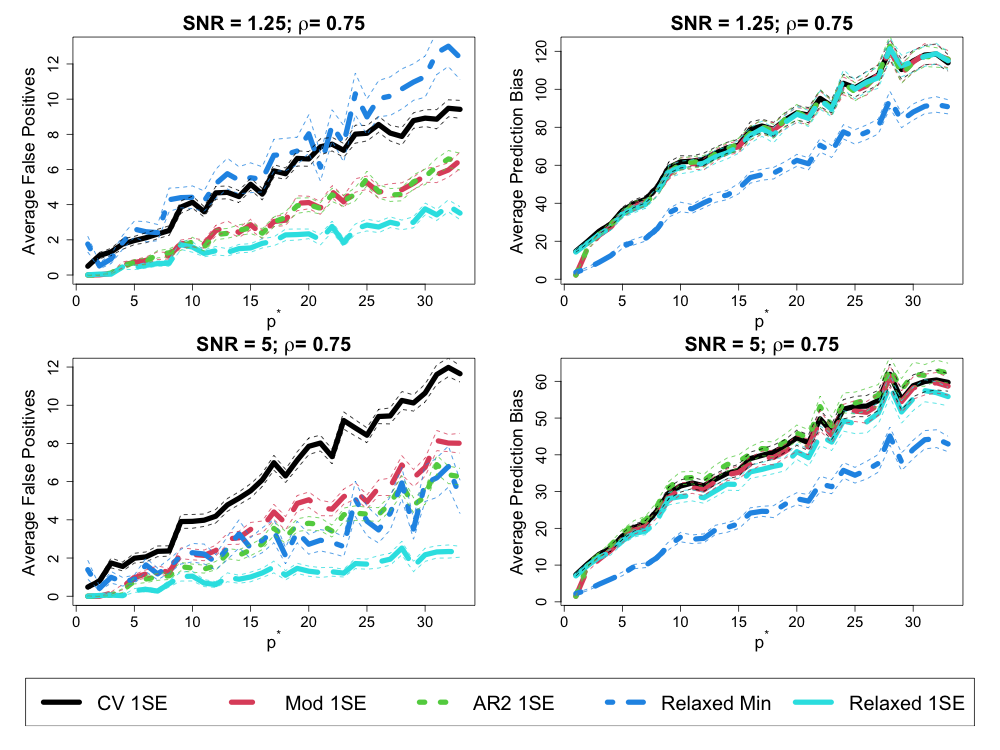}
	\caption{Average false positives and prediction error plots from 100 replications of simulated data with $n = 1000$ and $p = 100$ for both SNR $= 0.5$ and SNR $= 2$ with correlated predictors. Thin dotted lines represent the mean $\pm$ one standard error.}
	\label{fig:SimPlotLNCorrX}
\end{figure}

Finally, we evaluate the methods for limited data settings, letting $n = 100$, $p = \{100,800\}$, SNR $ = \{1.25, 5\}$ , and varying $p^*$. Once again, $\V{X}$ contained correlated columns and more methods are evaluated in Figure \ref{fig:SimPlot800CorrX} and Tables \ref{tab:n100FDRXCorr}, \ref{tab:n100PBXCorr}, and \ref{tab:n100FNXCorr}. Figure \ref{fig:SimPlotLNCorrX} and Tables \ref{tab:n1000FDRXCorr} through \ref{tab:n1000FNXCorr} give results from the case where $n = 1000$. Average false negatives are higher for all methods when predictors are correlated. Particularly for large $p^*$ all methods struggle to capture the true model, though the Relaxed Lasso seemingly performs the best. For the $n = 1000$ case, Mod 1SE, AR2 1SE, and Relaxed 1SE all have the smallest average FDRs so it seems that the new methods are most competitive with the Relaxed Lasso when sample sizes are large.

\input{n100XCorrFDRSNR}
\input{n100XCorrPBSNR}
\input{n100XCorrFNSNR}

\input{n1000XCorrFDRSNR}
\input{n1000XCorrPBSNR}
\input{n1000XCorrFNSNR}

\FloatBarrier
\subsection*{B.3. Simulation Study for Non-Convex Penalties}

This set of simulations examines the effect of the modification on the non-convex penalties, SCAD \citep{SCAD} and MC+ \citep{MinimaxConcavePenalty}. Both penalties are most commonly expressed in terms of its derivative which, for SCAD at a single point, $\beta_j$, is:
\begin{align}
	P'_\lambda(\beta_j) = \lambda_1 \Big[I(\beta_j \leq \lambda_1) + \frac{(\lambda_2\beta_j - \beta_j)_+}{(\lambda_2-1)\beta_j} I(\beta_j > \lambda_1) \Big] \ . \
\end{align}
The derivative of the MC+ penalty for a single coefficient, $\beta_j$, is:
\begin{align}
	P_\lambda'(\beta_j) = \begin{cases}
		\text{sign}(\beta_j) \Big(\lambda_1 - \frac{|\beta_j|}{\lambda_2} \Big) & \text{if} \hspace{5pt} \lvert \beta_j \rvert \leq  \lambda_1 \lambda_2\\
		0 & \text{otherwise}
	\end{cases} \ . \
\end{align}

\begin{figure}[h]
	\centering
	\includegraphics[width = 1 \textwidth]{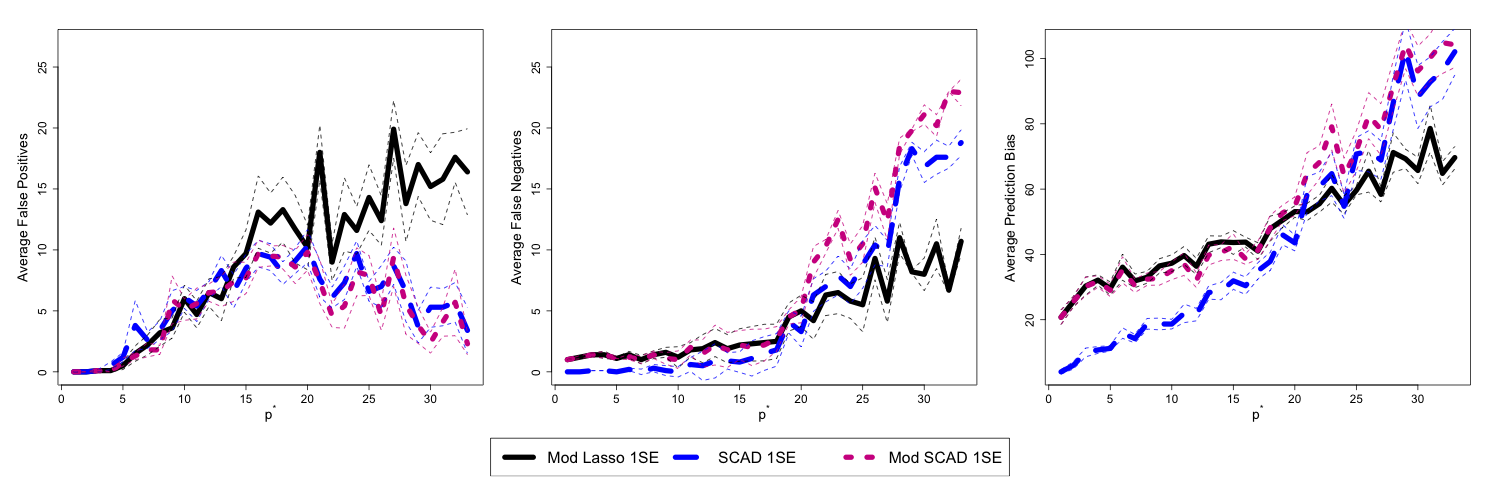}
	\caption{Average false positives, prediction error, and false negatives from 500 replications of simulated data with $n = 100$ and $p = 100$ for SNR = 5. Thin dotted lines represent the mean $\pm$ one standard error.}
	\label{fig:SCADPlot}
\end{figure}

\begin{figure}[h]
	\centering
	\includegraphics[width = 1 \textwidth]{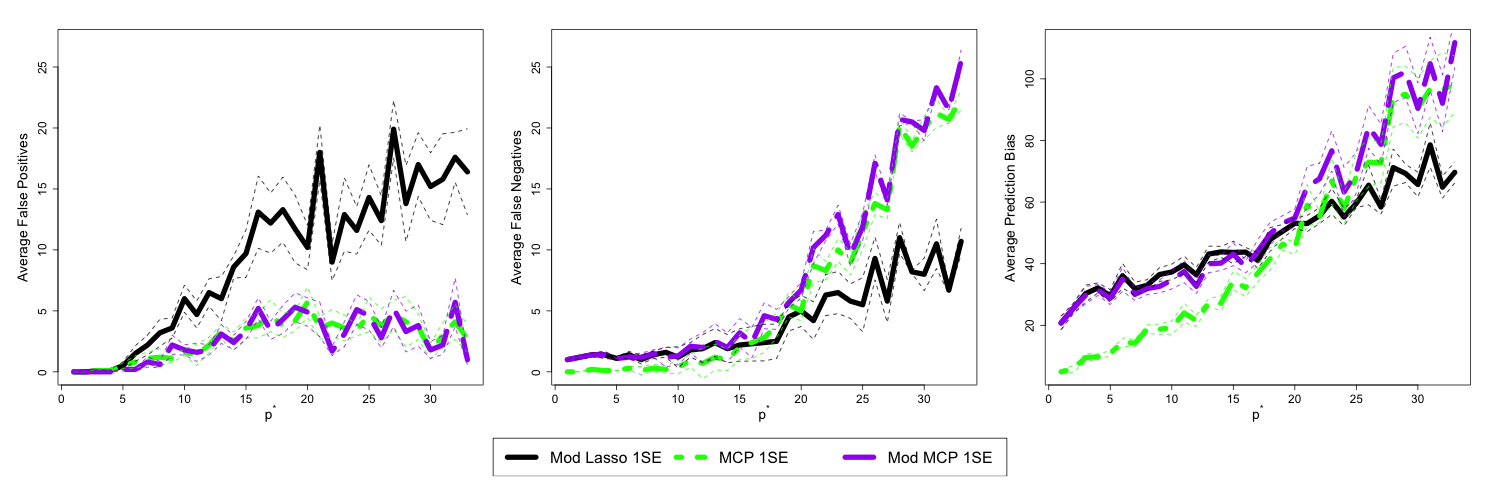}
	\caption{Average false positives, prediction error, and false negatives from 500 replications of simulated data with $n = 100$ and $p = 100$ for SNR = 5. Thin dotted lines represent the mean $\pm$ one standard error.}
	\label{fig:MCPPlot}
\end{figure}

We investigated SCAD with $\lambda_2 = 3.7$ and MC+ with $\lambda_2 = 3$, both with and without the $\alpha$-Modification using 10-fold CV with a 1SE rule.  We generated data from the following model:
\begin{equation}
	\V{y}_{n\times 1} = \beta^*_0  + \V{X}_{n \times p} \boldbeta_{p \times 1}^* + \boldsymbol{\epsilon}_{n \times 1}
	\label{eq:LinearModelIntercept}
\end{equation}
where $\V{X}$ was drawn from a Standard Normal distribution, $\beta^*_0$ and $\boldbeta^*$ were drawn from the same distribution as the Lasso simulations above, and SNR = \{1.25, 5\}. We include an intercept here because the \texttt{R} package \texttt{ncvreg} does not have include a no-intercept option. The optimization to find $\hat{\alpha}_\lambda$ is slightly more complicated with an intercept term. We now optimize the likelihood function over both $\alpha$ and $\beta_0$, the intercept. Figures \ref{fig:SCADPlot} and \ref{fig:MCPPlot} include the average False Positives, False Negatives, and Prediction Bias across increasing values of $p^*$ when $n = 100$ and $p = 100$ over 500 replications. Note the similarity between the two approaches.

Although average false positives for the non-convex penalties are lower than or equal to that of the $\alpha$-Modified Lasso, their average false negatives tend to be higher, particularly as $p^*$ increases. Prediction bias is slightly worse for $\alpha$-Modified SCAD and MC+ than unmodified, but similar to the $\alpha$-Modified Lasso for small $p^*$ and similar to unmodified for larger $p^*$. This is a fairly limited simulation study, capturing only one $n$ and $p$. Furthermore, for non-convex penalties, we no longer have a guarantee that $\hat{\alpha}_\lambda$ will always be greater than 1, so the bias-reduction aim of the modification is in question. Allowing the $\alpha$-Modification to compete with and enhance non-convex penalties may be the subject of future research.

\FloatBarrier
\subsection*{B.4. Simulation Study for Non-Gaussian Errors}

\begin{figure}[h]
	\centering
	\includegraphics[width = 0.75 \textwidth]{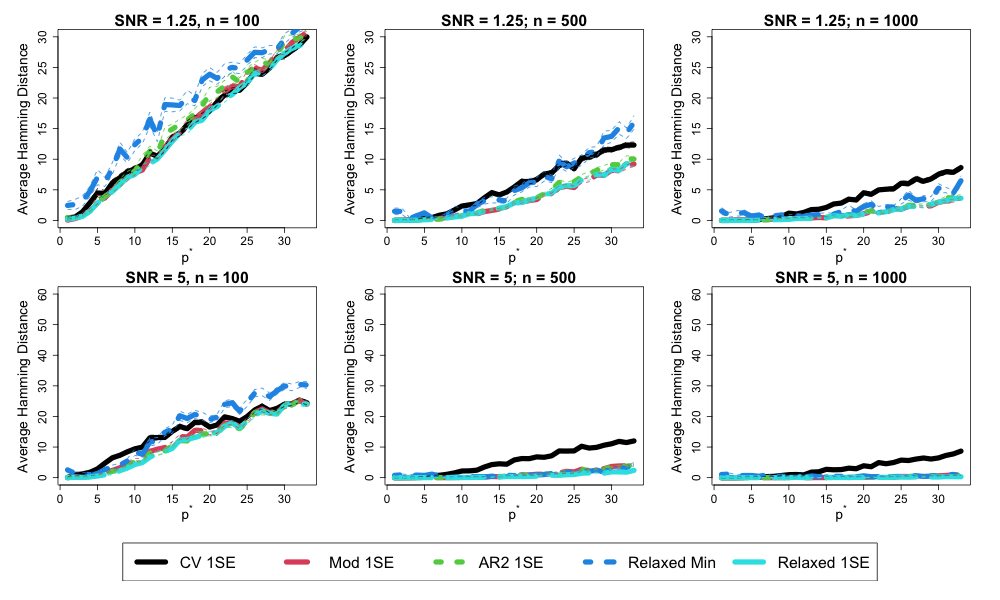}
	\caption{Average Hamming Distance from 100 replications of simulated data with $p = 100$ for both SNR $= 1.25$ and SNR $= 5$ independent Laplacian errors. Thin dotted lines represent the mean $\pm$ one standard error.}
	\label{fig:SimPlotLapHD}
\end{figure}

\begin{figure}[h]
	\centering
	\includegraphics[width = 0.75 \textwidth]{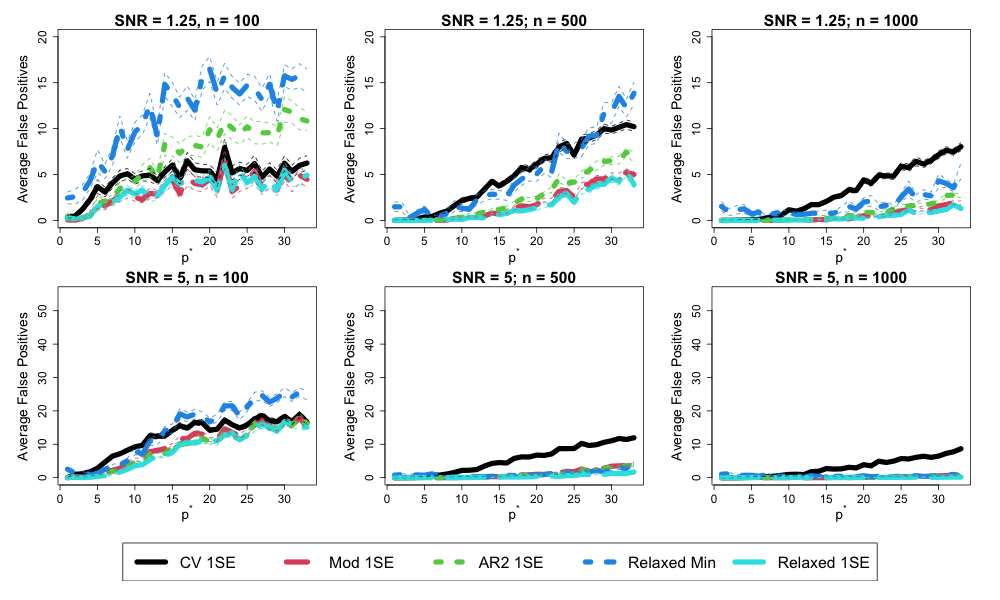}
	\caption{Average false positives from 100 replications of simulated data with $p = 100$ for both SNR $= 1.25$ and SNR $= 5$ independent Laplacian errors. Thin dotted lines represent the mean $\pm$ one standard error.}
	\label{fig:SimPlotLapFP}
\end{figure}

\begin{figure}[h]
	\centering
	\includegraphics[width = 0.75 \textwidth]{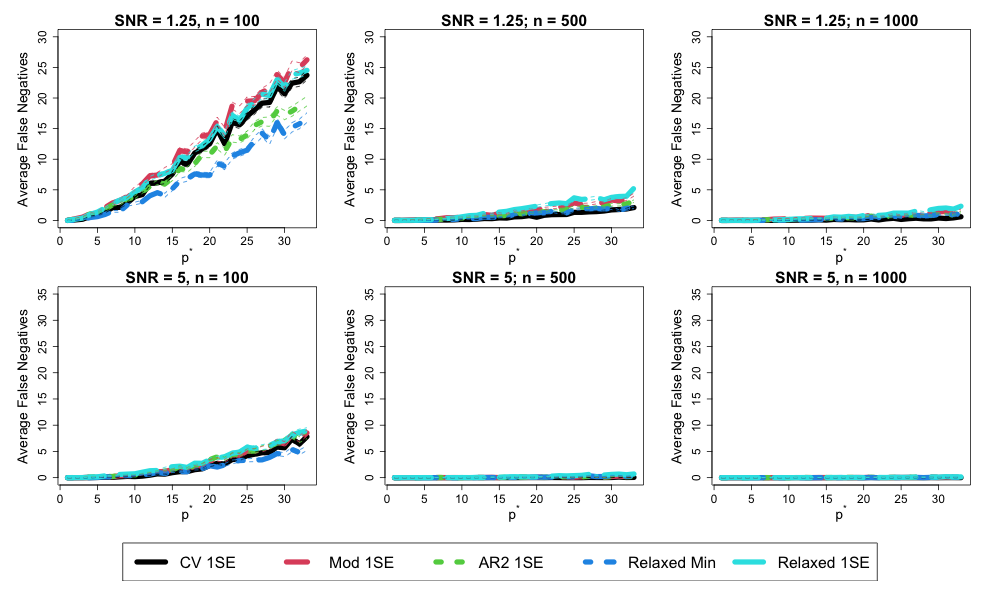}
	\caption{Average false negatives from 100 replications of simulated data with $p = 100$ for both SNR $= 1.25$ and SNR $= 5$ independent Laplacian errors. Thin dotted lines represent the mean $\pm$ one standard error.}
	\label{fig:SimPlotLapFN}
\end{figure}

\begin{figure}[h]
	\centering
	\includegraphics[width = 0.75 \textwidth]{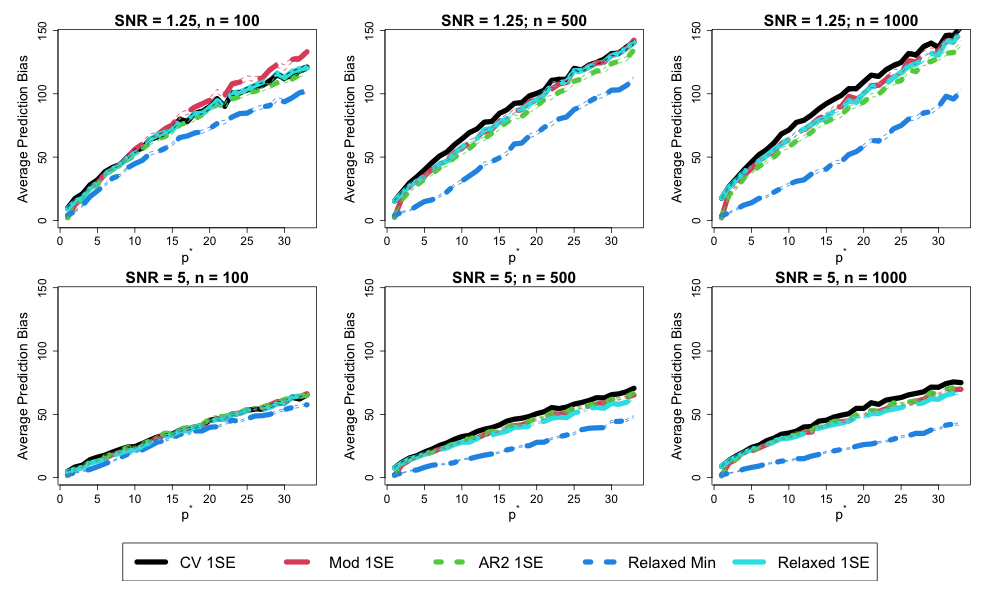}
	\caption{Average prediction bias from 100 replications of simulated data with $p = 100$ for both SNR $= 1.25$ and SNR $= 5$ independent Laplacian errors. Thin dotted lines represent the mean $\pm$ one standard error.}
	\label{fig:SimPlotLapPB}
\end{figure}

A final simulation study replicates the independent predictor setting found in the main document but elements of $\boldsymbol{\epsilon}$ are now independently drawn from a Laplace distribution with mean zero and scale parameter selected to maintain a consistent signal-to-noise ratio for each new data set. Tables \ref{fig:SimPlotLapHD} through \ref{fig:SimPlotLapPB} give the average Hamming Distance, False Positives, False Negatives, and Prediction Bias from these simulations.

\FloatBarrier
\section*{C. EMG Application Approximations and Results}
The model given in the main document does not involve raw data, so some manipulations are required in order to perform estimation. The response, velocity ($y_i$), must be estimated from hand position, $z_i$ and covariate functions in the model must be created from discrete observations. These tasks are performed exactly as described in \cite{SAFE}. 

Furthermore, the model also requires approximations for $\gamma_j$. We represent all $\gamma_j$ using: $\gamma_j(t,z) \approx \sum_{l=1}^L \sum_{m=1}^M \omega_l(t) \tau_m(z) \beta_{jlm}$ where $\omega_l(t)$ and $\tau_m(z)$ are the same univariate basis functions for all $\gamma_j$, and the $\beta_{jlm}$'s are basis coefficients that must be estimated. In matrix form, we have $\gamma_j(t,z) \approx \boldsymbol{\omega}^T(t) \V{B}_j \boldsymbol{\tau}(z)$ where $\boldsymbol{\omega}^T(t) = \big( \omega_1(t),  ...,  \omega_L(t) \big)$, $\boldsymbol{\tau}^T(z) = \big( \tau_1(z),  ...,  \tau_M(z) \big)$, and $\V{B}_j = \big( \beta_{jlm}\big)$ is the $L \times M$ coefficient matrix.

The integral in the model is approximated:
\begin{equation}
	\int_{-\delta}^0 X_{ij}(t)\gamma_j(t, z_i)dt \approx \bigg( \sum_{k=-\delta}^0 x_{(i+k)j} \boldsymbol{\omega(k)^T} \bigg) \V{B}_j \boldsymbol{\tau}(z_i) = \V{X}_{ij\omega}^T \V{B}_j \boldsymbol{\tau}(z_i).
\end{equation}
As noted in the main document, we use the penalized least squares approach to functional estimation proposed in \cite{Gertheiss2013} and adapted in \cite{SAFE}.

Tables \ref{tab:EMGFull2} through \ref{tab:EMGFull5} show the results from multiple stages of adaptive weighting, the SAFE procedure. After only one round, the differences in tuning parameter selection approaches disappear, giving the same results for APE, AR2, and Mod CV. The differences reappear again after three stages, but only for AR2 CV, which still overselects by a single coefficient for FR2, even after a fourth stage.

\begin{table}[ht]
\centering
\caption{Variable selection results for EMG finger movements after one stage of adaptive weighting. FP indicates the total number of false positives in the model and Size is the total number of EMG signals contained in the model.} 
\begin{tabular}{rr|ccccccc}
  \hline
 &  & FC1 & FC2 & FC3 & FR1 & FR2 & FR3 & Mean \\ 
  \hline
\multirow{3}{*}{APE} & FP & 0 & 0 & 0 & 0 & 1 & 1 & 0.33 \\ 
   & Size & 2 & 2 & 2 & 2 & 4 & 3 & 2.5 \\ 
  \multirow{3}{*}{AR2} & FP & 0 & 0 & 1 & 0 & 1 & 0 & 0.33 \\ 
   & Size & 2 & 2 & 3 & 2 & 4 & 2 & 2.5 \\ 
  \multirow{3}{*}{Mod} & FP & 0 & 0 & 0 & 0 & 1 & 1 & 0.33 \\ 
   & Size & 2 & 2 & 2 & 2 & 4 & 3 & 2.5 \\  
   \hline
\end{tabular}
\label{tab:EMGFull2}
\end{table}

\begin{table}[ht]
\centering
\caption{Variable selection results for EMG finger movements after two stages of adaptive weighting. FP indicates the total number of false positives in the model and Size is the total number of EMG signals contained in the model..} 
\begin{tabular}{rr|ccccccc}
  \hline
 &  & FC1 & FC2 & FC3 & FR1 & FR2 & FR3 & Mean \\ 
  \hline
\multirow{2}{*}{APE} & FP & 0 & 0 & 0 & 0 & 1 & 0 & 0.17 \\ 
   & Size & 2 & 2 & 2 & 2 & 4 & 2 & 2.33 \\ 
  \multirow{2}{*}{AR2} & FP & 0 & 0 & 0 & 0 & 1 & 0 & 0.17 \\ 
   & Size & 2 & 2 & 2 & 2 & 4 & 2 & 2.33 \\ 
  \multirow{2}{*}{Mod} & FP & 0 & 0 & 0 & 0 & 1 & 0 & 0.17 \\ 
   & Size & 2 & 2 & 2 & 2 & 4 & 2 & 2.33 \\ 
   \hline
\end{tabular}
\label{tab:EMGFull3}
\end{table}

\begin{table}[ht]
\centering
\caption{Variable selection results for EMG finger movements after three stages of adaptive weighting. FP indicates the total number of false positives in the model and Size is the total number of EMG signals contained in the model.} 
\begin{tabular}{rr|ccccccc}
  \hline
 &  & FC1 & FC2 & FC3 & FR1 & FR2 & FR3 & Mean \\ 
  \hline
\multirow{2}{*}{APE} & FP & 0 & 0 & 0 & 0 & 0 & 0 & 0 \\ 
   & Size & 2 & 2 & 2 & 2 & 3 & 2 & 2.17 \\ 
  \multirow{2}{*}{AR2}  & FP & 0 & 0 & 0 & 0 &  1 & 0 & 0.17 \\ 
   & Size & 2 & 2 & 2 & 2 &  4 & 2 & 2.33 \\ 
  \multirow{2}{*}{Mod}  & FP & 0 & 0 & 0 & 0 & 0 & 0 & 0 \\ 
   & Size & 2 & 2 & 2 & 2 & 3 & 2 & 2.17 \\ 
   \hline
\end{tabular}
\label{tab:EMGFull4}
\end{table}
\begin{table}[ht]
\centering
\caption{Variable selection results for EMG finger movements after the four stages of adaptive weighting recommended by \cite{SAFE}. FP indicates the total number of false positives in the model and Size is the total number of EMG signals contained in the model.}
\begin{tabular}{rr|ccccccc}
  \hline
 &  & FC1 & FC2 & FC3 & FR1 & FR2 & FR3 & Mean \\ 
  \hline
\multirow{2}{*}{APE} & FP & 0 & 0 & 0 & 0 & 0 & 0 & 0 \\ 
   & Size & 2 & 2 & 2 & 2 & 3 & 2 & 2.17 \\ 
  \multirow{2}{*}{AR2}  & FP & 0 & 0 & 0 & 0 & 1 & 0 & 0.17 \\ 
   & Size & 2 & 2 & 2 & 2 &  4 & 2 & 2.33 \\ 
  \multirow{2}{*}{Mod}  & FP & 0 & 0 & 0 & 0 & 0 & 0 & 0 \\ 
   & Size & 2 & 2 & 2 & 2 & 3 & 2 & 2.17 \\ 
   \hline
\end{tabular}
\label{tab:EMGFull5}
\end{table}

 \bibliographystyle{elsarticle-num}
 \bibliography{Refs}





\end{document}